\documentclass{aa}  

\usepackage{graphicx}
\usepackage{txfonts}
\usepackage{tablefootnote}

 \usepackage{lscape}
 \usepackage[dvipsnames]{xcolor}
\usepackage{txfonts}
\usepackage{amsmath}
\usepackage{amssymb}

\newcommand{\LO}{$\mathcal{L}_{0}$ }
\newcommand{\D}{$\mathcal{D}$ }
\begin{document}

   \title{Integral field spectroscopy supports atmospheric optics to reveal the finite outer scale of the turbulence}

\titlerunning{IFS to reveal the outer scale of the turbulence}
   \author{B. García-Lorenzo,
          \inst{1,2}
          \and D. Esparza-Arredondo,
          \inst{1,2}
          \and J.A. Acosta-Pulido,
          \inst{1,2}
          \and J.A. Castro-Almaz\'an
          \inst{1,2}
          }

    \institute{Instituto de Astrof\'isica de Canarias, C/ V\'ia L\'actea s/n, E-38205 La Laguna, Tenerife, Spain\\
              \email{begona.garcia@iac.es}
         \and
             Departamento de Astrof\'isica, Universidad de La Laguna, E-38200 La Laguna, Tenerife, Spain
             }

   \date{Received October 24, 2023; accepted , 2024}

  \abstract
   {The spatial coherence wavefront outer scale (\LO) characterizes the size of the largest turbulence eddies in Earth's atmosphere, determining low spatial frequency perturbations in the wavefront of the light captured by ground-based telescopes. Advances in adaptive optics (AO) techniques designed to compensate for atmospheric turbulence emphasize the crucial role of this parameter for the next generation of large telescopes.}
   {The motivation of this work is to introduce a novel technique for estimating \LO from seeing-limited integral field spectroscopic (IFS) data. This approach is based on the impact of a finite \LO on the light collected by the pupil entrance of a ground-based telescope.}
   {We take advantage of the homogeneity of IFS observations to generate band filter images spanning a wide wavelength range, enabling the assessment of image quality (IQ) at the telescope's focal plane. Comparing the measured wavelength-dependent IQ variation with predictions from \cite{2002Tokovinin} analytical approach offers valuable insights into the prevailing \LO parameter during the observations. We applied the proposed technique to observations from the Multi Unit Spectroscopic Explorer (MUSE) in the Wide Field Mode obtained at the Paranal Observatory.}
   {Our analysis successfully validates Tokovinin's analytical expression, which combines the seeing ($\epsilon_{0}$) and the \LO parameters, to predict the IQ variations with the wavelength in ground-based astronomical data. However, we observed some discrepancies between the measured and predictions of the IQ that are analyzed in terms of uncertainties in the estimated $\epsilon_{0}$ and dome-induced turbulence contributions.}
   {This work constitutes the empirical validation of the analytical expression for estimating IQ at the focal plane of ground-based telescopes under specific $\epsilon_{0}$ and finite \LO conditions. Additionally, we provide a simple methodology to characterize the \LO and dome-seeing ($\epsilon_{\mathrm{dome}}$) as by-products of IFS observations routinely conducted at major ground-based astronomical observatories.
}


   \maketitle
%
\nolinenumbers 
\section{Introduction}

Earth's atmospheric turbulence significantly affects light propagation through it, distorting both intensities and wavefronts. Consequently, ground-based astronomical observations often produce blurred images compared to those obtained from space using identical telescopes and instruments.

Atmospheric turbulence results from stochastic fluctuations in the refractive index attributed to temperature variations. The strength of this turbulence can be quantified using the refractive index structure parameter (C$_n^2$), which is a function of the position. The Kolmogorov model provides a satisfactory description of the statistical properties of atmospheric turbulence, assuming that it is homogeneous and isotropic \citep[e.g.,][]{1981Roddier}. This model pictures the turbulence as a cascade of energy, following a power-law distribution, from large- to small-scale turbulence eddies until dissipation. The Kolmogorov model applies only to the inertial range between inner ($\it{l}_{0}$) and outer ($\it{L}_{0}$) scales determined by the smallest and largest sizes of turbulence eddies. This model holds for any atmospheric turbulence layer that light passes through on its way to the pupil entrance of ground-based telescopes, each characterized by a C$_n^2$ and an $\it{L}_{0}$ depending on the local conditions at each layer height (h). Hence, the wavefront of the light reaching the telescope pupil will suffer perturbations from the distinct atmospheric turbulence layers. We can define the spatial coherence wavefront outer scale (\LO) as an equivalent outer scale determining the image quality (IQ) of any observation taken with that telescope \citep{1990Borgnino}. \LO is related to the $\it{L}_{0}$ of the atmospheric turbulence layers as follows \citep[e.g.,][]{2004aAbahamid}:

\begin{equation}
    \mathcal{L}_{0}  = \left( \frac{\int \it{L}_{0}(h)^{-1/3} C_{N}^{2}(h) dh}{\int C_{N}^{2}(h) dh} \right)^{-3} 
    \label{outer_scale}
\end{equation}

\noindent The \LO is a parameter independent of the wavelength of the observed light, representing a size and typically measured in meters. In many applications, equations are simplified by assuming infinite outer scales. Under this assumption, the full width at half maximum (FWHM) of the point spread function (PSF) in a long-exposure (LE) image, captured with a ground-based telescope limited by atmospheric turbulence, commonly referred to as the seeing-limited IQ ($\epsilon_{LE}$), can be expressed in terms of the strength of atmospheric turbulence as follows:

\begin{equation}    \mathrm{\epsilon_{LE}(\lambda) \approx \epsilon_{0}} (\lambda)  = \frac{0.976 \lambda}{r_{0}(\lambda)}
    \label{seeing}
\end{equation}

\noindent where $\epsilon_{0}$ is the seeing, $\lambda$ is the wavelength, and r${_0}$ is the Fried parameter, defined in terms of the refractive index structure constant profile with height (C$_{N}^{2}$(h)) as \citep{1965Fried}:

\begin{equation}
    \mathrm{ r{_0}} (\lambda)  = 0.185 \lambda^{6/5} (\cos{\zeta})^{3/5} \left( \int C_{N}^{2}(h) dh \right)^{-3/5} 
    \label{fried}
\end{equation}

\noindent with $\zeta$ being the observing zenith angle. We note that Eq. \ref{fried} implies that:
\begin{equation}
    \mathrm{\epsilon_{LE}(\lambda) \approx \epsilon_{0}} (\lambda)  \propto  \lambda^{-1/5}
    \label{seeing_p}
\end{equation}

However, interferometric measurements and the growing size of telescope pupils evidence the limitations of applying the Kolmogorov model for large distances (i.e., \LO$\approx\infty$), showing deviations from Eq. \ref{seeing_p} that reveal the finite nature of the \LO. Hence, with the increasing diameter of ground-based telescopes and adaptive optics (AO) developments, the relevance of \LO has increased in the last few years \citep[e.g.,][]{2020Fusco}. For example, the size of \LO strongly impacts the low-order modes in AO applications, showing a significant tilt attenuation even for pupil sizes much smaller than \LO. The validity of the Kolmogorov theory beyond the inertial range can be assessed using empirical models. These models describe the spectrum of the refractive-index fluctuations by means of the \LO parameter based on general physical considerations \citep[e.g.,][]{1995Voitsekhovich}. A widely used is the von Kármán model, where the \LO corresponds to the distance at which the structure-function of the phase fluctuations saturates \citep[see e.g., Fig. 2 in][]{1995Voitsekhovich}. 

Different techniques enable the determination of the magnitude of \LO either directly \citep[e.g., GSM,][]{2000Ziad} or indirectly \citep[e.g., from AO telemetry and applying Eq. \ref{outer_scale},][]{2017Guesalaga}, and both provide comparable results \citep[see][for a review]{2016Ziad}. At a telescope focal plane, the effect of a finite \LO is to smooth low spatial frequency perturbations, resulting in a reduction of image motion compared to the expected for an infinite \LO \citep[e.g.,][]{1991Winker}. As a result, the actual IQ improves over the $\epsilon_{0}$ by a factor that depends on both the \LO and the $\epsilon_{0}$. \cite{2002Tokovinin} proposed the following first-order approximation to $\epsilon_{LE}$:

 \begin{equation}
    \epsilon_{LE}(\lambda) \approx \epsilon_{0}(\lambda) \sqrt{1-2.183 \left( \frac{\mathrm{r_{0}} (\lambda)}{\mathcal{L}_{0}} \right)^{0.356}}
    \label{eq:ELE_lambda}
\end{equation}

\noindent This approach results from an analytical simplification of the turbulence statistics using the von Kármán model, and it seems to work with an uncertainty of about 1\% for $\mathcal{L}_{0}$/r${_0}$ > 20. Equation \ref{eq:ELE_lambda} was refined later to account for the pupil size relative to \LO, as follows \citep{ManualESO}.:

 \begin{equation}
\epsilon_{LE}(\lambda, \mathcal{D})  \approx \epsilon_{0}(\lambda) \sqrt{1-2.183\left( 1-\frac{1}{1 + 300 \times \frac{\mathcal{D}}{\mathcal{L}_{0}}} \right) \left( \frac{\mathrm{r_{0}} (\lambda)}{\mathcal{L}_{0}} \right)^{0.356}} 
    \label{eq:ELE_lambda_D}
\end{equation}

\noindent where \D is the telescope pupil diameter. Previous studies have adopted these analytical expressions to analyze the deviation of IQ from $\epsilon_{0}$ using numerical simulations \citep{2010Martinez} and through simultaneous visible and infrared images taken with different instruments \citep{2007Tokovinin}. Moreover, \cite{2010Floyd} derived an average turbulence \LO of 25 m for Las Campanas observatory by applying Eq. \ref{eq:ELE_lambda} to a large series of IQ measurements from band filter images of stars. Currently, first-class astronomical observatories commonly use Eq. \ref{eq:ELE_lambda} (or \ref{eq:ELE_lambda_D}, depending on the telescope) to predict the IQ of their queue observations using the $\epsilon_{0}$ provided by Differential Image Motion Monitors \citep[DIMMs,][]{1990Sarazin}, and assuming a typical \LO for the site. However, to our knowledge, the actual $\epsilon_{LE}$ behavior with wavelength has not been verified empirically to follow such analytical approaches.

In this work, we take advantage of integral field spectroscopic observations limited by atmospheric turbulence to track the wavelength behavior of the IQ at the telescope focal plane. We compare our findings with the analytical model predictions proposed by \cite{2002Tokovinin} to determine the prevailing \LO during observations. We applied the methodology to data from the Multi Unit Spectroscopic Explorer (MUSE\footnote{https://www.eso.org/sci/facilities/develop/instruments/muse.html}) at the Paranal Observatory, empirically validating the analytical model predictions with wavelength. Thanks to the spectral range covered by MUSE, our analysis reveals the prevailing $\epsilon_{0}$ and \LO parameters best matching the IQ measurements. We explore different conditions, including those influenced by dome-induced turbulence.

The structure of this article is as follows. In Sect. \S\ref{telescopes_turbulence}, we discuss the factors influencing IQ at the focal plane of a ground-based telescope, with special attention to dome-induced turbulence. Section \S\ref{IFS_section} introduces integral field spectroscopy (IFS) as a valuable technique for assessing the \LO during observations. In Sect. \S\ref{application_MUSE}, we apply the proposed methodology to data cubes observed with MUSE, analyzing different scenarios. Finally, Sect. \S\ref{conclusions} summarizes the main findings and conclusions.


\section{Telescopes and turbulence contributors}
\label{telescopes_turbulence}
The IQ at the focal plane of an instrument attached to a ground-based telescope results from various factors, including the telescope/instrument optics and stability, and the air turbulence conditions during observations. Through this work, we use the FWHM of the PSF to quantify the IQ, although other metrics might also be employed, such as the Strehl ratio in AO applications. The design and operation of modern telescopes try to minimize optical and stability aspects, while advanced AO systems primarily compensate for atmospheric turbulence and small optical aberrations. However, air turbulence will depend on environmental conditions both inside and outside of the telescope enclosure. We identify the external turbulence to the atmospheric turbulence, assuming Eqs. \ref{eq:ELE_lambda} and \ref{eq:ELE_lambda_D} (see Appendix \ref{appenIQ} for a comparison of predictions from these equations) describe its effect on IQ. The influence of turbulence within the telescope enclosure is complex and challenging to parameterize. We explore it further in the following subsection. Hereafter, we assume an optimal optical design of the telescope+instrument system providing negligible image degradation in comparison to air turbulence.

\subsection{The dome-seeing}
\label{dome-seeing}

Telescope enclosures protect from external factors such as adverse weather conditions, but they can also induce turbulence that degrades the quality of the images, the so-called dome-seeing ($\epsilon_{\mathrm{dome}}$). While $\epsilon_{\mathrm{dome}}$ is recognized as a variable source of IQ degradation, only recent efforts actively promote its monitoring, even aiming for real-time quantification \citep{2019Lai, 2021Stubbs, 2023Osborn}. However, there remains a limited quantitative understanding of $\epsilon_{\mathrm{dome}}$, particularly concerning its wavelength behavior.

The non-Kolmogorov regime of the dome turbulence combines thermal inversion and convection, and its strength will depend on factors such as the dome design, the inside-outside temperature gradient, the temperature of the primary mirror surface, the outside wind, the orientation of the dome window, etc \citep{Woolf1979, 1987Henry, 2023Munro}. We can define $\tilde{r}_{0}$ as an equivalent parameter to Fried's parameter, which will be a power law of the wavelength and the non-Kolmogorov refractive index structure constant ($\tilde{C}_{N}^{2}$) for the dome turbulence:

\begin{equation}
    \mathrm{\tilde{r}{_0}} (\lambda)  = \left( \frac{16.7}{\cos{\zeta}\times\lambda^{2}}\int \tilde{C}_{N}^{2} dh \right)^{1/(2-\gamma)} 
    \label{fried_general}
\end{equation}

\noindent with the power law index in the range \(3 < \gamma < 4\) \citep{2007Toselli}. We note that Eq. \ref{fried_general} reduces to the Kolmogorov case (i.e., Eq. \ref{fried}) when \(\gamma = \frac{11}{3}\approx3.67\). Assuming that the $\epsilon_{\mathrm{dome}}$ varies in proportion to the relationship between $\epsilon_{0}$ and r$_{0}$ in Eq. \ref{seeing}, and taking into account Eq. \ref{fried_general}:

\begin{equation}
    \epsilon_{\mathrm{dome}} (\lambda) \propto \frac{\lambda}{\mathrm{\tilde{r}{_0}} (\lambda)}  \ \stackrel{(\ref{fried_general})}{\implies} \epsilon_{\mathrm{dome}} (\lambda) \propto \lambda^{(\gamma-4)/(\gamma-2)} 
    \label{dome_seeing_general}
\end{equation}

Considering the dome's contribution to turbulence as an additional atmospheric layer, we can linearly combine the dome-induced turbulence using the corresponding refractive index structure constants. Nevertheless, the resulting image blur from dome-turbulence could only be added to atmospheric $\epsilon_{0}$ using a 5/3-power law within the constraints of the Kolmogorov regime \citep{Woolf1979, 1991Racine}. However, the effect of $\epsilon_{\mathrm{dome}}$ on the focal plane is typically represented as a Gaussian blur \cite[e.g.,][]{2018Bustos}, making a quadratic contribution to the overall IQ:

\begin{equation}
    [\mathrm{IQ}(\lambda)]^{2} =  [\epsilon_{\mathrm{LE}}(\lambda)]^{2} + [\epsilon_{\mathrm{dome} }(\lambda)]^{2}
    \label{IQ_general}
\end{equation}

\begin{figure*}
\begin{center}
\includegraphics[trim={0cm 2.cm 1cm 1.75cm},clip,width=9.15cm, angle=180]{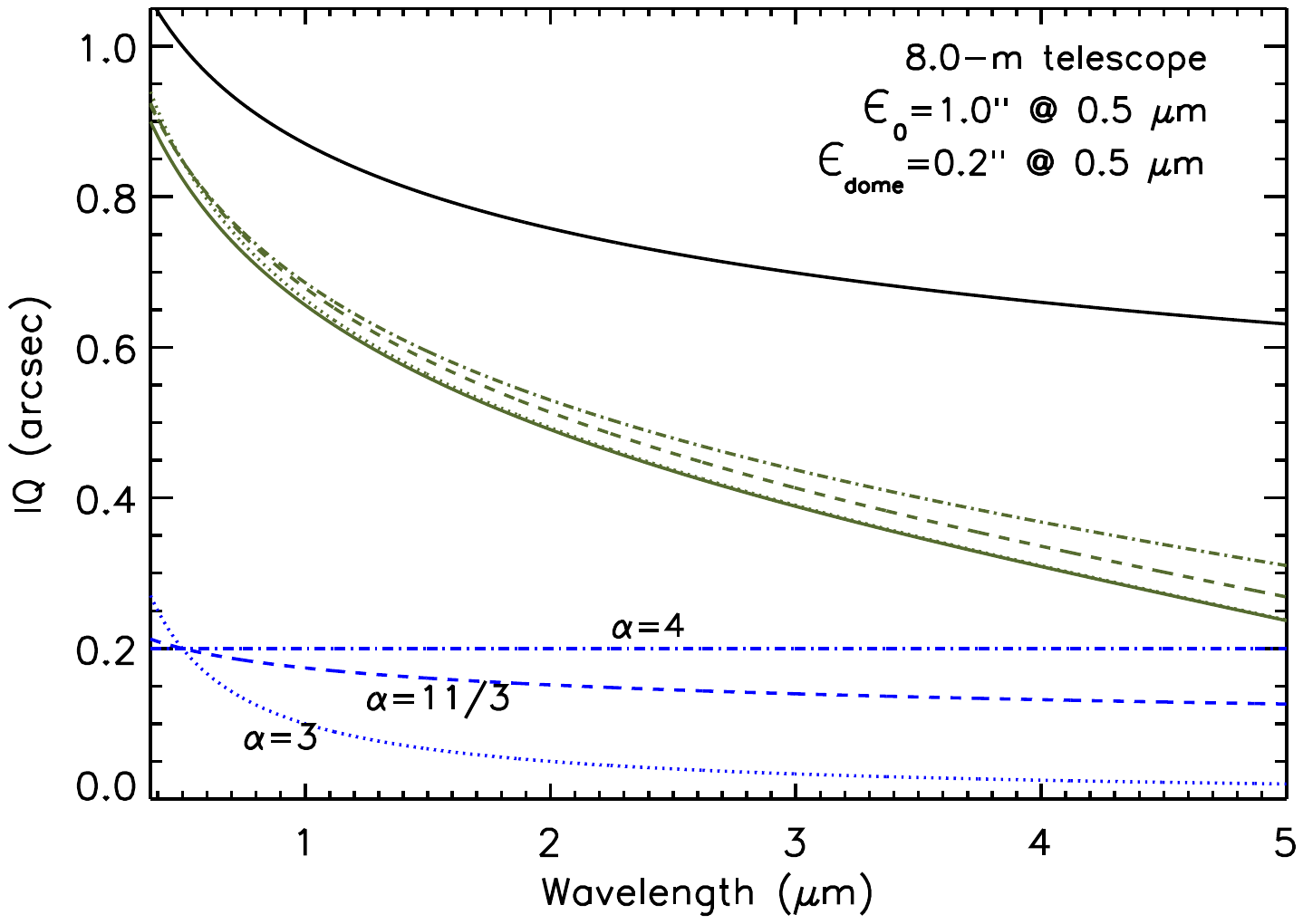}
\includegraphics[trim={0cm 2.cm 1cm 1.75cm},clip,width=9.15cm, angle=180]{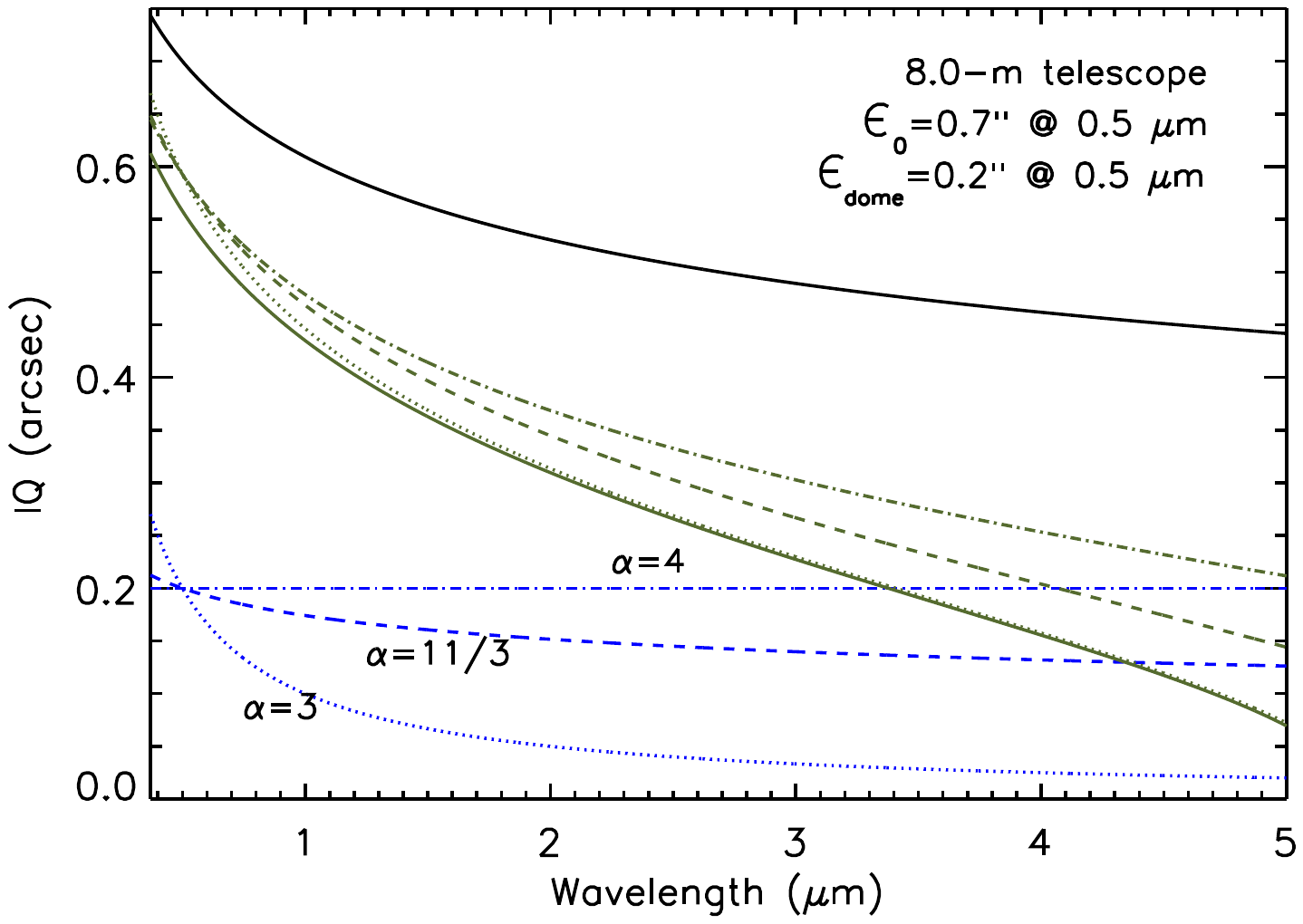}
\includegraphics[trim={0cm 2.cm 1cm 1.75cm},clip,width=9.15cm, angle=180]{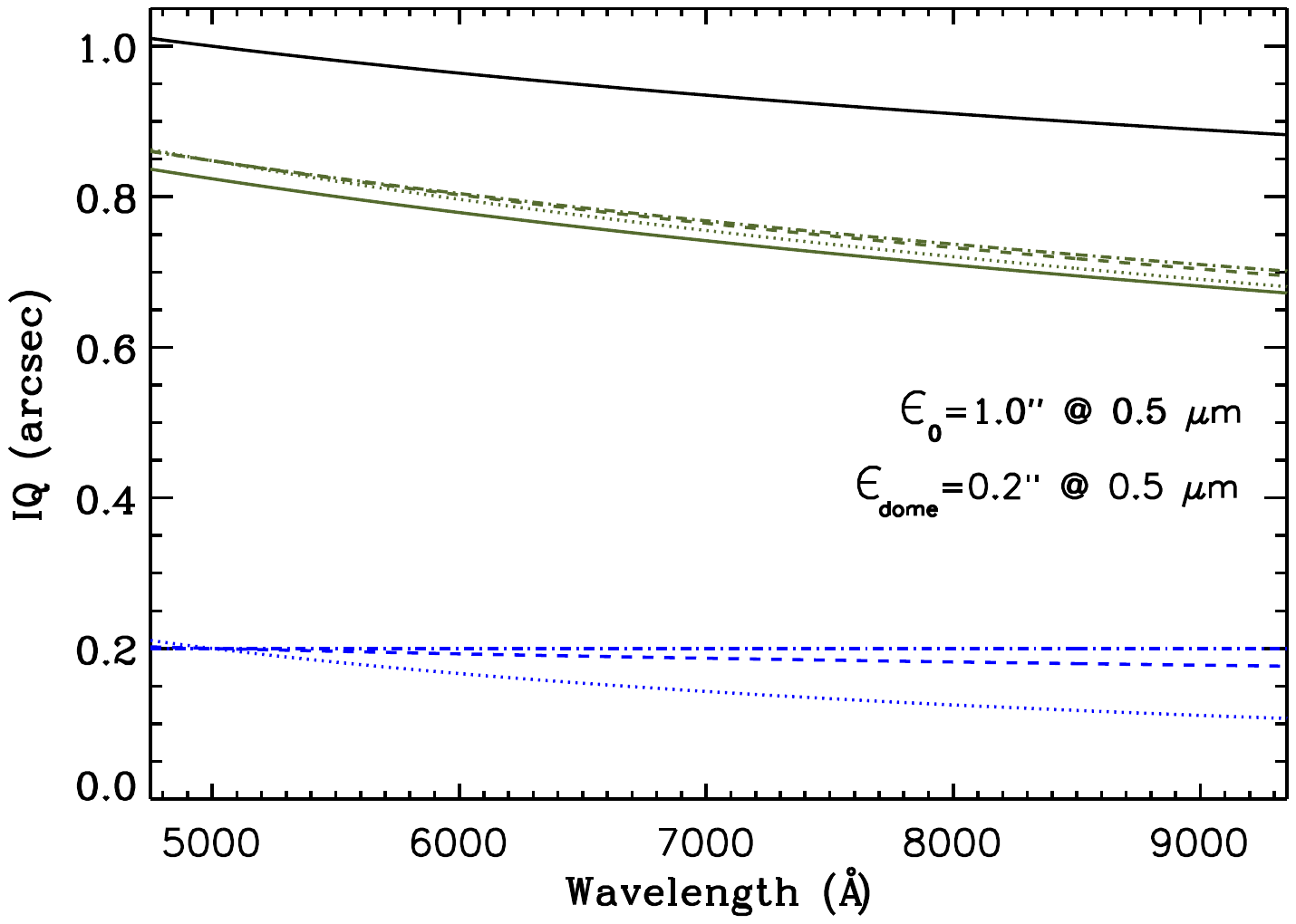}
\includegraphics[trim={0cm 2.cm 1cm 1.75cm},clip,width=9.15cm, angle=180]{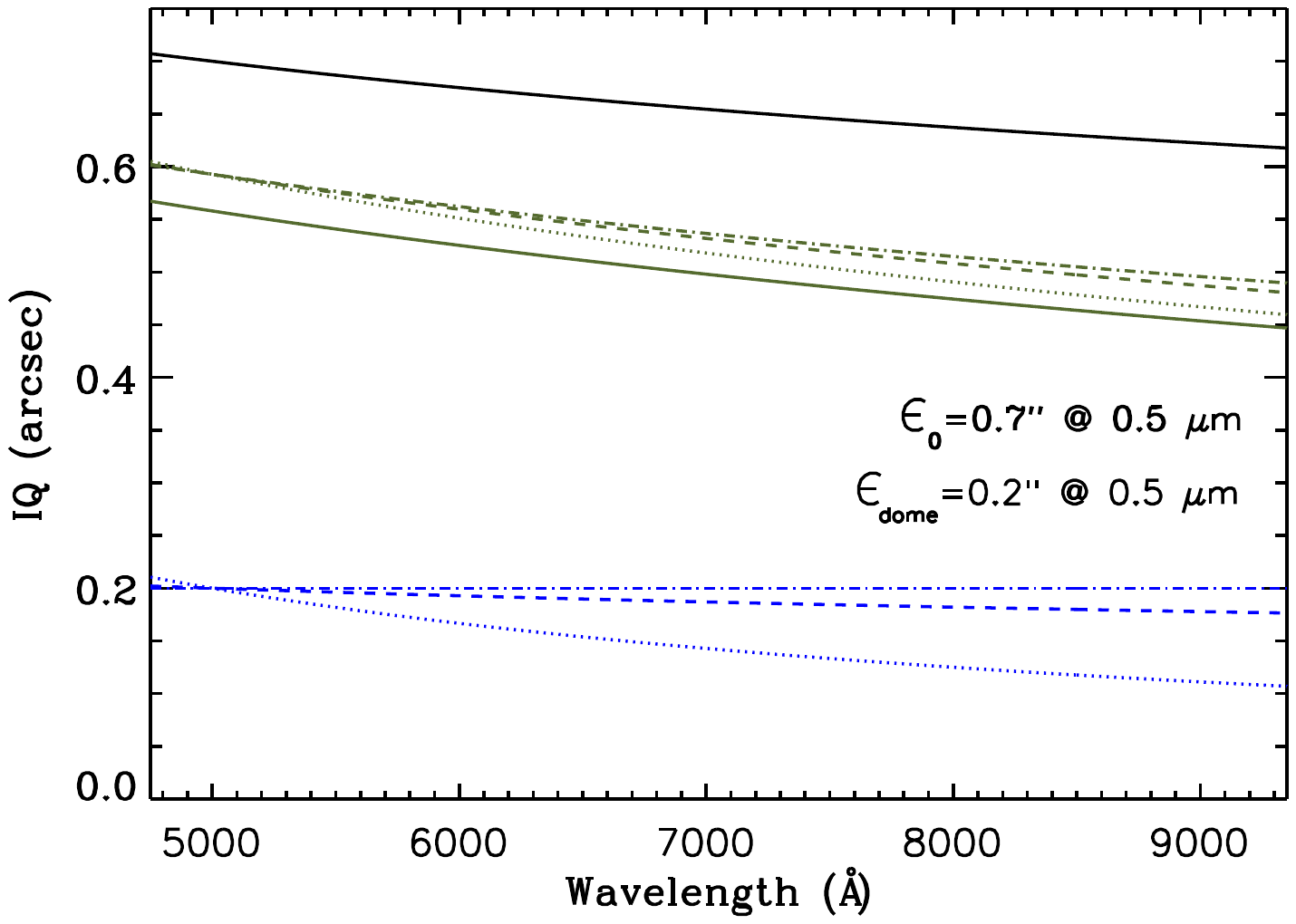}

 \caption{Estimation of the IQ at the focal plane of an 8-m telescope as a function of the wavelength under a $\epsilon_{0}$ of 1.0 (left) and 0.7 (right) arcsecs at 5000 \AA, considering four different wavelength behaviors of the dome-turbulence (blue lines) contribution: (1) $\epsilon_{\mathrm{dome}}$=0 (solid green line), and (2, 3, and 4) $\epsilon_{\mathrm{dome}}$=0.2 arcsecs at 5000 \AA, with a power-law wavelength dependence of index $\gamma$=3 (2, dotted lines), $\gamma$=11/3 (3, dashed lines), and $\gamma$=4 (4, dotted-dashed lines). We assumed a typical \LO value of 22 m. The black solid curve follows the $\epsilon_{0}$ wavelength dependence ($\epsilon_{0} \propto \lambda^{-1/5}$). The upper panels depict the general spectral range covered by instruments in ground-based optical-infrared observatories. The lower panels specifically showcase the spectral range of the MUSE integral field spectrograph, the selected instrument for implementing the proposed methodology on real data.}
 
\label{dome_seeing_lambda}
\end{center}
\end{figure*}

\noindent  We note that we have considered a negligible degradation of the IQ due to instrumental factors, adopting such a quadratic approach hereafter. Figure \ref{dome_seeing_lambda} presents examples of the quadratic contribution to IQ for the extreme values of the power law $\gamma$-index of the non-Kolmogorov dome-induced turbulence, assuming a $\epsilon_{\mathrm{dome}}$ of 0.2 arcsecs at 5000 \AA. We also include the $\gamma=\frac{11}{3}$ corresponding to the Kolmogorov regime. The dome contribution remains constant with wavelength as the $\gamma$ index approaches 4, whereas it becomes inversely proportional to wavelength as $\gamma$ approaches 3. As shown in Fig. \ref{dome_seeing_lambda}, the IQ will consistently exceed that associated with atmospheric turbulence at any wavelength. Furthermore, its wavelength dependence deviates from the natural atmospheric turbulence behavior (described by Eqs. \ref{eq:ELE_lambda} and \ref{eq:ELE_lambda_D}), which is particularly noticeable for the smallest $\gamma$ value. 

Large telescopes invest significant efforts to mitigate the occurrence of dome-induced turbulence (i.e., $\epsilon_{\mathrm{dome}} \approx 0$), thereby ensuring seeing-limited conditions, and consequently, IQ $\approx \epsilon_{\mathrm{LE}}$. Then, we can reasonably assume negligible or minimal $\epsilon_{\mathrm{dome}}$ contributions at any wavelength for large telescopes. Nonetheless, the monitoring of $\epsilon_{\mathrm{dome}}$ at distinct wavelengths will largely benefit the characterization of the actual IQ at the focal plane.

\section{Using Integral Field Spectroscopy to reveal \LO}
\label{IFS_section}

IFS is a powerful technique that allows for the simultaneous recording of spatial and spectral information over a field. Since its first applications in astronomy in the 1980s and 1990s, IFS rapidly grew in popularity during the first decade of the 21st century. IFS has become a widely used technique in both ground-based and space-borne astronomy enabling many scientific programs, from studies in our solar system to high-redshifted objects of cosmological interest. Some of the most advanced IFS instruments in operation at the largest ground-based telescopes include the MUSE spectrograph on the Very Large Telescope (VLT) in Chile, the Gemini Multi-Object Spectrograph (GMOS\footnote{http://www.gemini.edu/instrumentation/gmos/}) on the Gemini Observatory in Chile and Hawaii, the Keck Cosmic Web Imager (KCWI\footnote{https://www2.keck.hawaii.edu/inst/kcwi/}) on the Keck Observatory in Hawaii, and the {\it Multi-Espectrógrafo en GTC de Alta Resolución para Astronomía} (MEGARA\footnote{http://www.gtc.iac.es/instruments/megara/megara.php}) on the Gran Telescopio Canarias on the {\it Roque de los Muchachos} Observatory in La Palma island (Spain).

The basic instrumental concept behind IFS is to spatially split the telescope focal plane into several sub-apertures, rebinning them to create a pseudo-slit to feed on long-slit spectrographs. Different instrumental approaches allow achieving IFS \citep[see e.g., Fig. 1 in][]{2006Allington-smith}, all of them resulting in obtaining the flux density (F) over a sky area in three dimensions (3D): right ascension ($\alpha$), declination ($\delta$), and wavelength ($\lambda$) (i.e., F($\alpha$,$\delta$,$\lambda$)), under the same instrumental and atmospheric conditions. For this work, the F($\alpha$,$\delta$,$\lambda$) data cube should be viewed as a collection of narrow-band filter images at different wavelengths obtained simultaneously, ensuring data homogeneity \citep[see e.g., Fig. 2.1 in][]{Harrison2016}. This set of images enables us to empirically assess the variation of IQ with wavelength by characterizing the PSF using any point-like source in the observed field of view \citep[for the initial exploration of this concept see][]{2023Garcia-Lorenzo}. Fitting the $\epsilon_{LE}$($\lambda$) to the curve described by Eq. \ref{eq:ELE_lambda} (or \ref{eq:ELE_lambda_D} when applicable) will yield the \LO and $\epsilon_{0}$ parameters during observations.

Hence, we recognize IFS instruments operating under atmospheric turbulence-limited conditions as valuable tools for validating the first-order analytical approximation to estimate $\epsilon_{LE}$($\lambda$) and for assessing the prevailing \LO during IFS observations. For that, low-spectral resolution IFS covering a broad spectral range will be preferable to a high-spectral resolution IFS sampling a short wavelength range.

The F($\alpha$,$\delta$,$\lambda$) data cube will be characterized by a certain number of spatial and spectral pixels (spaxels) of $\Delta\alpha$, $\Delta\delta$, and $\Delta\lambda$ sizes. $\Delta\alpha$ and $\Delta\delta$ should achieve Nyquist or better sampling for the typical FWHM of the atmospheric turbulence-limited PSF at the observing site (i.e., $\epsilon_{LE}$($\lambda$) > 2$\Delta\alpha$ \& 2$\Delta\delta$). Along the spectral direction, the critical point will be the detection of point-like sources with a signal-to-noise ratio (S/N) larger than a threshold to characterize the PSF properly. To achieve a minimum S/N threshold, IFS enables the easy recovery of band-filter images (I$_{\mathrm{BW}_j}$) by summing N consecutive slices of the data cube within a chosen wavelength range, as follows:


 \begin{equation}
I_{{\mathrm{BW}j}}(\alpha,\delta) = \sum_{i-N}^{i+N}F(\alpha,\delta,\lambda_{i}), \hspace{0.15cm} \\ |\lambda_{i}-\lambda_{j}| \leq \frac{N\Delta\lambda}{2}
    \label{eq:band}
\end{equation}

\noindent where $\lambda_{j}$ and BW$_{j}$=N$\Delta\lambda$ are the central wavelength and width of the desired filter, respectively. However, the final IQ of the recovered I$_{{\mathrm{WB}_j}}$ image may differ from the actual IQ of the particular slide image F($\alpha,\delta,\lambda_{j}$) due to instrumental behavior, the observed target characteristics, or a combination of both factors (see Sect. \S\ref{filter-impact} in appendix \ref{appenB}). It is important to note that we assume a filter with constant transmission within the selected wavelength band. In any different case, Eq. \ref{eq:band} should be multiplied by the specific transmission function. The characteristics of the point-like source will determine the maximum number and wavelength distribution of I$_{\mathrm{BW}}$. In some cases, the number of bands can equal the total spectral channels in the IFS data cube, such as when a bright star allows for characterizing the PSF at each channel in the IFS field of view. Conversely, the number of bands can be few (or even just one), as in IFS observations of active galaxies, where the PSF can be imaged in isolation from the host galaxy using specific spectral lines unique to the point-like active nucleus \citep[e.g.,][]{2016Husemann, 2024Esparzaarredondo}.

Finally, to determine the optimal \LO and $\epsilon_{0}$ values that best fit the IQ($\lambda$) measurements, we might fit the curve described by Eq. \ref{eq:ELE_lambda} (or \ref{eq:ELE_lambda_D} when applicable). Different fitting approaches may be considered based on the intended use of the derived parameters and the specific characteristics of the measurements, including the number of empirical data points. It is recommended to experiment with various methods and assess the fit quality using appropriate metrics, such as the coefficient of determination or residual analysis. 

While a global fitting algorithm considering all IQ measurements at once is the most suitable and elegant procedure, in this study we adopt a straightforward fitting approach due to its versatility in accommodating different scenarios (see Sect \S\ref{application_MUSE}), ranging from a low to a high number of measurements, as the two cases mentioned above.



\section{Application to real IFS observations}
\label{application_MUSE}

To apply the proposed methodology to actual data, we identified the MUSE instrument \citep{2010Bacon} installed on the Nasmyth B platform of the European Southern Observatory (ESO) VLT Unit Four (UT4) at the Paranal Observatory\footnote{https://www.eso.org/public/spain/teles-instr/paranal-observatory/}.
MUSE is an integral field spectrograph providing spectral information in the range of 4800 to 9300 \AA\ with a mean resolution of 3000. In the seeing-limited mode (Wide-field mode), MUSE samples a one-squared arcmin field of view with 0.2 arcsecs spatial pixels. The VLT UT4 has a $\sim$30-m-high dome, and the observing floor is $\sim$10 m above ground level\footnote{https://www.eso.org/sci/facilities/paranal/telescopes/ut.html}. At about 100 m to the North of the UT4 building, a robotic DIMM monitors the turbulence conditions at Paranal on top of a 7-m-high tower \citep{ManualESO}. Due to the difference in altitude between the DIMM tower and the Nasmyth VLT platform, we expect a slightly better $\epsilon_{0}$ at the MUSE focal station ($\Delta\epsilon_{\mathrm{h}}$), with a projected improvement of about $\Delta\epsilon_{\mathrm{h}}\sim$0.05 arcsecs compared to the DIMM measurements \citep[see Figure 15 in][]{2020Butterley}, assuming a negligible dome turbulence contribution.

\begin{table}
	\centering
	\caption{Selection criteria used in the ESO Science Portal for identifying processed data cubes from IFS observations.}
	\label{selection_criteria}
	\begin{tabular}{lr} 
	\hline \\
 {\bf Instrument, Mode} & MUSE, WFM-NOAO \\
 {\bf Sensitivity (AB mag)}       &     $\geq$ 19  \\          
    {\bf Date} & 1 Jan 2017 - 31 May 2023       \\
     {\bf Total exposure time (s) } & 100-200   \\
	{\bf Sky resolution (arcsec)} & $\leq$ 1  \\
                                      &   \\
 \hline
\end{tabular}
\end{table}

\begin{table*}
    \tabcolsep 2.5pt
	\centering
	\caption{Parameter for the MUSE processed data cubes downloaded from the ESO Science Portal.}
	\label{night_parameters}
	\begin{tabular}{ccccccccccc} 
	\hline \\
 {\bf ID} & {\bf PI} & {\bf Field} & {\bf Obs. date} & {\bf UTC} & {\bf IT (s)} & {\bf $T_{mm}$} & {\bf $T_{in-en}$} & {\bf $T_{out-en}$} & {\bf Air mass} & {\bf $\epsilon_{DIMM}$} \\
    (1) & (2) & (3) & (4) & (5) & (6) & (7) & (8) & (9) & (10) & (11) \\
   0102.C-0589(A) & Vogt, F. &  HD90177a          & 29 Dec 2018 & 05:58:47 & 180 & 12.38 &  14.88 &  12.70   & 1.40 & 0.47$\pm$0.10  \\
   0102.C-0589(A) & Vogt, F. &  HD90177b          & 29 Dec 2018 & 06:40:43 & 180 & 12.18 &  14.83 &  12.48   & 1.31 & 0.39$\pm$0.05  \\
   0102.C-0589(A) & Vogt, F. &  HD90177c          & 29 Dec 2018 & 07:23:27 & 180 & 12.08 &  14.75 &  12.37   & 1.25 & 0.51$\pm$0.04  \\
   0102.C-0589(A) & Vogt, F. &  HD90177d          & 29 Dec 2018 & 08:04:58 & 105 & 11.98 &  14.66 &  12.22   & 1.23 & 0.51$\pm$0.02 \\
   106.21GS.001   & Ivanov, V. & OGLEII           & 20 Feb 2021 & 08:54:08 & 140 & 10.78 &  13.18 &  11.62   & 1.56 & 0.41$\pm$0.02  \\

    \hline
	\end{tabular}
 \tablefoot{ (1) ESO program and run identification code, (2) principal investigator, (3) observed field/object, (4) observing date (day month year), (5) date time (UTC), (6) integration time (s), (7) main mirror temperature (in $^{\circ}$C), (8) temperature inside the enclosure (in $^{\circ}$C). This temperature corresponds to the average of the values provided for sensors at different locations, (9) observatory ambient temperature (in $^{\circ}$C), (10) average air mass of MUSE observations, (11) average and standard deviation of DIMM $\epsilon_{0}$ measurements in arcsecs during MUSE observations. Temperatures and air mass for each night were retrieved from the headers of the MUSE data cubes.}
\end{table*}

We used the extensive ESO science archive{\footnote{http://archive.eso.org/cms.html}} to collect MUSE data. This archive offers thousands of reduced seeing-limited MUSE data cubes, ready for analysis. We note that any point-like source within the field of view of these data cubes can be a potential candidate for retrieving the prevailing atmospheric parameters during observations. To apply the method described in Sect. \S\ref{IFS_section} to real data, we systematically searched on the ESO science portal for processed data using the selection criteria outlined in Table \ref{selection_criteria} to constrain the number of potential data cubes to analyze. This search yielded a total of 718 data cubes, of which 389 were identified as being obtained in seeing-limited mode (labeled as WFM-NOAO in the MUSE data cubes header). About 95.9\% of them correspond to sky exposures, and 2.8\% to unfocused calibration stars or extended objects, data cubes that are useless for the purpose of this work. Eventually, we concluded with five (1.3\%) MUSE data cubes of two fields showing a few stars (see Figs. \ref{MUSE_fields} in Sect. \S\ref{appenA} of appendix \ref{appenB}). Table \ref{night_parameters} presents relevant parameters for the retrieved MUSE data cubes. Observations for the HD90177 field (four data cubes) come from the same observing night, acquired within an interval of around two and a half hours (see Table \ref{night_parameters}).

We used the source detection tool within the {\it \emph{photutils Astropy}} package\footnote{https://photutils.readthedocs.io/en/stable/detection.html} to retrieve the positions of stars in the field-of-view of the selected MUSE data. For the subsequent analysis, we focused on non-saturated bright stars isolated from companions within a radius of at least 5 arcsecs. Then, we used the MUSE Python Data Analysis Framework \citep[MPDAF\footnote{https://mpdaf.readthedocs.io/en/latest/},][]{2019Piqueras} to recover band-filter images of 8 arcsec$^{2}$ field-of-view centered on each selected star. For this, we designed a set of 100 \AA \ wide bandpass filters uniformly distributed along the MUSE spectral axis, starting with the bluest at 4850 \AA. The resulting 44 filter-band images for each star depict the seeing-limited MUSE PSF in the spectral coverage of those bands. We selected the analytical elliptical Moffat profile defined in Eq. \ref{equation_Moffat} to model our band-filter images. We ensure a zero background in the Moffat profile by subtracting any sky background level from the filter-band images before performing the fitting.

\begin{equation}
    EM1 = \frac{AM}{\left[1+A(x-x_0)^2 + B(y-y_0)^2 + C(x-x_0)(y-y_0)\right]^{\beta}}
    \label{equation_Moffat}
\end{equation}

In Eq. \ref{equation_Moffat}, AM and $\beta$ are the amplitude and the power index of the Moffat model centered on the star image peak at the field-of-view coordinates (x$_{0}$, y${_0}$). Furthermore, the coefficients A, B, and C are defined as follows:

\begin{equation*}
    A  = \left(\frac{cos\theta}{\sigma_1}\right) ^2 + \left(\frac{sin \theta}{\sigma_2}\right)^2 \rm{ ; \,  } 
\end{equation*}

    \begin{equation*}
    B = \left(\frac{sin\theta}{\sigma_1}\right)^2 + \left(\frac{cos\theta}{\sigma_2}\right)^2 \rm{ ; and } 
\end{equation*}
\begin{equation*}
     C = 2 sin \theta cos \theta \left(\frac{1}{\sigma_1^2} - \frac{1}{\sigma_2^2}\right)
\end{equation*}

\noindent where $\theta$ defines the orientation of the Moffat model with respect to the image axes, while $\sigma_1$ and $\sigma_2$ represent the characteristic widths of the profile along the major and minor axes, respectively, defining the ellipticity of the Moffat model. The Moffat function serves as a general model encompassing the Gaussian and Lorentzian functions, approaching each depending on the values of its parameters \citep[e.g.,][]{2019Fetick, 2001Trujillo}. Specifically, for large values of $\beta$, the Moffat function converges to a Gaussian profile, while for values close to unity, it tends toward a Lorentzian profile.

We used a modified version of the Moffat2D model, incorporating an elliptical Moffat instead of a circular one. This model was introduced in the \emph{modeling} module within the \emph{Astropy} package, enabling the determination of optimal parameters for fitting the Moffat model to the band-filter images recovered from MUSE data cubes. The FWHM of the PSF along the major and minor axes for each band-filter image is then determined from the Moffat parameters as follows:
\begin{equation}
 FWHM_{n} = 2\times\sigma_{n}\times\sqrt{2^{(1/\beta)}-1} , \, \mathrm{with} \ n=1,2
 \label{FWHM_PSF}
\end{equation}

Previous works, based on the analysis of stars, found that a circular Moffat profile with a power index of $\beta\approx2.5$ effectively reproduces the MUSE PSF across wavelengths \citep[e.g.,][]{2015Bacon, 2016Husser} in its seeing-limited mode. For other MUSE modes, including AO compensation, the PSF core is also approximated using a Moffat function \citep[][]{2019Fetick, 2020Fusco}.

We explored four distinct model configurations analyzing the residuals resulting from the subtraction of Moffat models from star images: (1) $\sigma_1=\sigma_2$ and $\beta$ as a free parameter, (2) $\sigma_1=\sigma_2$ with a fixed $\beta$ of 2.5, (3) both $\sigma_1$ and $\sigma_2$, as well as $\beta$, as free parameters, and (4) $\sigma_1$ and $\sigma_2$, as free parameters with $\beta$ fixed at 2.5 \citep[see][]{2024Esparzaarredondo}. Our findings show that residual levels remain comparable whether the $\beta$ parameter is constrained to 2.5 or not. However, we obtained smaller residuals for an elliptical Moffat profile, even when $\sigma_1-\sigma_2$ is significantly smaller than the spatial sampling. Subsequently, as a reference for analyzing the wavelength-dependent variation of MUSE IQ, we adopt the average FWHM derived from the $\sigma_{1,2}$ parameters obtained with configuration (4) of the Moffat model. Section \S\ref{appenA} of appendix \ref{appenB} presents the Moffat model parameters obtained for each individual star analyzed in this study.

\begin{figure*}
\begin{center}
\includegraphics[trim={3cm 13.cm 2.5cm 2.25cm},clip,width=9.cm]{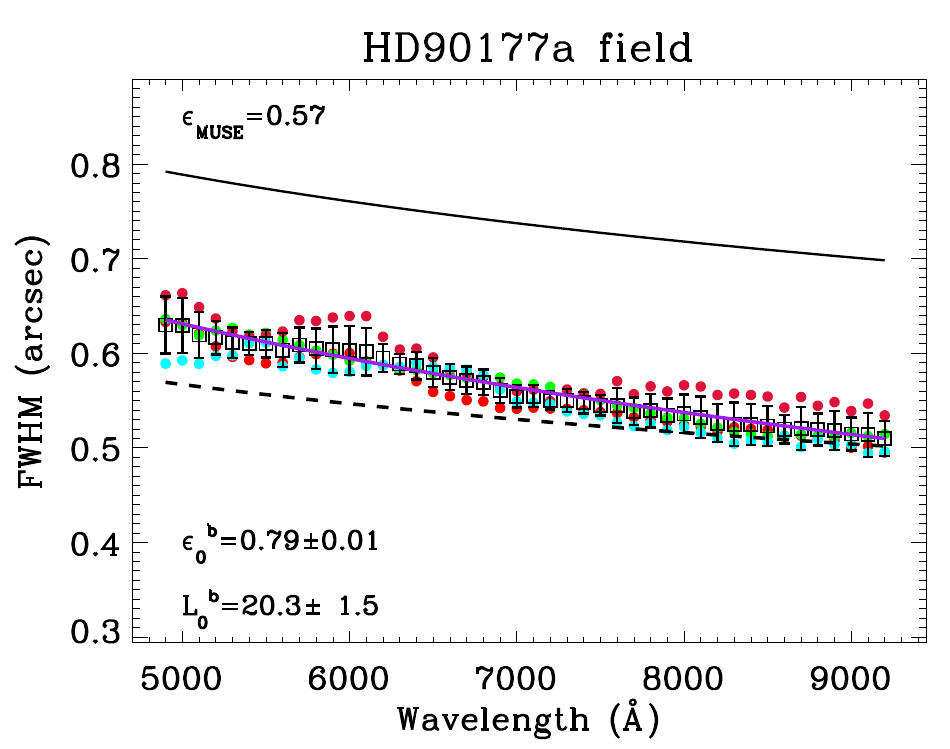}
\includegraphics[trim={3cm 13.cm 2.5cm 2.25cm},clip,width=9.cm]{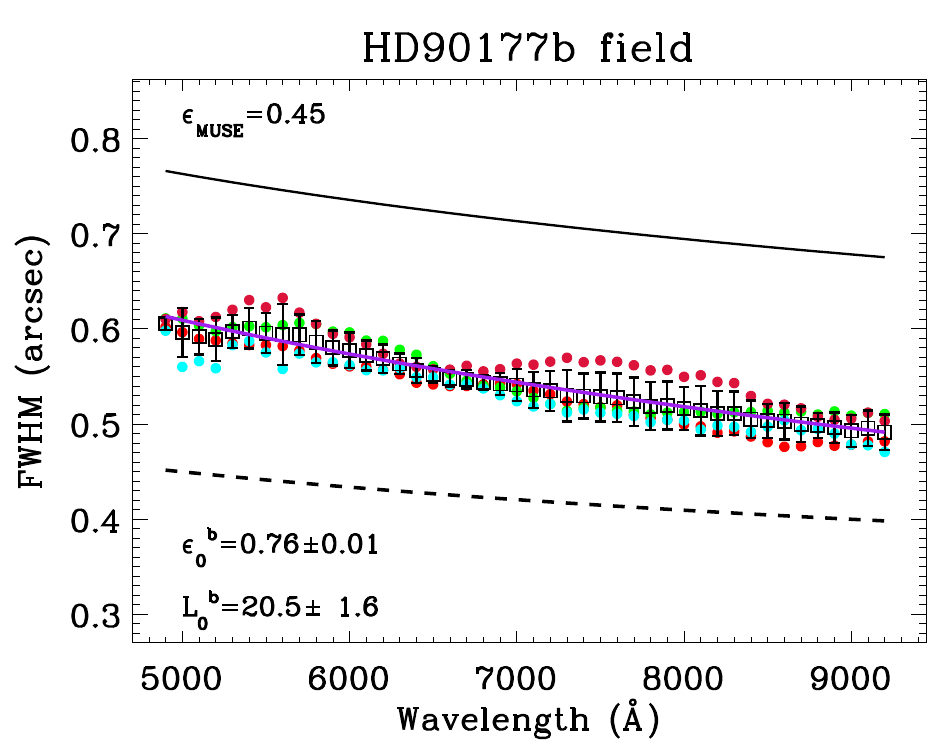}
\includegraphics[trim={3cm 13.cm 2.5cm 2.25cm},clip,width=9.cm]{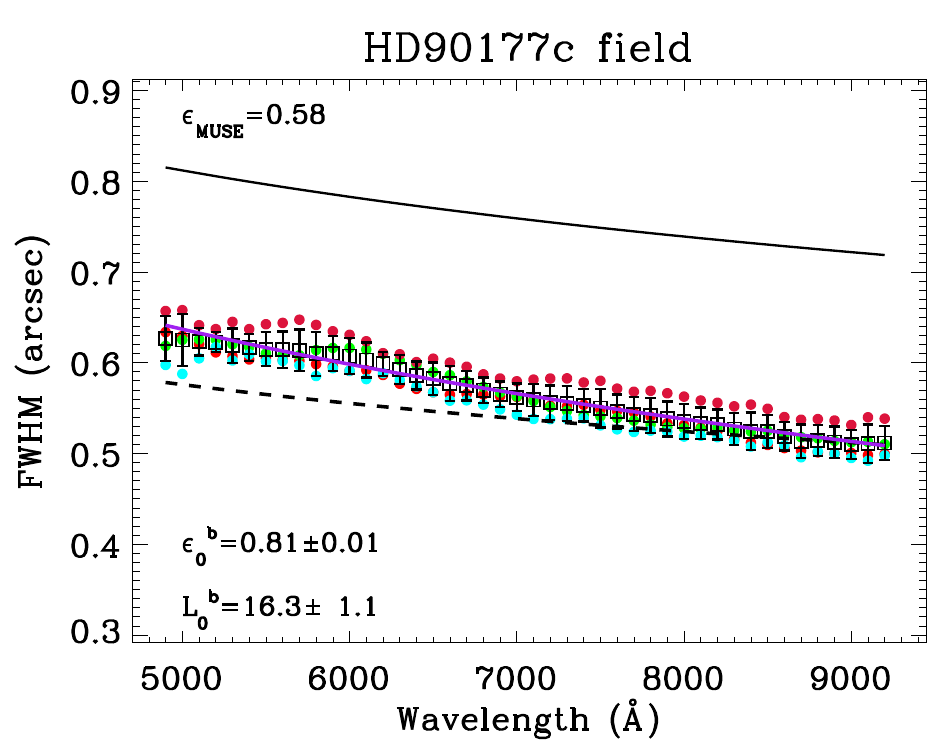}
\includegraphics[trim={3cm 13.cm 2.5cm 2.25cm},clip,width=9.cm]{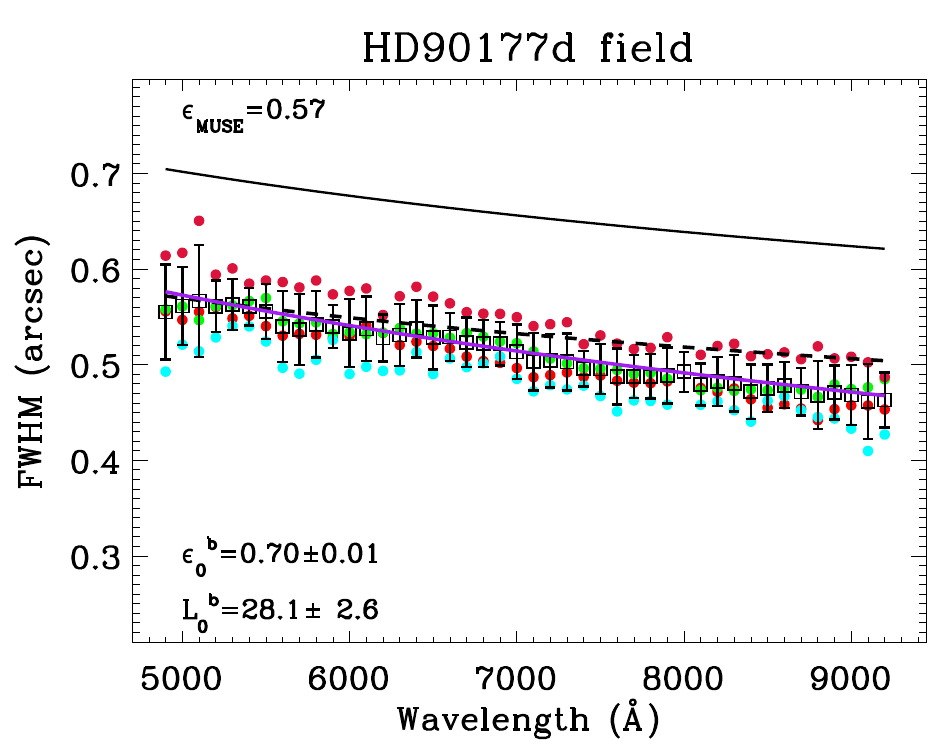}
\includegraphics[trim={3cm 13.cm 2.5cm 2.25cm},clip,width=9cm]{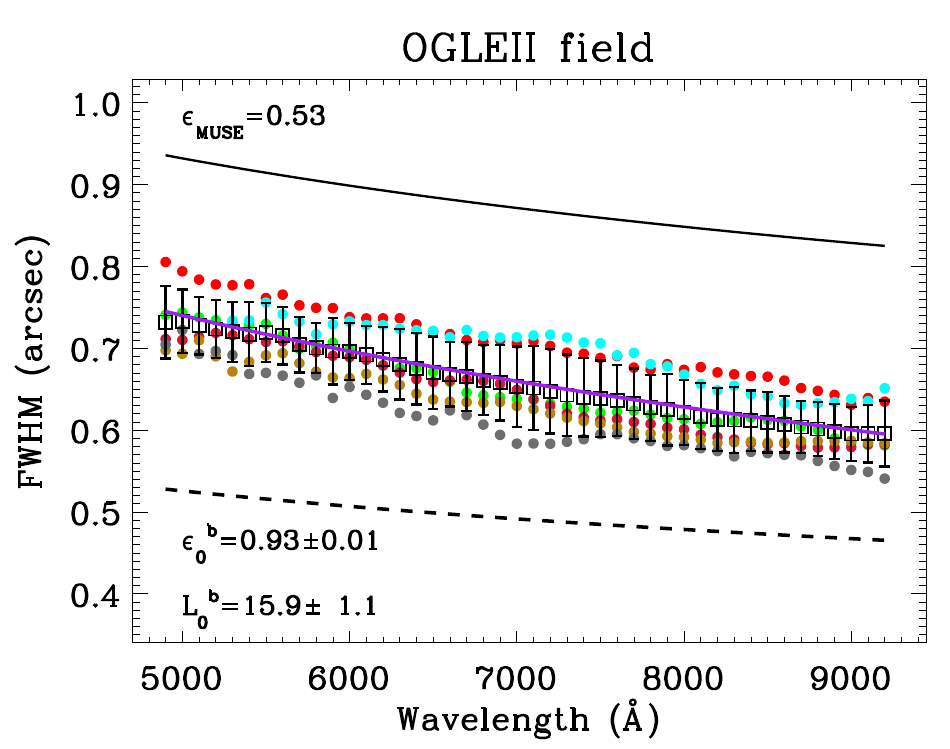}
 \caption{Wavelength dependence of IQ at the MUSE instrument's focal plane (IQ$_{\mathrm{MUSE}}$), analyzed using point-like sources in the MUSE fields indicated in each panel (see Table \ref{night_parameters}). Color dots represent IQ$_{\mathrm{MUSE}}$ values derived from Moffat model fits for 44 band-filter images obtained from distinct stars within each MUSE field (individual plots available in Sect. \S\ref{appenA} of appendix \ref{appenB}). Black open squares with error bars denote the average and standard deviation of measured IQ$_{\mathrm{MUSE}}$ across all stars at each wavelength. The black-dashed line shows the $\lambda^{-1/5}$ behavior of the $\epsilon_{0}$ along the MUSE observation direction ($\epsilon_{\mathrm{MUSE}}$ in arcsecs, top-left corner), derived from average DIMM values during MUSE observations (see Table \ref{night_parameters}) using Eq. \ref{seeing_zenith}. 
 The black-solid and purple-solid lines represent the $\epsilon_{0}$ and $\epsilon_{LE}$ best-fitting the measurements obtained for the $\epsilon_{0}^{b}$ (in arcsecs) and $\mathcal{L}_{0}^{b}$ (in meters) parameters in the bottom-left corner of each plot (details in text).}

\label{field_stars_juntas}
\end{center}
\end{figure*}

Since the size and wavelength variation of IQ obtained from stars at distinct positions within the MUSE field-of-view are comparable within uncertainties, we computed the average IQ for each MUSE field (IQ${\mathrm{MUSE}}$($\lambda$), see Fig. \ref{field_stars_juntas}) to compare it with the behavior predicted by Eq. \ref{eq:ELE_lambda}. The standard deviation of these averaged IQ was taken as the uncertainty for IQ${\mathrm{MUSE}}$($\lambda$). From the ESO ambient conditions database for Paranal\footnote{http://archive.eso.org/wdb/wdb/asm/dimm\_paranal/form}, we retrieved the DIMM measurements recorded on the corresponding nights and computed the DIMM average value ($\epsilon_{\mathrm{DIMM}}$) during the MUSE observations (see Table \ref{night_parameters}). We estimated the natural $\epsilon_{0}$ affecting the selected MUSE data cubes ($\epsilon_{\mathrm{MUSE}}$, listed in Table \ref{tabla_parameters_global}) by compensating for the expected improvement due to the DIMM-MUSE platform difference in height and projecting the result to the zenith angle ($\zeta$) for the observations (Table \ref{night_parameters}), that is:

\begin{equation}
    \epsilon_{\mathrm{MUSE}} =  \left(\frac{[\epsilon_{\mathrm{DIMM}}]^{5/3} - [\Delta\epsilon_{\mathrm{h} }]^{5/3}}{\cos\zeta}\right)^{3/5}
    \label{seeing_zenith}
\end{equation}

For two of the five MUSE fields (i.e., HD90177b, and OGLEII), the measured IQs$_{\mathrm{MUSE}}$ consistently exceed $\epsilon_{0}$ at any wavelength (Fig. \ref{field_stars_juntas}), while for the other three fields (i.e., HD90177a, HD90177c and HD90177d), IQs are comparable or slightly smaller across regions of the analyzed spectral range. This scenario is consistent with either an infinite \LO with some contribution from dome-induced turbulence or a finite \LO with a significant dome-induced blur. Previous works have reported a finite \LO at the Paranal observatory, with a typical value of about 22 m \citep[e.g.,][]{2016Ziad}, supporting the second as the most plausible option. 
Furthermore, the measured IQ$_{\mathrm{MUSE}}$($\lambda$) follows a similar curve to the predicted $\epsilon_{LE}$ through Eq. \ref{eq:ELE_lambda}, although we observe some slope discrepancies for the analyzed fields. Such discrepancies may suggest that the estimated $\epsilon_{\mathrm{MUSE}}$ may not accurately represent the atmospheric turbulence affecting MUSE observations, indicating a non-negligible difference between the reported $\epsilon_{0}$ by the DIMM and the actual $\epsilon_{0}$ along the MUSE observation direction. Additionally, as discussed in Sect. \S\ref{dome-seeing}, a non-negligible $\epsilon_{\mathrm{dome}}$ contribution may also contribute to these slope discrepancies.

Fortunately, the empirical IQs$_{\mathrm{MUSE}}$ measurements enable us to estimate the optimal combination of $\epsilon_{0}$ and \LO values that best align with the observed data by fitting the analytical curve derived by \cite{2002Tokovinin}.
Various fitting strategies may be employed for this purpose. However, the choice of a fitting method for implementing the proposed method to a specific instrument should be made after analyzing each particular case. The selected fitting method should align with the nature of the measurements, considering factors such as the number of empirical data points, sampling, spectral range, uncertainties, etc. It should also align with the intended use of the derived parameters. In this study, we adopt a straightforward numerical approach, varying parameters to minimize residuals with the observed IQs$_{\mathrm{MUSE}}$.

In particular, we used an iterative process involving variations of the input $\epsilon_{0}$ within the analytical approach (i.e., Eq. \ref{eq:ELE_lambda}) across a range of possible values for the $\epsilon_{0}$ along the MUSE observing direction. For our analysis, we explored $\epsilon_{0}$ values from $\epsilon_{\mathrm{MUSE}}-0.2$ to $\epsilon_{\mathrm{MUSE}}+0.5$ arcsecs, with 0.01 arcsecs steps. For each specific $\epsilon_{0}$ within this range, we determined the unknown parameter \LO by computing $\mathcal{L}_{0i}$ for each $\lambda_{i}$ to match the measured IQ$_{\mathrm{MUSE}}$($\lambda_{i}$) and the predicted $\epsilon_{LE}$($\lambda_{i}$). The best-fit \LO value and its uncertainty are obtained as the average and standard deviation of these $\mathcal{L}_{0i}$ values. To account for the uncertainties in IQ$_{\mathrm{MUSE}}$($\lambda_{i}$), we perform a Monte Carlo (MC) simulation randomly varying IQ$_{\mathrm{MUSE}}$($\lambda_{i}$) within their uncertainty range many times (a thousand times for this work). The average and standard deviation of all these MC realizations were adopted as the observed \LO ($\mathcal{L}_{0}^{b}$) and its uncertainty for that specific $\epsilon_{0}$. After considering all combinations of $\epsilon_{0}$ and $\mathcal{L}_{0}^{b}$ , we adopted the best-fit parameters (i.e., $\epsilon_{0}^{b}$ and $\mathcal{L}_{0}^{b}$) as those yielding the minimal residual with empirical IQ measurements.

Before analyzing the empirical measurement of IQ$_{\mathrm{MUSE}}$ to retrieve the combination of $\epsilon_{0}$ and \LO that best matches the observations, we may consider a potential blur contribution from turbulence inside the telescope dome or assume 100\% atmospheric seeing-limited conditions, implying negligible dome-induced turbulence. We explore these two scenarios for the analyzed MUSE observed fields in the following subsections.

\subsection{Analysis assuming negligible dome-induced turbulence}
\label{negligible_dome}

Under the assumption of a negligible dome turbulence contribution to the observed IQ, we seek to determine the optimal combination of $\epsilon_{0}$ and $\mathcal{L}_{0}$ predicting the best matching with the IQ$_{\mathrm{MUSE}}$ measurements. To achieve this, we used an iterative process involving variations of the input $\epsilon_{0}$ within the Eq. \ref{eq:ELE_lambda} analytical approach, following the methodology described in the preceding section. Finally, we adopt the parameters providing the minimal residuals between the predicted and measured IQ curves ($\epsilon_{0}^{b}$ and $\mathcal{L}_{0}^{b}$ in Fig. \ref{field_stars_juntas}).

The derived $\epsilon_{0}^{b}$ and $\mathcal{L}_{0}^{b}$ parameters (see Fig. \ref{field_stars_juntas} and Table \ref{tabla_parameters_global}) for the selected MUSE fields are consistent with the statistical distribution of these parameters at the Paranal observatory \citep{2016Ziad, 2023Otarola}.
We found that the obtained $\epsilon_{0}^{b}$ are on average 0.26 arcsec larger than those estimated from DIMM measurements, with a minimum and maximum of 0.13 and 0.40 arcsecs for HD90177d and OGLEII fields, respectively. It is important to note that MUSE observations were obtained at a higher airmass than typical for DIMM measurements. This, combined with the difference in observing directions between DIMM and MUSE, may contribute to these discrepancies. Interestingly, comparisons of $\epsilon_{0}$ measurements from different atmospheric characterization instruments have found around 0.2 arcsecs of a discrepancy, attributed to the specific instrument locations on the Paranal observatory plateau \citep[e.g.,][]{2018Osborn}. Consequently, the parameters obtained using the proposed methodology for the analyzed MUSE fields are compatible with purely seeing-limited observations, well within the statistical behavior of atmospheric turbulence at the Paranal Observatory.

\subsection{Analysis assuming dome turbulence contribution}
\label{with_dome}
As discussed in Sect. \S\ref{dome-seeing}, when the contribution of dome-induced turbulence becomes significant, it not only broadens the images but can also influence the wavelength variation, depending on the nature of the dome's local turbulence. In these cases, a direct measure of $\epsilon_{\mathrm{dome}}$ at different wavelengths is desirable to properly isolate the broadening in the images due to the atmosphere and apply the proposed methodology in this work to retrieve \LO.

Unfortunately, we could not find a specific analysis of the $\epsilon_{\mathrm{dome}}$ for the VLT UT4 telescope or any direct measurements of the dome contribution to turbulence degradation on the selected nights. However, \cite{1991Racine} found that the principal factors influencing the dome turbulence are the primary mirror and the enclosure temperatures, with contributions described as follows:

\begin{equation}
    \epsilon_{\mathrm{mm}} \approx  0.4\times(T_{mm}-T_{in-en})^{6/5} \  \mathrm{,and}
    \label{dome_mm}
\end{equation}
\begin{equation}
    \epsilon_{\mathrm{en}} \approx  0.1\times(T_{in-en}-T_{out-en})^{6/5}
    \label{dome_en}
\end{equation}

\noindent where $T_{mm}$ is the primary mirror temperature, and $T_{in-en}$ and $T_{out-en}$ are the temperatures inside and outside the telescope enclosure, respectively, assuming that $T_{in-en}$ is the same at any location inside the dome. These contributions were empirically determined at an effective wavelength of 7000 \AA \ \citep{1991Racine}. Hereafter, we assume that the Paranal VLT UT4 exhibits an equivalent behavior. Table \ref{night_parameters} presents the recorded temperatures for the observation nights of the analyzed MUSE data cubes. Table \ref{tabla_parameters_global} shows the estimates for dome-induced turbulence contribution during MUSE observations assuming Eqs. \ref{dome_mm} and \ref{dome_en} apply to VLT UT4. On these nights, the temperature of the main mirror of the telescope was lower than that inside the dome, preventing any turbulence from it, that is $\epsilon_{\mathrm{mm}}$=0. However, the air within the dome was warmer than the outside environment, potentially leading to turbulence and contributing to the blur in the observed data. Then, we adopt the estimated $\epsilon_{\mathrm{en}}$ as the total $\epsilon_{\mathrm{dome}}$ contribution to the measured IQ at a wavelength of 7000 \AA \ (i.e., $\epsilon_{\mathrm{dome}}(7000$ \AA$)=\epsilon_{\mathrm{en}}$).

Equation \ref{dome_seeing_general} allows us to extend the $\epsilon_{\mathrm{dome}}$ approach from 7000 \AA \ to any wavelength. Given the lack of a predefined power-law index $\gamma$, we treated it as a free parameter in the fitting procedure. As described in Sect. \S\ref{negligible_dome}, we performed an iterative process involving the variations of the input $\epsilon_{0}$ and the $\gamma$ power-law index. Initially, we allowed $\epsilon_{0}$ to vary within a range spanning from $\epsilon_{MUSE}-0.2$ to $\epsilon_{MUSE}+0.5$ arcsecs, with steps of 0.01 arcsecs. For each specific $\epsilon_{0}$ value within this range, we followed these steps: (1) variations of the $\gamma$ parameter in the range of 3 to 4, with steps of 0.01, to compute the dome turbulence contribution; (2) for each $\gamma$ value, subtracting the resulting dome turbulence contribution from the measured IQ$_{\mathrm{MUSE}}$($\lambda_{i}$) at each wavelength to infer the $\epsilon_{LE}$($\lambda_{i}$) using Eq. \ref{IQ_general}; (3) determining the corresponding \LO from each $\epsilon_{LE}$($\lambda_{i}$), following the same methodology as described in previous sections. Subsequently, the combination of parameters ($\epsilon_{0}$, $\gamma$, \LO) yielding minimal residuals between the predicted (Eq. \ref{eq:ELE_lambda}) and measured IQ curves was adopted.

As in the previous section, we assess the uncertainties in the derived $\epsilon_{0}$, $\gamma$, and \LO parameters using a MC approach. We randomly perturbed each IQ$_{\mathrm{MUSE}}$($\lambda_{i}$) within its uncertainty range for one thousand iterations, resulting in one thousand different sets of ($\epsilon_{0}$, $\gamma$, \LO) parameters. Finally, we calculated the average and standard deviation of these sets to determine the optimal combination that minimizes the residuals and best fits the observed data (see Table \ref{tabla_parameters_global}).

\begin{figure*}
\begin{center}

\includegraphics[trim={3cm 13.cm 2.5cm 2.25cm},clip,width=9.15cm]{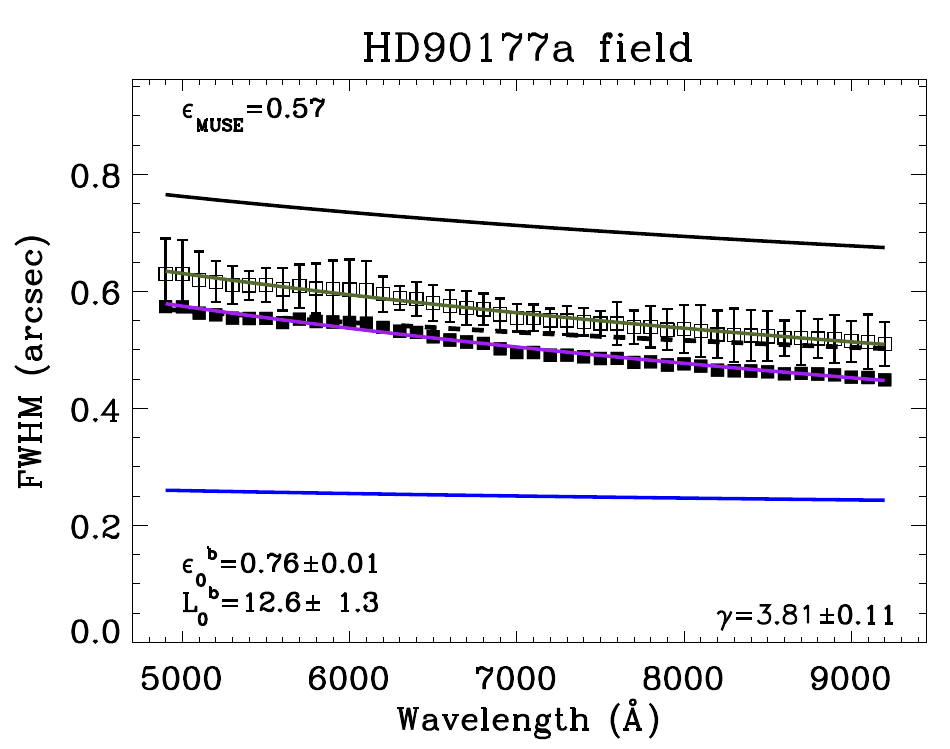}
\includegraphics[trim={3cm 13.cm 2.5cm 2.25cm},clip,width=9.15cm]{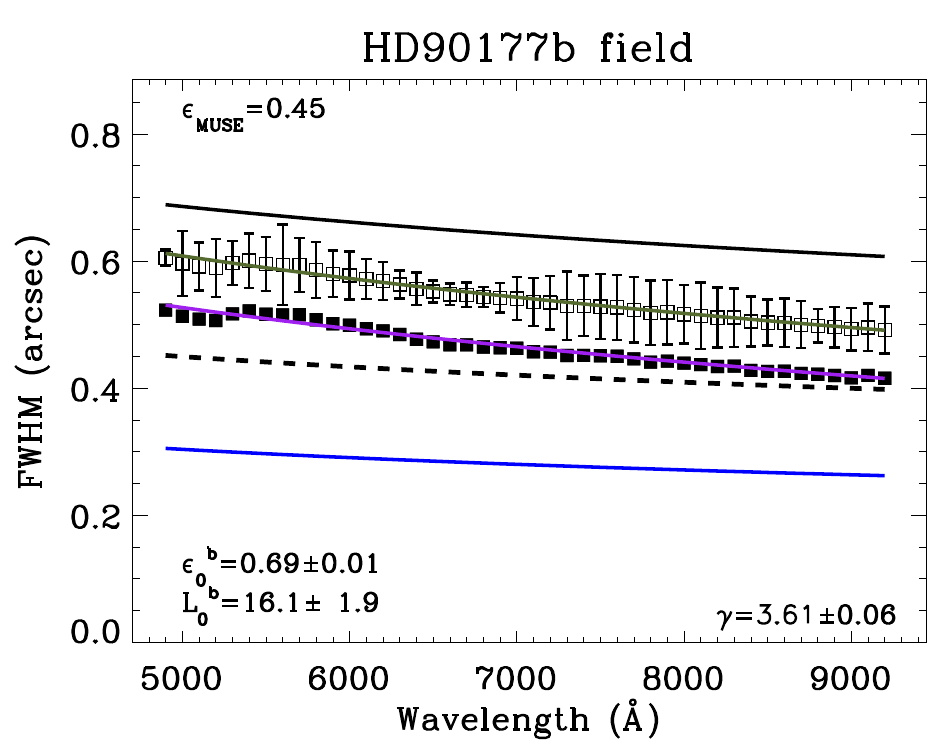}
\includegraphics[trim={3cm 13.cm 2.5cm 2.25cm},clip,width=9.15cm]{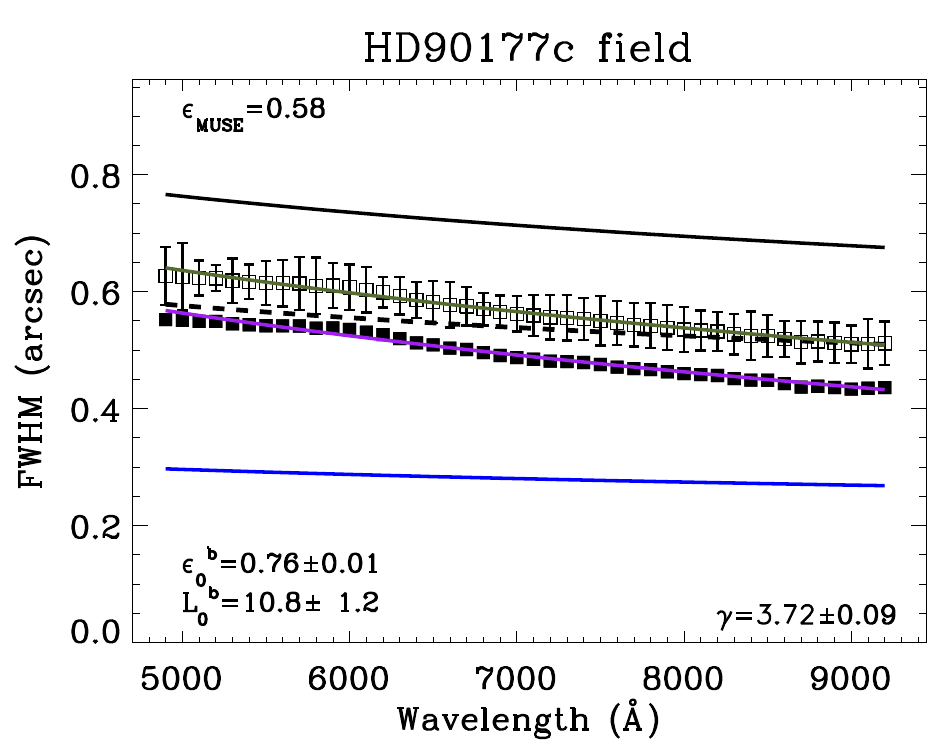}
\includegraphics[trim={3cm 13.cm 2.5cm 2.25cm},clip,width=9.15cm]{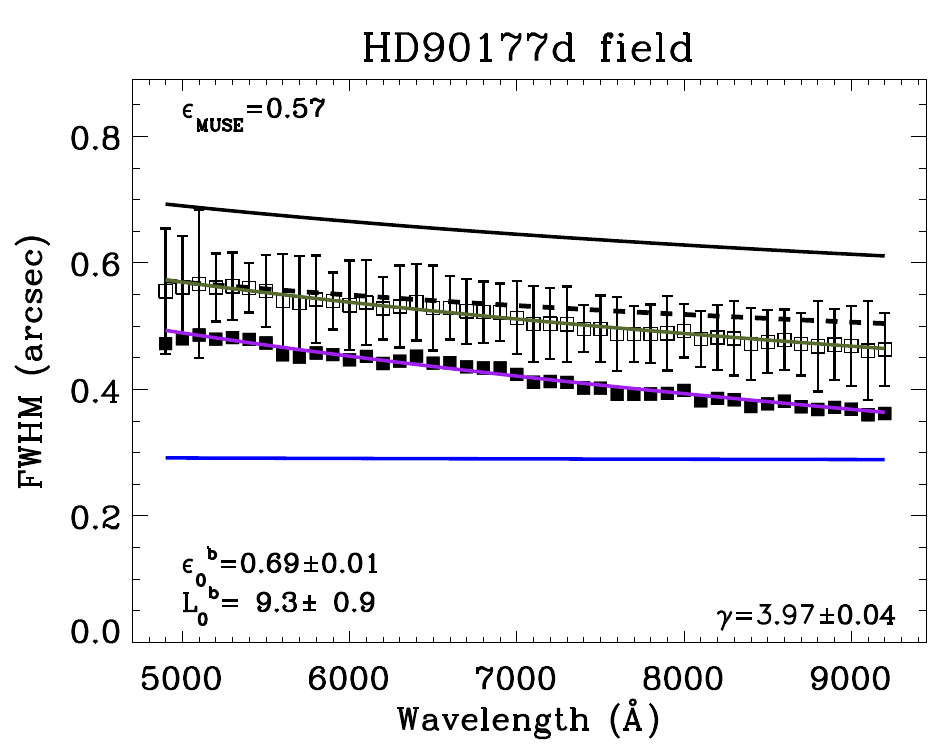}
\includegraphics[trim={3cm 13.cm 2.5cm 2.25cm},clip,width=9.15cm]{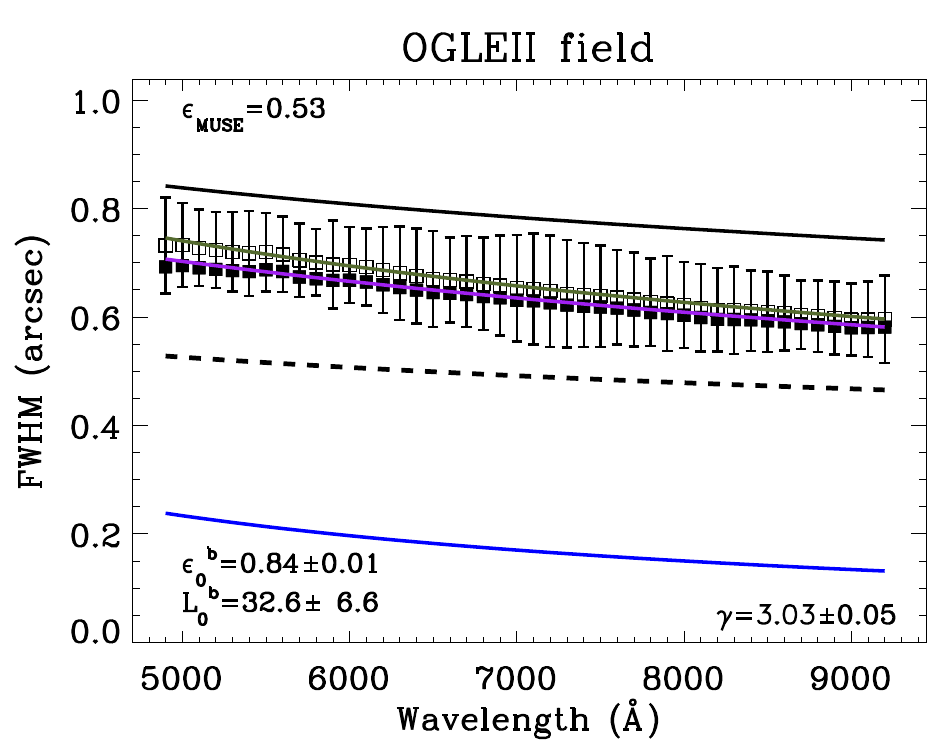}
 \caption{Same as Fig. \ref{field_stars_juntas}, but considering the dome turbulence contribution. Black-open squares represent the average IQ$_{\mathrm{MUSE}}$ across each field, the same as in Fig. \ref{field_stars_juntas}. The black-filled squares show the IQ broadening attributed solely to atmospheric turbulence (i.e., IQ$_{\mathrm{MUSE}}$ after quadratically subtracting the estimated dome-turbulence contribution at each $\lambda$). The blue solid line corresponds to the estimated dome-turbulence contribution, following a power-law dependence on wavelength with the $\gamma$ index indicated in the bottom right corner. The green line corresponds to the quadratic sum of the predicted atmospheric (purple line) and dome (blue line) turbulence blur, fitting the measured MUSE IQs.}

\label{field_stars_juntas_dome}
\end{center}
\end{figure*}

Figure \ref{field_stars_juntas_dome} presents the results of this procedure applied to each of the selected MUSE data cubes. The obtained $\epsilon_{0}^{b}$ and $\mathcal{L}_{0}^{b}$ through considering a $\epsilon_{\mathrm{dome}}$ contribution to the actual MUSE IQ are again consistent with their statistical distribution at the site. Differences between $\epsilon_{0}^{b}$ and the expected from DIMM measurements along the MUSE pointing are within the typical discrepancies of $\epsilon_{0}$ measurements at different locations on the Paranal plateau \citep[e.g.,][]{2018Osborn}.

We found some discrepancies in the estimated $\mathcal{L}_{0}^{b}$ for the four MUSE data cubes obtained on the same night in an interval of a few hours (see Table \ref{night_parameters}). Previous studies found significant variations in the \LO parameter on time scales comparable to fluctuations in other atmospheric parameters, such as $\epsilon_{0}$ \citep[e.g.,][]{2016Ziad}.

On the night with four data cubes of the same field (see Table \ref{night_parameters}), the analysis of the IQ$_{\mathrm{MUSE}}$ reveals dome-induced turbulence that varies with wavelength according to a power law, with an index larger than the Kolmogorov regime for three of them (i.e., HD90177a, HD90177c, and HD90177d), while it is within the Kolmogorov regime for the HD90177b data cube. The $\epsilon_{\mathrm{dome}}$ parameters derived from these four MUSE data cubes indicate a subtle evolution of dome-induced turbulence throughout the night. This evolution primarily stems from changes in its power-law regime, while remaining nearly constant in terms of intensity at the reference wavelength (see Table \ref{tabla_parameters_global}). For the other field observed on a different night, the power law index was smaller than the Kolmogorov $\gamma=\frac{11}{3}$ index.

The obtained $\epsilon_{0}^{b}$ and $\mathcal{L}_{0}^{b}$ parameters considering the dome turbulence contribution also align with the statistical distribution of these parameters at the Paranal observatory (see Fig. \ref{field_stars_juntas} and Table \ref{tabla_parameters_global}).


\begin{table*}
	\centering
	\caption{Atmospheric parameters derived from the measured IQ at different wavelengths from MUSE data cubes.}
	\label{tabla_parameters_global}
	\begin{tabular}{c|c|ccc|cccc|} 
        & {\bf DIMM-seeing} &  \multicolumn{3}{c|}{\bf Negligible} & \multicolumn{4}{c}{\bf Non-negligible} \\ [3pt] 
        &  {\bf along MUSE direction} &  \multicolumn{3}{c|}{\bf dome-seeing} & \multicolumn{4}{c}{\bf dome-seeing} \\ [3pt]
        \hline
	 &  &  &  & & & & & \\ [0pt]
	{\bf MUSE Field} & $\epsilon_{MUSE}$  & $\epsilon_{0}^{b}$ & $\mathcal{L}_{0}^{b}$ & $\epsilon_{dome}$ &  $\epsilon_{0}^{b}$ & $\mathcal{L}_{0}^{b}$ & $\epsilon_{dome}$ & $\gamma$ \\ [3pt]
	(1) & (2) &(3) & (4) & (5) & (6) & (7) & (8) & (9)  \\ [3pt]
    \hline 
	 &  & &  &  &  & & &  \\ [0pt]
HD90177a    &   0.57  & 0.79 &  20.3$\pm$1.5   & 0.00 &   0.76 & 12.6$\pm$1.3 & 0.25 & 3.81$\pm$0.11 \\ [3pt]
HD90177b    &   0.45  & 0.76 &  20.5$\pm$1.6   & 0.00 &   0.69 & 16.1$\pm$1.9 & 0.28 & 3.61$\pm$0.06 \\ [3pt]
HD90177c    &   0.58  & 0.81 &  16.3$\pm$1.1   & 0.00 &   0.76 & 10.8$\pm$1.2 & 0.28 & 3.72$\pm$0.09 \\ [3pt]
HD90177d    &   0.57  & 0.70 &  28.1$\pm$2.6   & 0.00 &   0.69 & 9.3$\pm$0.9  & 0.29 & 3.97$\pm$0.04 \\ [3pt]
OGLEII      &   0.53  & 0.93 &  15.9$\pm$1.1   & 0.00 &   0.84 & 32.6$\pm$6.6 & 0.17 & 3.03$\pm$0.05 \\ [3pt]

 \hline
	\end{tabular}
	\tablefoot{Columns correspond to: (1) MUSE observed field. (2) $\epsilon_{0}$ (in arcsecs) along the MUSE observing direction estimated from the DIMM value corrected from the Nasmyth platform altitude respecting DIMM and airmass. (3) and (4) Combination of $\epsilon_{0}$ and \LO values minimizing residuals when comparing the predicted $\epsilon_{LE}$ using Eq. \ref{eq:ELE_lambda} with these values with the measured IQ at any wavelength, assuming a negligible dome turbulence contribution (5). (6) and (7) Combination of $\epsilon_{0}$ and \LO values minimizing residuals when comparing the predicted $\epsilon_{LE}$ using Eq. \ref{eq:ELE_lambda} with the measured IQ$_{MUSE}$ at any wavelength, considering the contribution of the $\epsilon_{\mathrm{dome}}$. Uncertainties for columns (3) and (6) are consistently 0.01 arcsecs, corresponding to the $\epsilon_{0}$ steps in the iterative process, with units in arcsecs. Units for columns (4) and (7) are meters. (8) Estimated turbulence contribution (in arcsecs) of the dome enclosure at the reference wavelength of 7000 \AA \ \citep{1991Racine}. (9) Power-law index for the wavelength dependence of the $\epsilon_{\mathrm{dome}}$.}
\end{table*}

\section{Conclusions}
\label{conclusions}

This work introduces a novel approach using IFS to measure the spatial coherence wavefront \LO prevailing during seeing-limited observations. This method takes advantage of the capability of IFS techniques to provide filter-band images over a wide wavelength range, all obtained under the same atmospheric and instrumental conditions. The homogeneity of IFS data enables the tracing of IQ variation with wavelength at the focal plane, directly linked to the prevailing $\epsilon_{0}$ and \LO characterizing the atmospheric turbulence conditions. We applied the proposed technique to MUSE observations obtained at the Paranal Observatory, exploring two situations: one assuming negligible dome-induced turbulence and the other considering the $\epsilon_{\mathrm{dome}}$ contribution. Through this analysis, the methodology accurately derived essential atmospheric parameters such as $\epsilon_{0}$ and \LO, while also revealing the wavelength dependence of the dome-induced turbulence, providing valuable insights into atmospheric conditions and instrument performance. The main findings and conclusions of this work can be summarized as follows.

\begin{itemize}
 \item[$\bullet$] IFS is a powerful tool for astronomical observations, enabling simultaneous spatial and spectral datasets over wide fields and wavelength ranges.
    \item[$\bullet$] The homogeneity of data cubes resulting from IFS observations provides a collection of filter-band images that allow for comprehensive analysis of IQ variations with wavelength. The S/N of each filter-band image is crucial for precise IQ assessment.
    \item[$\bullet$] IFS enables the empirical verification of the analytical approximation to IQ variations with wavelength (Eqs. \ref{eq:ELE_lambda} and \ref{eq:ELE_lambda_D}), derived from simplified models of atmospheric turbulence theory. We accomplished this validation using IFS data obtained with the MUSE instrument at the Paranal Observatory.
    \item[$\bullet$] The MUSE PSF is well-described by a Moffat Model with the power-law index set to a value of 2.5 at any wavelength. We adopt the Moffat model FWHM as the metric for assessing the MUSE IQ.
\item[$\bullet$] We analyze the IQ$_{\mathrm{MUSE}}$ wavelength variation to estimate the \LO parameter under two different scenarios: 
    \begin{enumerate}
        \item Both the \LO and the $\epsilon_{0}$ as free parameters with negligible dome-induced turbulence contribute to MUSE IQ.
           \begin{itemize}
            \item An iterative approach led to determining the optimal $\epsilon_{0}$ and \LO parameters ($\epsilon_{0}^{b}$ and $\mathcal{L}_{0}^{b}$) matching the predicted and the measured IQ$_{\mathrm{MUSE}}$($\lambda$) with minimal residuals.
            \item Both $\epsilon_{0}^{b}$ and $\mathcal{L}_{0}^{b}$ parameters derived for the analyzed MUSE fields are in agreement with the statistical behavior of these parameters at the Paranal Observatory, supporting that MUSE provides predominantly seeing-limited observations.
            \item The differences between the retrieved $\epsilon_{0}^{b}$ and the MUSE seeing estimated from DIMM are within typical discrepancies of $\epsilon_{0}$ measurements at various locations on the Paranal plateau with different instruments.
        \end{itemize}
        \item  \LO and $\epsilon_{0}$ as free parameters, and considering a non-negligible dome-induced turbulence, estimated using the approach by \cite{1991Racine}, with a power-law dependence on wavelength and the power-law index as a free parameter.
                   \begin{itemize}
            \item An iterative process involving variations in the $\epsilon_{0}$ and the power-law index controlling the wavelength behavior of the $\epsilon_{\mathrm{dome}}$ was performed to identify optimal $\epsilon_{0}^{b}$, $\mathcal{L}_{0}^{b}$, and $\gamma$ parameters, minimizing residuals between predicted and measured IQ$_{\mathrm{MUSE}}$($\lambda$).
            \item  The derived $\epsilon_{0}^{b}$ and $\mathcal{L}_{0}^{b}$ parameters for this scenario also align with the statistical turbulence behavior at the Paranal Observatory.
            \item Direct measurements of $\epsilon_{\mathrm{dome}}$ at various wavelengths are crucial for accurately isolating atmospheric broadening and determining \LO.
        \end{itemize}

    \end{enumerate}
\end{itemize}

Seeing-limited integral field spectrographs commonly operate in ground-based observatories. As a by-product of these observations, the \LO parameter could be routinely obtained and also facilitate the tracking of dome-induced turbulence. A reassessment of observing strategies for IFS observations, especially during the observation of calibration stars, may be beneficial to implement the proposed methodology for site characterization purposes.

\begin{acknowledgements}
      Based on data obtained from the ESO science archive facility with DOI(s): https://doi.org/10.18727/archive/41.
      We thank the referee and David Mouillet for their valuable feedback. The authors acknowledge support from the Spanish Ministry of Science and Innovation through the Spanish State Research Agency (AEI-MCINN/10.13039/501100011033) through grants "Participation of the Instituto de Astrofísica de Canarias in the development of HARMONI: D1 and Delta-D1 phases with references PID2019-107010GB-100 and PID2022-140483NB-C21 and the Severo Ochoa Program 2020-2023 (CEX2019-000920-S). 

\end{acknowledgements}

%
%

\bibliographystyle{aa}
\bibliography{aa48364-23_astro_ph.bib}

\begin{thebibliography}{44}
\expandafter\ifx\csname natexlab\endcsname\relax\def\natexlab#1{#1}\fi

\bibitem[{{Abahamid} {et~al.}(2004){Abahamid}, {Jabiri}, {Vernin},
  {Benkhaldoun}, {Azouit}, \& {Agabi}}]{2004aAbahamid}
{Abahamid}, A., {Jabiri}, A., {Vernin}, J., {et~al.} 2004, \aap, 416, 1193

\bibitem[{{Allington-Smith}(2006)}]{2006Allington-smith}
{Allington-Smith}, J. 2006, New Astronomy Reviews, 50, 244

\bibitem[{{Bacon} {et~al.}(2010){Bacon}, {Accardo}, {Adjali}, {Anwand},
  {Bauer}, {Biswas}, {Blaizot}, {Boudon}, {Brau-Nogue}, {Brinchmann},
  {Caillier}, {Capoani}, {Carollo}, {Contini}, {Couderc}, {Daguis{\'e}},
  {Deiries}, {Delabre}, {Dreizler}, {Dubois}, {Dupieux}, {Dupuy}, {Emsellem},
  {Fechner}, {Fleischmann}, {Fran{\c{c}}ois}, {Gallou}, {Gharsa}, {Glindemann},
  {Gojak}, {Guiderdoni}, {Hansali}, {Hahn}, {Jarno}, {Kelz}, {Koehler},
  {Kosmalski}, {Laurent}, {Le Floch}, {Lilly}, {Lizon}, {Loupias}, {Manescau},
  {Monstein}, {Nicklas}, {Olaya}, {Pares}, {Pasquini}, {P{\'e}contal-Rousset},
  {Pell{\'o}}, {Petit}, {Popow}, {Reiss}, {Remillieux}, {Renault}, {Roth},
  {Rupprecht}, {Serre}, {Schaye}, {Soucail}, {Steinmetz}, {Streicher}, {Stuik},
  {Valentin}, {Vernet}, {Weilbacher}, {Wisotzki}, \& {Yerle}}]{2010Bacon}
{Bacon}, R., {Accardo}, M., {Adjali}, L., {et~al.} 2010, in Society of
  Photo-Optical Instrumentation Engineers (SPIE) Conference Series, Vol. 7735,
  Ground-based and Airborne Instrumentation for Astronomy III, ed. I.~S.
  {McLean}, S.~K. {Ramsay}, \& H.~{Takami}, 773508

\bibitem[{{Bacon} {et~al.}(2015){Bacon}, {Brinchmann}, {Richard}, {Contini},
  {Drake}, {Franx}, {Tacchella}, {Vernet}, {Wisotzki}, {Blaizot}, {Bouch{\'e}},
  {Bouwens}, {Cantalupo}, {Carollo}, {Carton}, {Caruana}, {Cl{\'e}ment},
  {Dreizler}, {Epinat}, {Guiderdoni}, {Herenz}, {Husser}, {Kamann}, {Kerutt},
  {Kollatschny}, {Krajnovic}, {Lilly}, {Martinsson}, {Michel-Dansac},
  {Patricio}, {Schaye}, {Shirazi}, {Soto}, {Soucail}, {Steinmetz}, {Urrutia},
  {Weilbacher}, \& {de Zeeuw}}]{2015Bacon}
{Bacon}, R., {Brinchmann}, J., {Richard}, J., {et~al.} 2015, \aap, 575, A75

\bibitem[{Borgnino(1990)}]{1990Borgnino}
Borgnino, J. 1990, Appl. Opt., 29, 1863

\bibitem[{{Bustos} \& {Tokovinin}(2018)}]{2018Bustos}
{Bustos}, E. \& {Tokovinin}, A. 2018, in Society of Photo-Optical
  Instrumentation Engineers (SPIE) Conference Series, Vol. 10700, Ground-based
  and Airborne Telescopes VII, ed. H.~K. {Marshall} \& J.~{Spyromilio}, 107000Q

\bibitem[{{Butterley} {et~al.}(2020){Butterley}, {Wilson}, {Sarazin},
  {Dubbeldam}, {Osborn}, \& {Clark}}]{2020Butterley}
{Butterley}, T., {Wilson}, R.~W., {Sarazin}, M., {et~al.} 2020, MNRAS, 492, 934

\bibitem[{{Esparza-Arredondo} {et~al.}(2024){Esparza-Arredondo},
  {Garc\'{\i}a-Lorenzo}, {Acosta-Pulido}, \& {et al.}}]{2024Esparzaarredondo}
{Esparza-Arredondo}, D., {Garc\'{\i}a-Lorenzo}, B., {Acosta-Pulido}, J., \& {et
  al.} 2024, A\&A, in preparation

\bibitem[{{F{\'e}tick} {et~al.}(2019){F{\'e}tick}, {Fusco}, {Neichel},
  {Mugnier}, {Beltramo-Martin}, {Bonnefois}, {Petit}, {Milli}, {Vernet},
  {Oberti}, \& {Bacon}}]{2019Fetick}
{F{\'e}tick}, R.~J.~L., {Fusco}, T., {Neichel}, B., {et~al.} 2019, \aap, 628,
  A99

\bibitem[{{Floyd} {et~al.}(2010){Floyd}, {Thomas-Osip}, \&
  {Prieto}}]{2010Floyd}
{Floyd}, D. J.~E., {Thomas-Osip}, J., \& {Prieto}, G. 2010, \pasp, 122, 731

\bibitem[{{Fried}(1965)}]{1965Fried}
{Fried}, D.~L. 1965, Journal of the Optical Society of America (1917-1983), 55,
  1427

\bibitem[{{Fusco} {et~al.}(2020){Fusco}, {Bacon}, {Kamann}, {Conseil},
  {Neichel}, {Correia}, {Beltramo-Martin}, {Vernet}, {Kolb}, \&
  {Madec}}]{2020Fusco}
{Fusco}, T., {Bacon}, R., {Kamann}, S., {et~al.} 2020, \aap, 635, A208

\bibitem[{{García-Lorenzo} {et~al.}(2023){García-Lorenzo},
  {Esparza-Arredondo}, {Acosta-Pulido}, \&
  {Castro-Almazán}}]{2023Garcia-Lorenzo}
{García-Lorenzo}, B., {Esparza-Arredondo}, D., {Acosta-Pulido}, J., \&
  {Castro-Almazán}, J. 2023, in COAT2023: Communications and Observations
  through Atmospheric Turbulence, COAT2023: Communications and Observations
  through Atmospheric Turbulence

\bibitem[{{Gonneau} {et~al.}(2020){Gonneau}, {Lyubenova}, {Lan{\c{c}}on},
  {Trager}, {Peletier}, {Arentsen}, {Chen}, {Coelho}, {Dries},
  {Falc{\'o}n-Barroso}, {Prugniel}, {S{\'a}nchez-Bl{\'a}zquez}, {Vazdekis}, \&
  {Verro}}]{2020Gonneau}
{Gonneau}, A., {Lyubenova}, M., {Lan{\c{c}}on}, A., {et~al.} 2020, \aap, 634,
  A133

\bibitem[{{Guesalaga} {et~al.}(2017){Guesalaga}, {Neichel}, {Correia},
  {Butterley}, {Osborn}, {Masciadri}, {Fusco}, \& {Sauvage}}]{2017Guesalaga}
{Guesalaga}, A., {Neichel}, B., {Correia}, C.~M., {et~al.} 2017, \mnras, 465,
  1984

\bibitem[{Harrison(2016)}]{Harrison2016}
Harrison, C.~M. 2016, Integral Field Spectroscopy and Spectral Energy
  Distributions (Cham: Springer International Publishing), 37--46

\bibitem[{{Henry} {et~al.}(1987){Henry}, {Heyer}, {Cecil}, {Barnes}, \&
  {Cheigh}}]{1987Henry}
{Henry}, J.~P., {Heyer}, I., {Cecil}, G., {Barnes}, B., \& {Cheigh}, F. 1987,
  \pasp, 99, 1354

\bibitem[{{Husemann} {et~al.}(2016){Husemann}, {Bennert}, {Scharw{\"a}chter},
  {Woo}, \& {Choudhury}}]{2016Husemann}
{Husemann}, B., {Bennert}, V.~N., {Scharw{\"a}chter}, J., {Woo}, J.-H., \&
  {Choudhury}, O.~S. 2016, \mnras, 455, 1905

\bibitem[{{Husser} {et~al.}(2016){Husser}, {Kamann}, {Dreizler}, {Wendt},
  {Wulff}, {Bacon}, {Wisotzki}, {Brinchmann}, {Weilbacher}, {Roth}, \&
  {Monreal-Ibero}}]{2016Husser}
{Husser}, T.-O., {Kamann}, S., {Dreizler}, S., {et~al.} 2016, A\&A, 588, A148

\bibitem[{{Lai} {et~al.}(2019){Lai}, {Withington}, {Laugier}, \&
  {Chun}}]{2019Lai}
{Lai}, O., {Withington}, J.~K., {Laugier}, R., \& {Chun}, M. 2019, \mnras, 484,
  5568

\bibitem[{{Martinez} {et~al.}(2010){Martinez}, {Kolb}, {Tokovinin}, \&
  {Sarazin}}]{2010Martinez}
{Martinez}, P., {Kolb}, J., {Tokovinin}, A., \& {Sarazin}, M. 2010, \aap, 516,
  A90

\bibitem[{{Mehner} {et~al.}(2021){Mehner}, {Janssens}, {Agliozzo}, {de Wit},
  {Boffin}, {Baade}, {Bodensteiner}, {Groh}, {Mahy}, \& {Vogt}}]{2021Mehner}
{Mehner}, A., {Janssens}, S., {Agliozzo}, C., {et~al.} 2021, \aap, 655, A33

\bibitem[{{Munro} {et~al.}(2023){Munro}, {Hansen}, {Travouillon}, {Grosse}, \&
  {Tokovinin}}]{2023Munro}
{Munro}, J., {Hansen}, J., {Travouillon}, T., {Grosse}, D., \& {Tokovinin}, A.
  2023, Journal of Astronomical Telescopes, Instruments, and Systems, 9, 017004

\bibitem[{{Osborn} \& {Alaluf}(2023)}]{2023Osborn}
{Osborn}, J. \& {Alaluf}, D. 2023, \mnras, 525, 1936

\bibitem[{{Osborn} {et~al.}(2018){Osborn}, {Wilson}, {Sarazin}, {Butterley},
  {Chac{\'o}n}, {Derie}, {Farley}, {Haubois}, {Laidlaw}, {LeLouarn},
  {Masciadri}, {Milli}, {Navarrete}, \& {Townson}}]{2018Osborn}
{Osborn}, J., {Wilson}, R.~W., {Sarazin}, M., {et~al.} 2018, \mnras, 478, 825

\bibitem[{{Otarola}(2023)}]{2023Otarola}
{Otarola}, A. 2023, in COAT2023: Communications and Observations through
  Atmospheric Turbulence, COAT2023: Communications and Observations through
  Atmospheric Turbulence

\bibitem[{{Piqueras} {et~al.}(2019){Piqueras}, {Conseil}, {Shepherd}, {Bacon},
  {Leclercq}, \& {Richard}}]{2019Piqueras}
{Piqueras}, L., {Conseil}, S., {Shepherd}, M., {et~al.} 2019, in Astronomical
  Society of the Pacific Conference Series, Vol. 521, Astronomical Data
  Analysis Software and Systems XXVI, ed. M.~{Molinaro}, K.~{Shortridge}, \&
  F.~{Pasian}, 545

\bibitem[{{Racine} {et~al.}(1991){Racine}, {Salmon}, {Cowley}, \&
  {Sovka}}]{1991Racine}
{Racine}, R., {Salmon}, D., {Cowley}, D., \& {Sovka}, J. 1991, \pasp, 103, 1020

\bibitem[{{Roddier}(1981)}]{1981Roddier}
{Roddier}, F. 1981, Progess in Optics, 19, 281

\bibitem[{{S{\'a}nchez-Bl{\'a}zquez} {et~al.}(2006){S{\'a}nchez-Bl{\'a}zquez},
  {Peletier}, {Jim{\'e}nez-Vicente}, {Cardiel}, {Cenarro},
  {Falc{\'o}n-Barroso}, {Gorgas}, {Selam}, \&
  {Vazdekis}}]{2006Sanchez-Blazquez}
{S{\'a}nchez-Bl{\'a}zquez}, P., {Peletier}, R.~F., {Jim{\'e}nez-Vicente}, J.,
  {et~al.} 2006, \mnras, 371, 703

\bibitem[{{Sarazin}(2017)}]{ManualESO}
{Sarazin}, M. 2017, Astronomical Site Monitor Data User Manual (ESO-281474),
  ESO-281474, 29

\bibitem[{{Sarazin} \& {Roddier}(1990)}]{1990Sarazin}
{Sarazin}, M. \& {Roddier}, F. 1990, \aap, 227, 294

\bibitem[{{Skidmore} {et~al.}(2009){Skidmore}, {Els}, {Travouillon}, {Riddle},
  {Sch{\"o}ck}, {Bustos}, {Seguel}, \& {Walker}}]{2009Skidmore}
{Skidmore}, W., {Els}, S., {Travouillon}, T., {et~al.} 2009, \pasp, 121, 1151

\bibitem[{{Stubbs}(2021)}]{2021Stubbs}
{Stubbs}, C.~W. 2021, \mnras, 508, 3936

\bibitem[{{Tokovinin}(2002)}]{2002Tokovinin}
{Tokovinin}, A. 2002, \pasp, 114, 1156

\bibitem[{{Tokovinin} {et~al.}(2007){Tokovinin}, {Sarazin}, \&
  {Smette}}]{2007Tokovinin}
{Tokovinin}, A., {Sarazin}, M., \& {Smette}, A. 2007, \mnras, 378, 701

\bibitem[{{Toselli} {et~al.}(2007){Toselli}, {Andrews}, {Phillips}, \&
  {Ferrero}}]{2007Toselli}
{Toselli}, I., {Andrews}, L.~C., {Phillips}, R.~L., \& {Ferrero}, V. 2007, in
  Society of Photo-Optical Instrumentation Engineers (SPIE) Conference Series,
  Vol. 6551, Atmospheric Propagation IV, ed. C.~Y. {Young} \& G.~C.
  {Gilbreath}, 65510E

\bibitem[{{Trujillo} {et~al.}(2001){Trujillo}, {Aguerri}, {Cepa}, \&
  {Guti{\'e}rrez}}]{2001Trujillo}
{Trujillo}, I., {Aguerri}, J.~A.~L., {Cepa}, J., \& {Guti{\'e}rrez}, C.~M.
  2001, \mnras, 328, 977

\bibitem[{{V{\'a}zquez Rami{\'o}} {et~al.}(2012){V{\'a}zquez Rami{\'o}},
  {Vernin}, {Mu{\~n}oz-Tu{\~n}{\'o}n}, {Sarazin}, {Varela}, {Trinquet},
  {Delgado}, {Fuensalida}, {Reyes}, {Benhida}, {Benkhaldoun}, {Garc{\'\i}a
  Lambas}, {Hach}, {Lazrek}, {Lombardi}, {Navarrete}, {Recabarren}, {Renzi},
  {Sabil}, \& {Vrech}}]{2012Hector}
{V{\'a}zquez Rami{\'o}}, H., {Vernin}, J., {Mu{\~n}oz-Tu{\~n}{\'o}n}, C.,
  {et~al.} 2012, \pasp, 124, 868

\bibitem[{{Voitsekhovich}(1995)}]{1995Voitsekhovich}
{Voitsekhovich}, V.~V. 1995, Journal of the Optical Society of America A, 12,
  1346

\bibitem[{{Winker}(1991)}]{1991Winker}
{Winker}, D.~M. 1991, Journal of the Optical Society of America A, 8, 1568

\bibitem[{{Woolf}(1979)}]{Woolf1979}
{Woolf}, N. 1979, \pasp, 91, 523

\bibitem[{{Ziad}(2016)}]{2016Ziad}
{Ziad}, A. 2016, in Society of Photo-Optical Instrumentation Engineers (SPIE)
  Conference Series, Vol. 9909, Adaptive Optics Systems V, ed. E.~{Marchetti},
  L.~M. {Close}, \& J.-P. {V{\'e}ran}, 99091K

\bibitem[{{Ziad} {et~al.}(2000){Ziad}, {Conan}, {Tokovinin}, {Martin}, \&
  {Borgnino}}]{2000Ziad}
{Ziad}, A., {Conan}, R., {Tokovinin}, A., {Martin}, F., \& {Borgnino}, J. 2000,
  \ao, 39, 5415

\end{thebibliography}

\begin{appendix} 

\onecolumn
\section{Comparison of the two analytical functions approaching the image quality }
\label{appenIQ}

\begin{figure*}[h]
\begin{center}
\includegraphics[trim={1.75cm 1.75cm 1.75cm 1.75cm},clip,width=6cm,angle=180]{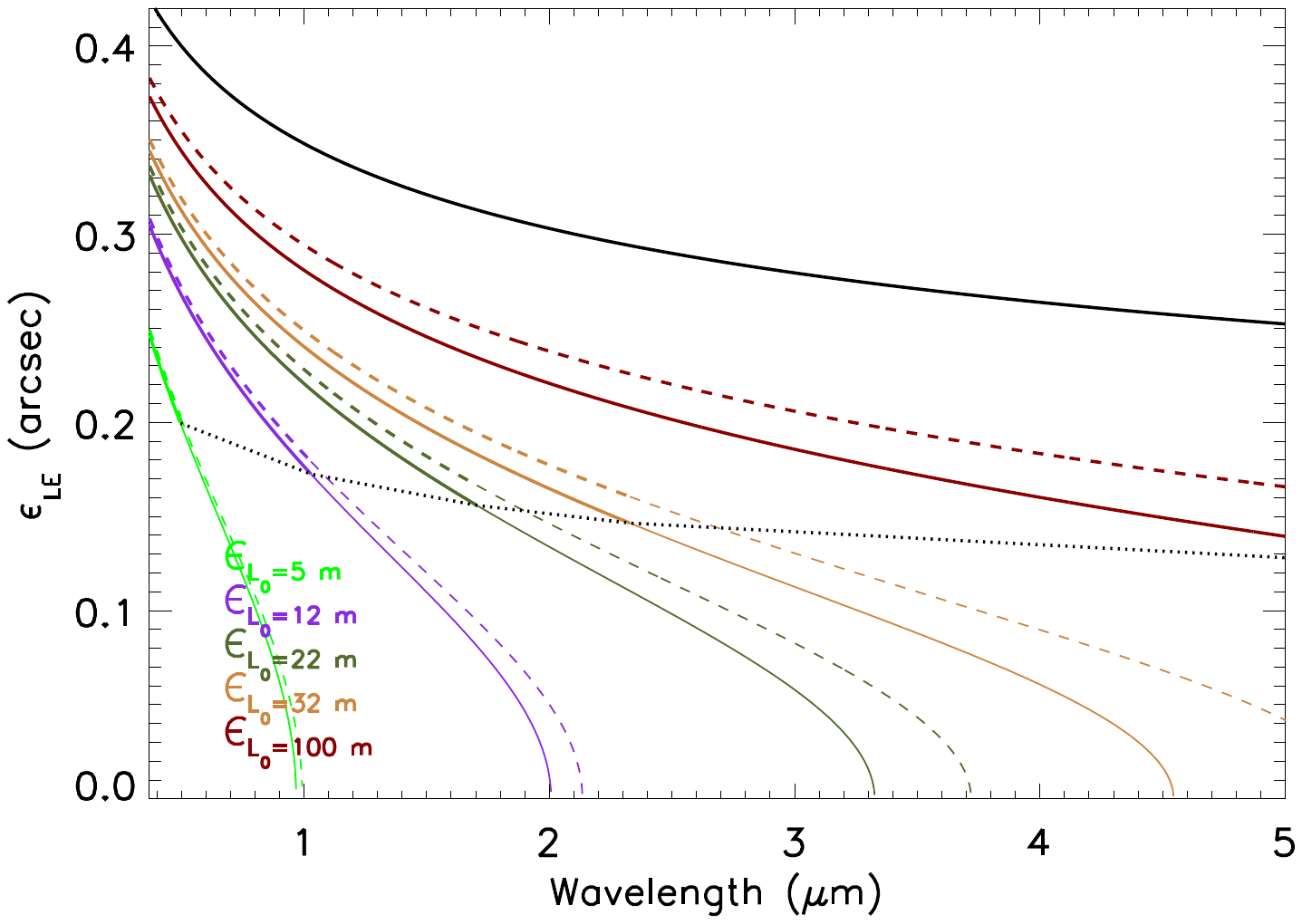}
\includegraphics[trim={1.75cm 1.75cm 1.75cm 1.75cm},clip,width=6cm,angle=180]{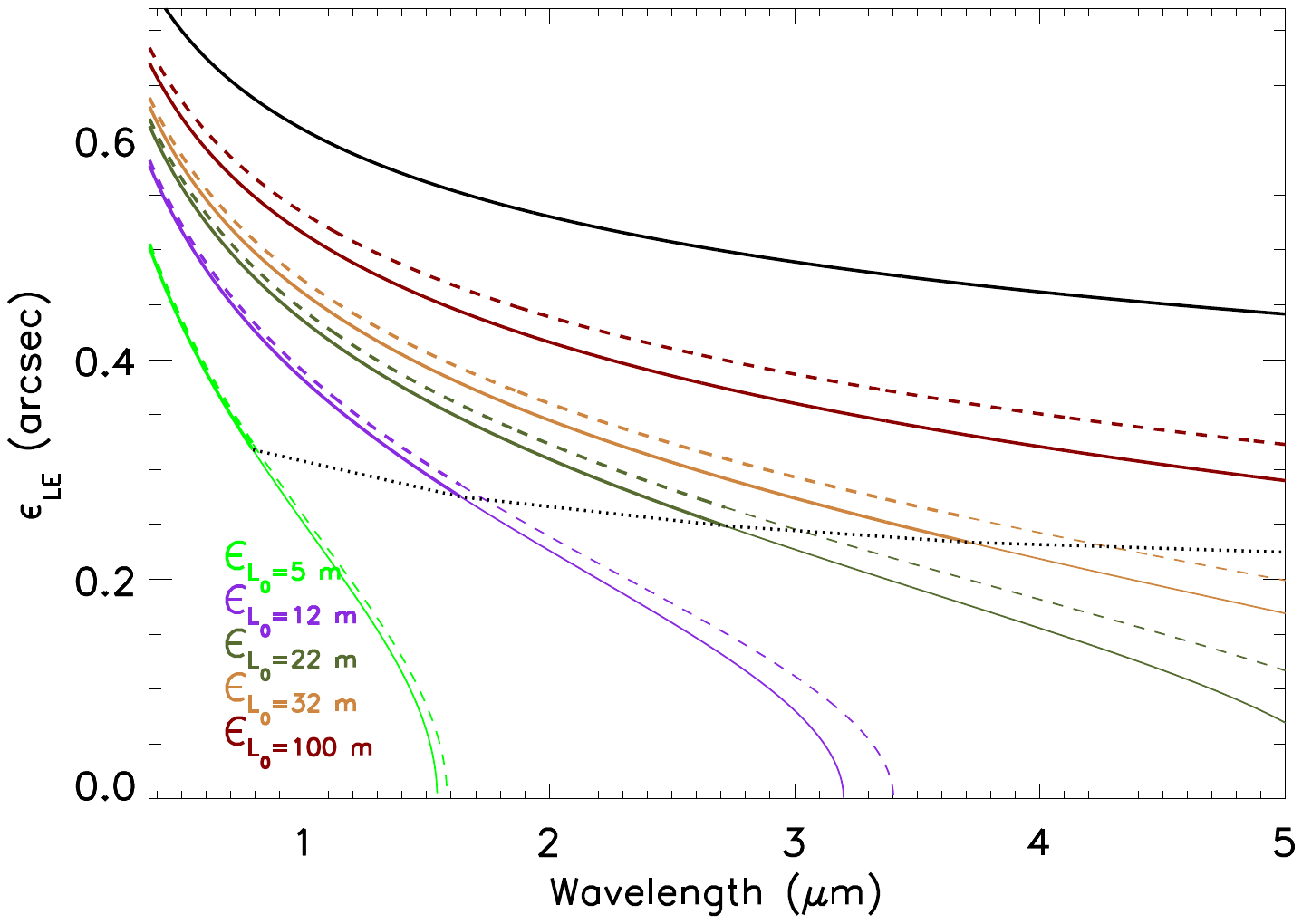}
\includegraphics[trim={1.75cm 1.75cm 1.75cm 1.75cm},clip,width=6cm,angle=180]{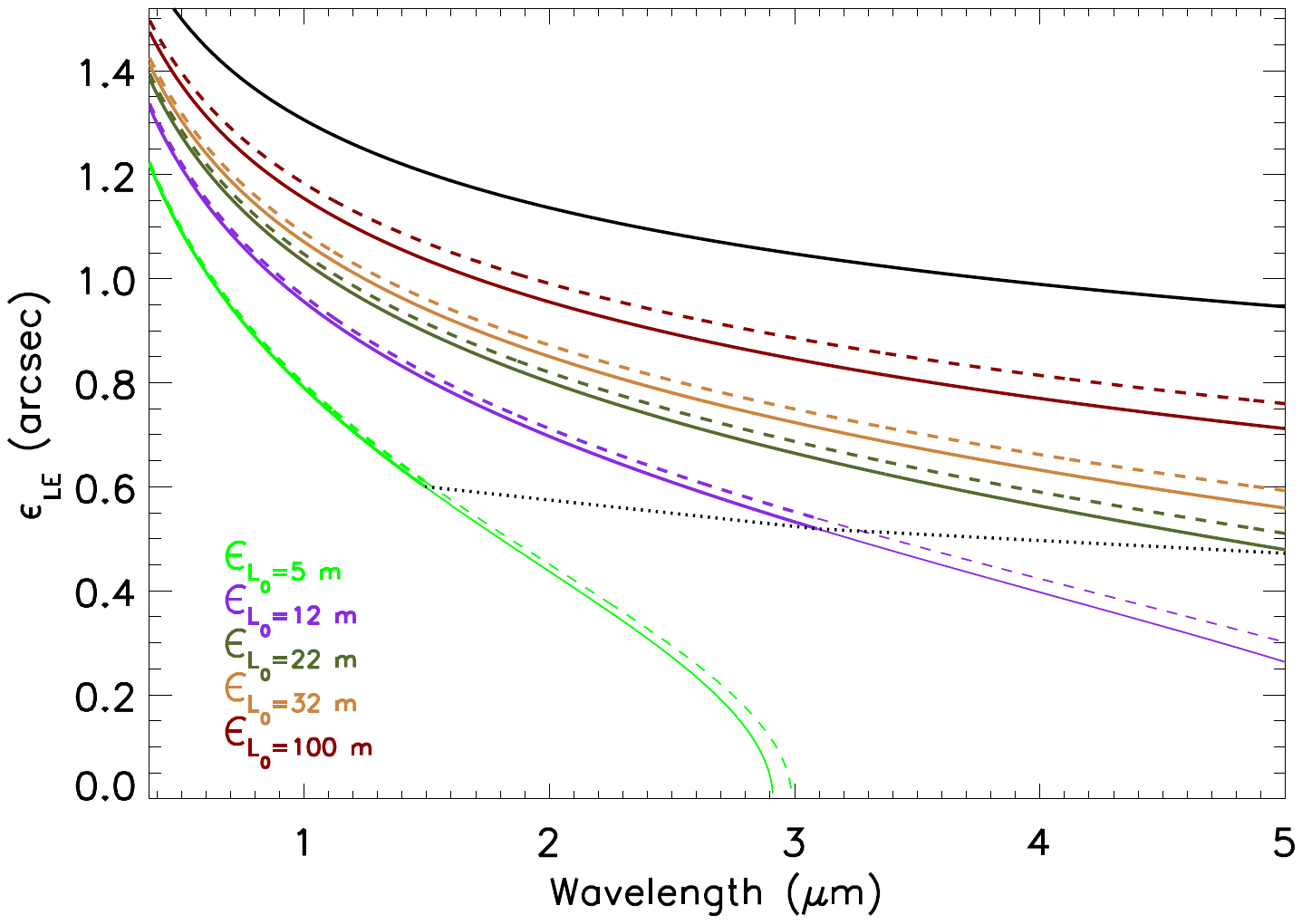}
\includegraphics[trim={1.75cm 1.75cm 1.75cm 1.75cm},clip,width=6cm,angle=180]{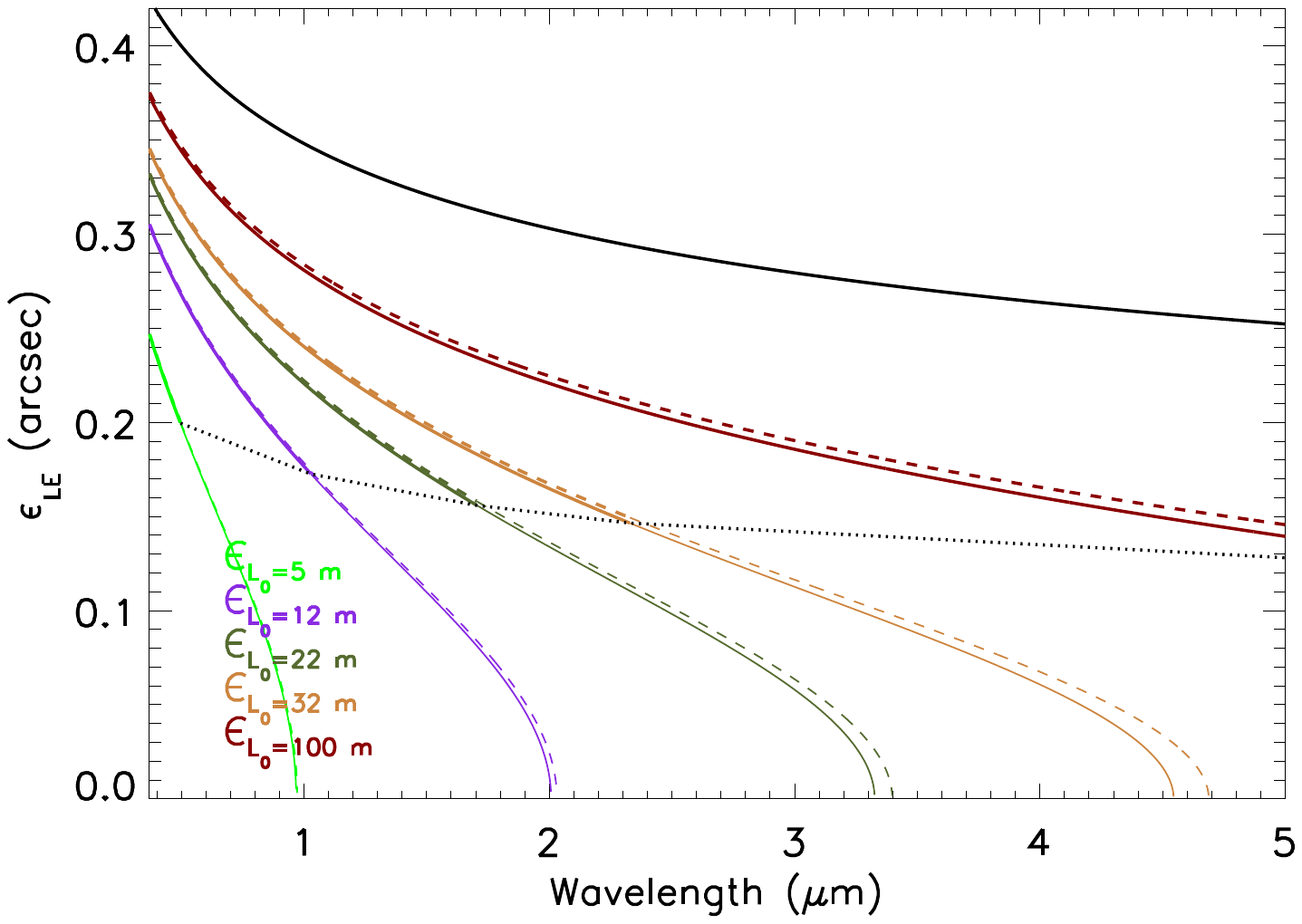}
\includegraphics[trim={1.75cm 1.75cm 1.75cm 1.75cm},clip,width=6cm,angle=180]{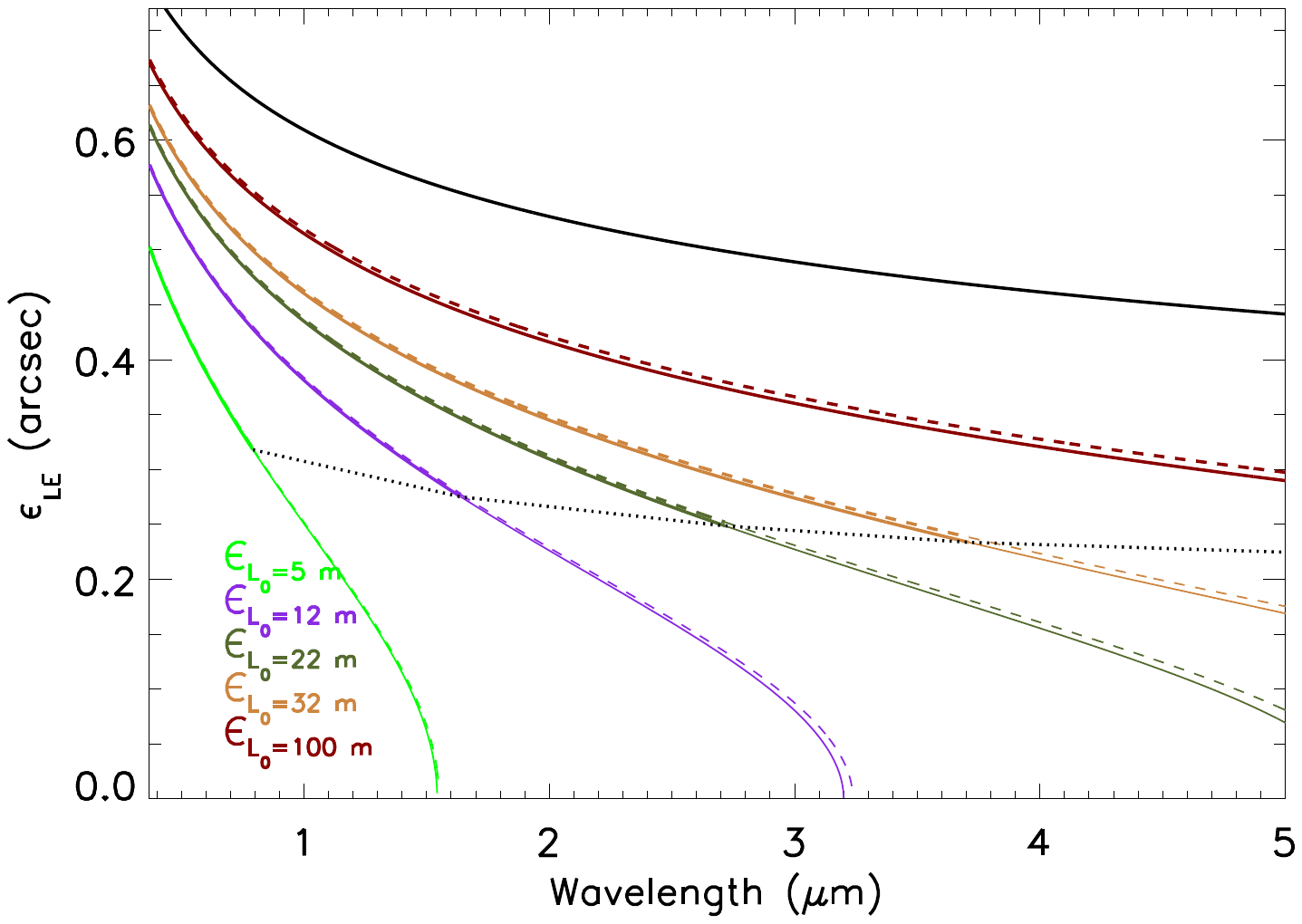}
\includegraphics[trim={1.75cm 1.75cm 1.75cm 1.75cm},clip,width=6cm,angle=180]{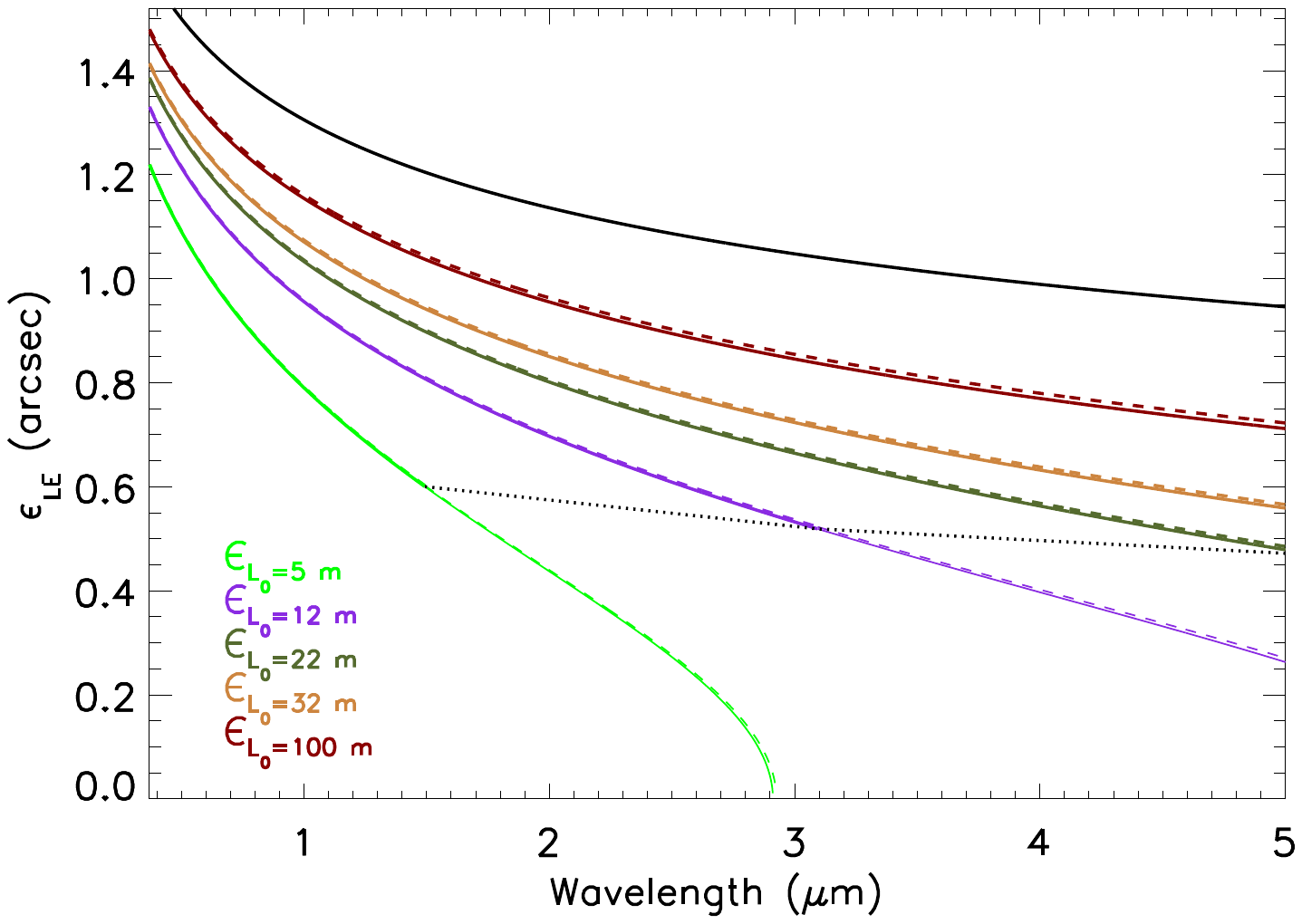}
\includegraphics[trim={1.5cm 1.75cm 1.75cm 1.75cm},clip,width=6cm,angle=180]{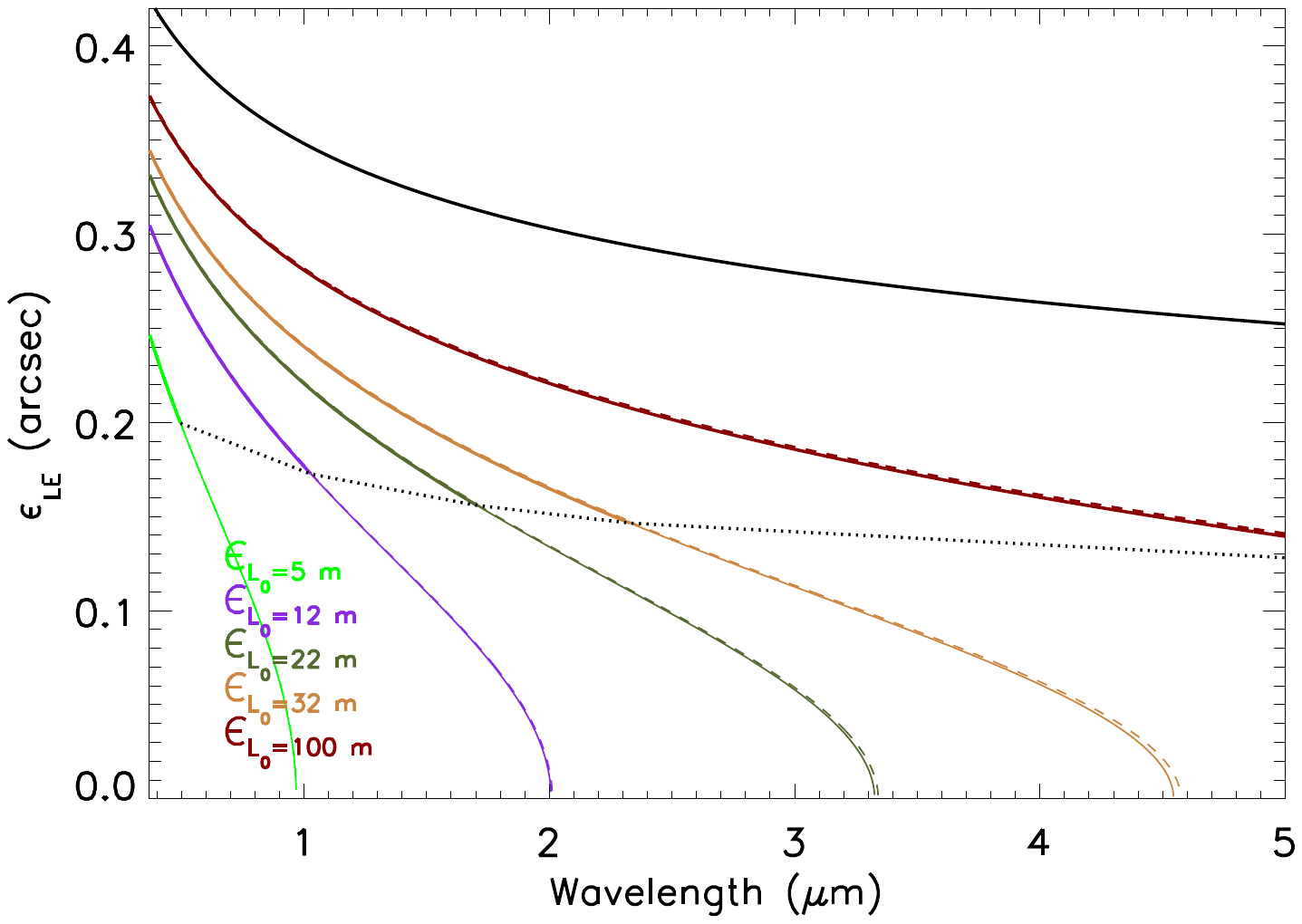}
\includegraphics[trim={1.75cm 1.75cm 1.75cm 1.75cm},clip,width=6cm,angle=180]{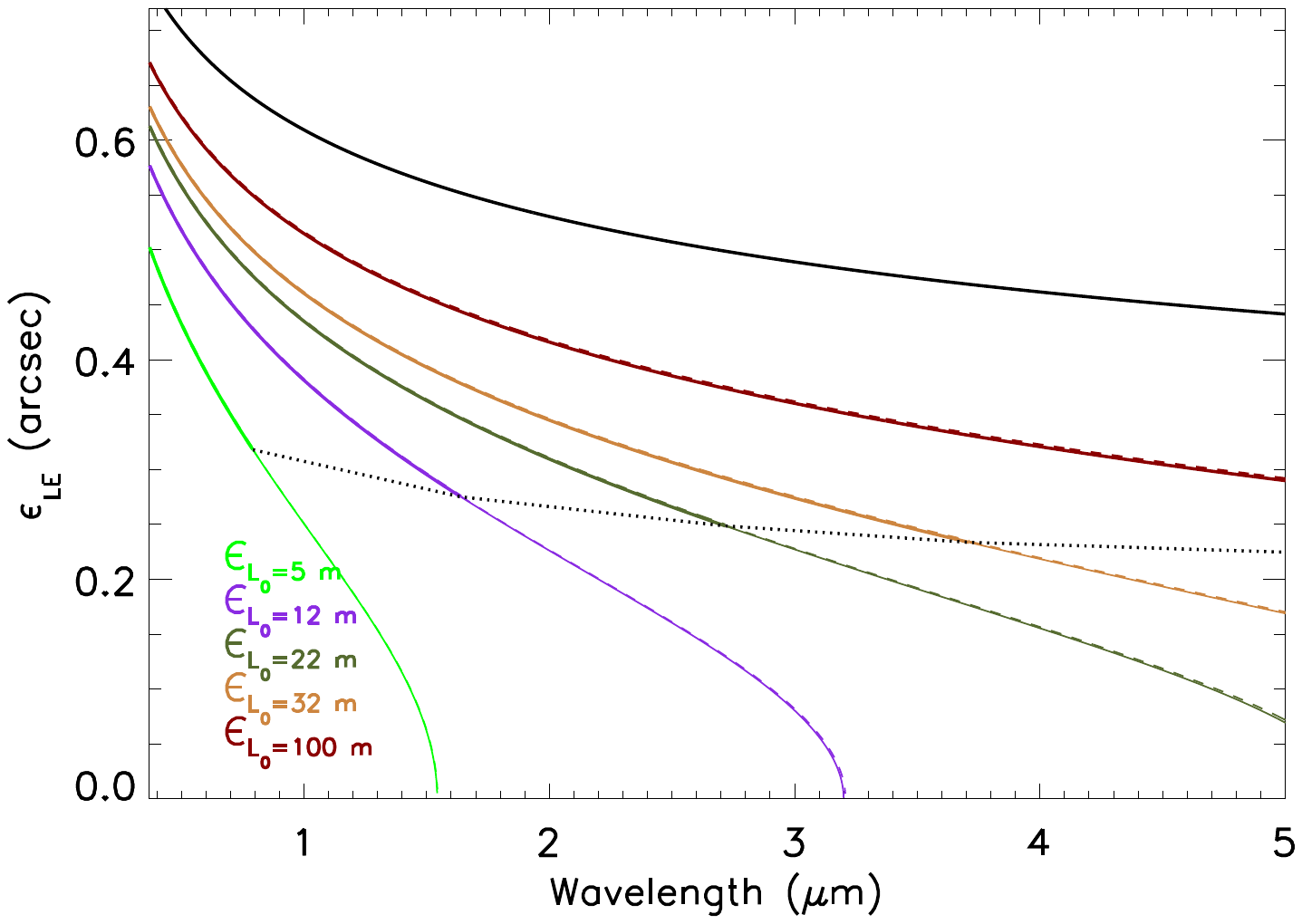}
\includegraphics[trim={1.75cm 1.75cm 1.75cm 1.75cm},clip,width=6cm,angle=180]{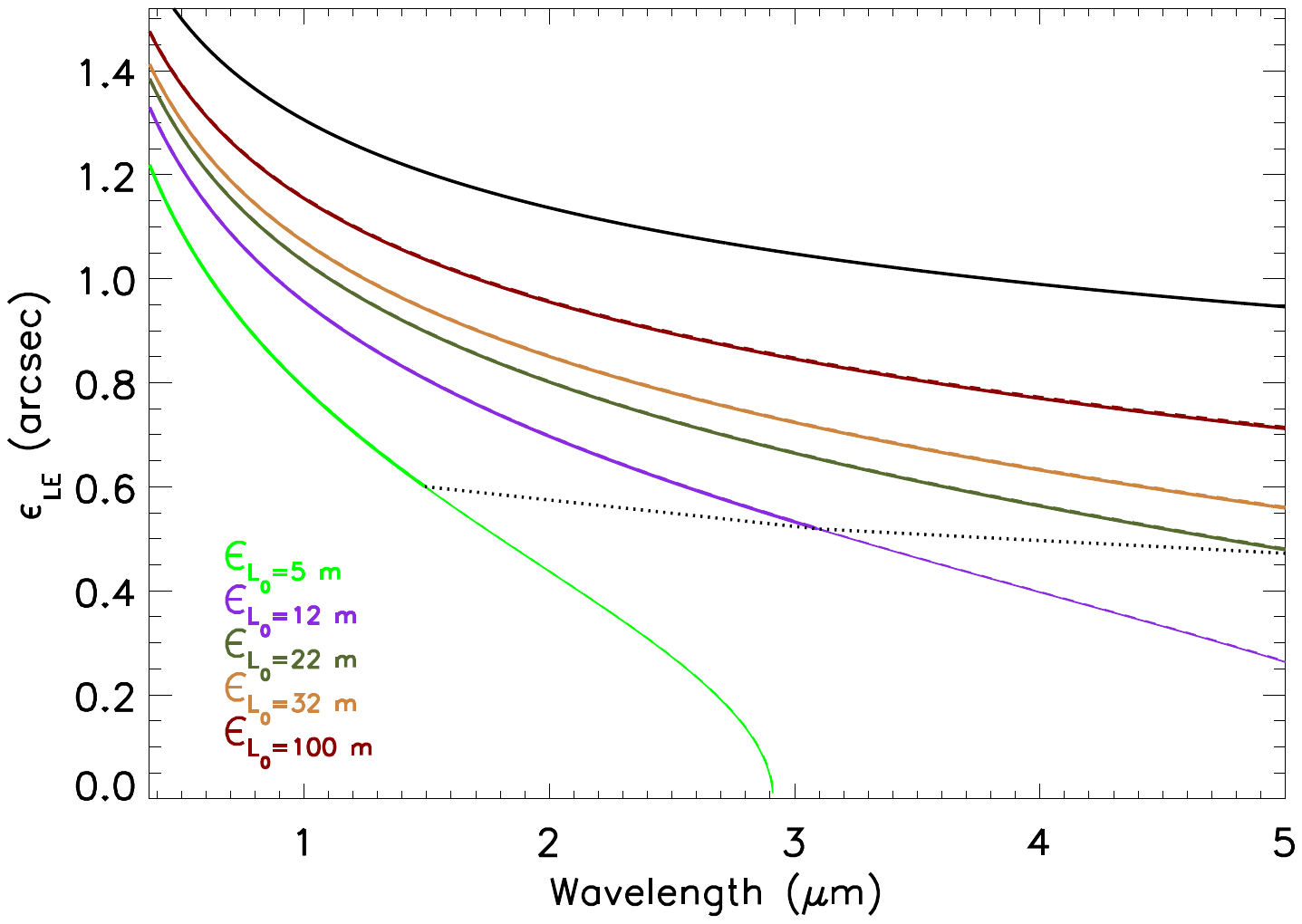}
 \caption{Predicted image quality for a long-exposure image (e.g., $\epsilon_{LE}$) of a point-like source obtained using ground-based telescopes with diameters of (top) 1.5 m, (middle) 8 m, and (bottom) 39 m, as a function of the wavelength. Panels correspond to different $\epsilon_{0}$ conditions at 5000 \AA: (left panels) 0.4 arcsecs, (central panels), 0.7 arcsecs, and (right panels) 1.5 arcsec. Each curve corresponds to the indicated value of the \LO of the turbulence in meters: \LO=100 $\rightarrow$ red, \LO=32$\rightarrow$ orange, \LO=22 $\rightarrow$ olive green, \LO=12 $\rightarrow$ purple, and \LO=5 $\rightarrow$ lime green}. The black curve follows the $\epsilon_{0}$ wavelength dependence ($\epsilon{0} (\lambda) \propto \lambda^{-1/5}$), corresponding to \LO=$\infty$. Solid curves correspond to the first-order approach to $\epsilon{LE}$ following Eq. \ref{eq:ELE_lambda}, while dashed curves to the second-order approach considering the telescope diameter (Eq. \ref{eq:ELE_lambda_D}). The dotted line delineates the $\mathcal{L}_{0}$/r${_0}$>20 boundary, above which errors are constrained within $\pm1$\%.
 
\label{IQ_seeing}
\end{center}
\end{figure*}

\begin{figure*}[h]
\begin{center}
\includegraphics[trim={1.75cm 1.75cm 1.75cm 1.75cm},clip,width=6cm,angle=180]{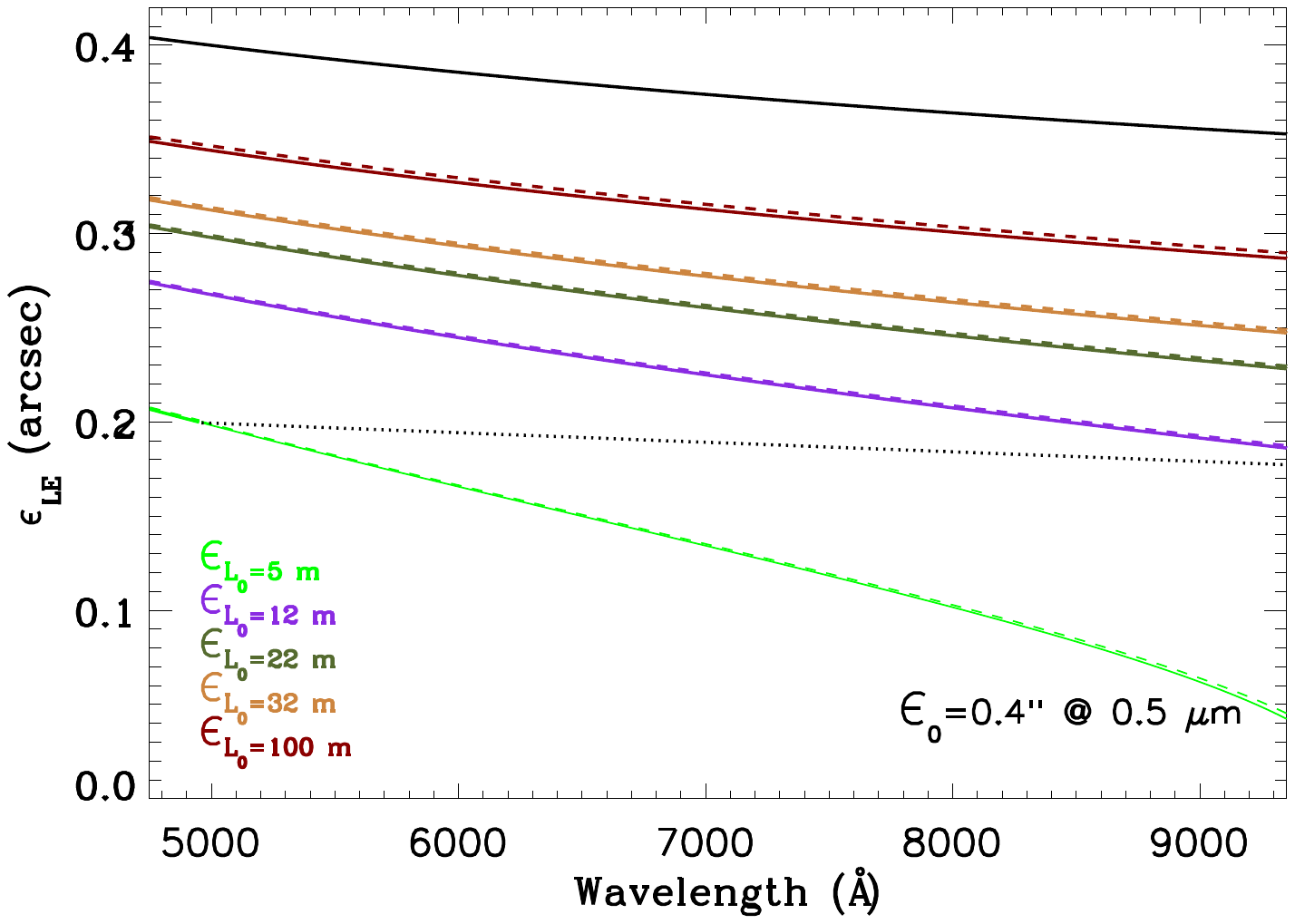}
\includegraphics[trim={1.75cm 1.75cm 1.75cm 1.75cm},clip,width=6cm,angle=180]{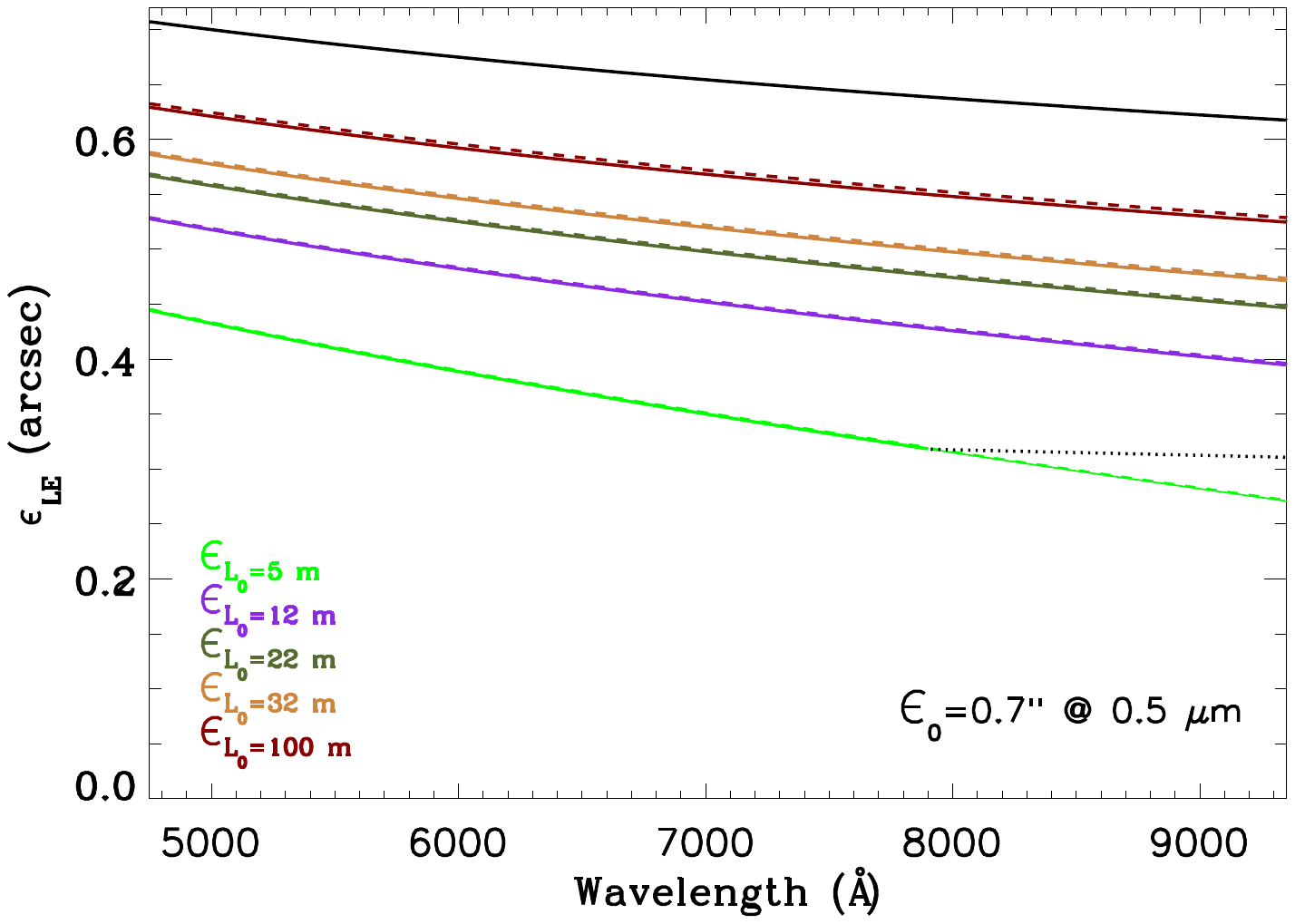}
\includegraphics[trim={1.75cm 1.75cm 1.75cm 1.75cm},clip,width=6cm,angle=180]{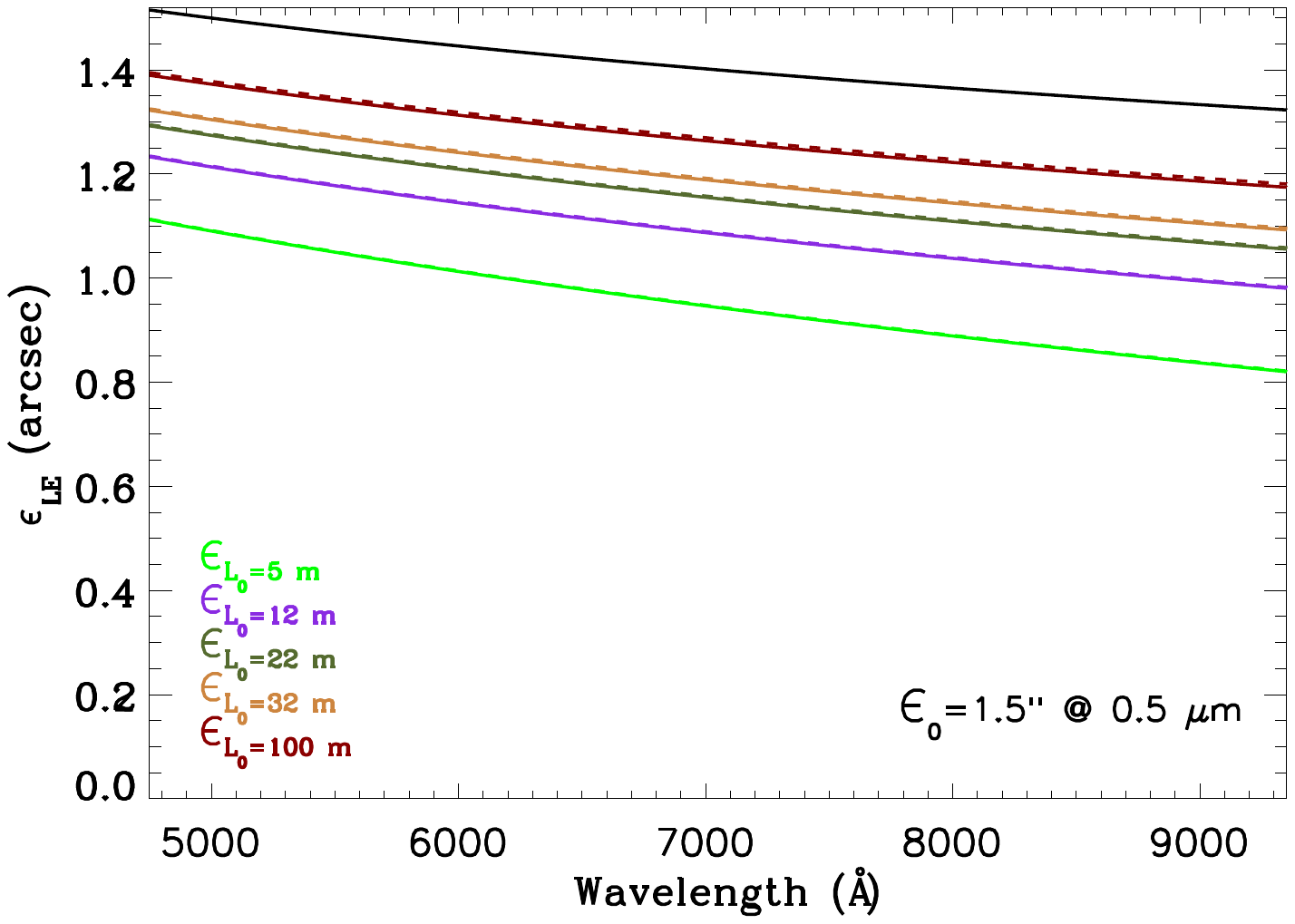}
 \caption{Same as Fig. \ref{IQ_seeing}, but for the parameters of the MUSE instrument, the IFS installed at the 8-m VLT operating in the optical range between 4750 and 9350 \AA.}
\label{IQ_seeing_MUSE}
\end{center}
\end{figure*}

We have computed the predicted IQ as a function of wavelength considering various telescope diameters and turbulence conditions (Fig. \ref{IQ_seeing}) using Eqs. \ref{eq:ELE_lambda} and \ref{eq:ELE_lambda_D}, and tracking the atmospheric conditions where $\mathcal{L}_{0}$/r$_{0}$ > 20. Figure \ref{IQ_seeing} shows that both equations provide similar assessments for the IQ at the focal plane of telescopes. Differences become more notable for decreasing telescope diameter and increasing the size of the \LO parameter of the turbulence.
As a general trend, it is important to note that these approaches apply only to wavelengths within the range of 3700 to 9300 \AA \ for \LO $\leq$ 12 m and excellent $\epsilon_{0}$ conditions, to maintain relative error within $\pm$1\%. For typical $\epsilon_{0}$ \citep[e.g.,][]{2009Skidmore, 2012Hector} and \LO \citep[e.g.,][]{2016Ziad} at astronomical sites, the applicability without exceeding the $\pm$1\% error extends up to 15000 \AA . In the case of poor atmospheric turbulence conditions, these approaches are usable with $\pm$1\% uncertainty up to 3 $\mu$m. Figure \ref{IQ_seeing_MUSE} zooms in on the predicted IQ($\lambda$) for the spectral range of the MUSE instrument, the data of which are used in this work to illustrate the applicability of the proposed method for estimating the \LO of turbulence from IFS observations.

\begin{figure*}
\begin{center}
\includegraphics[width=\textwidth,angle=0]{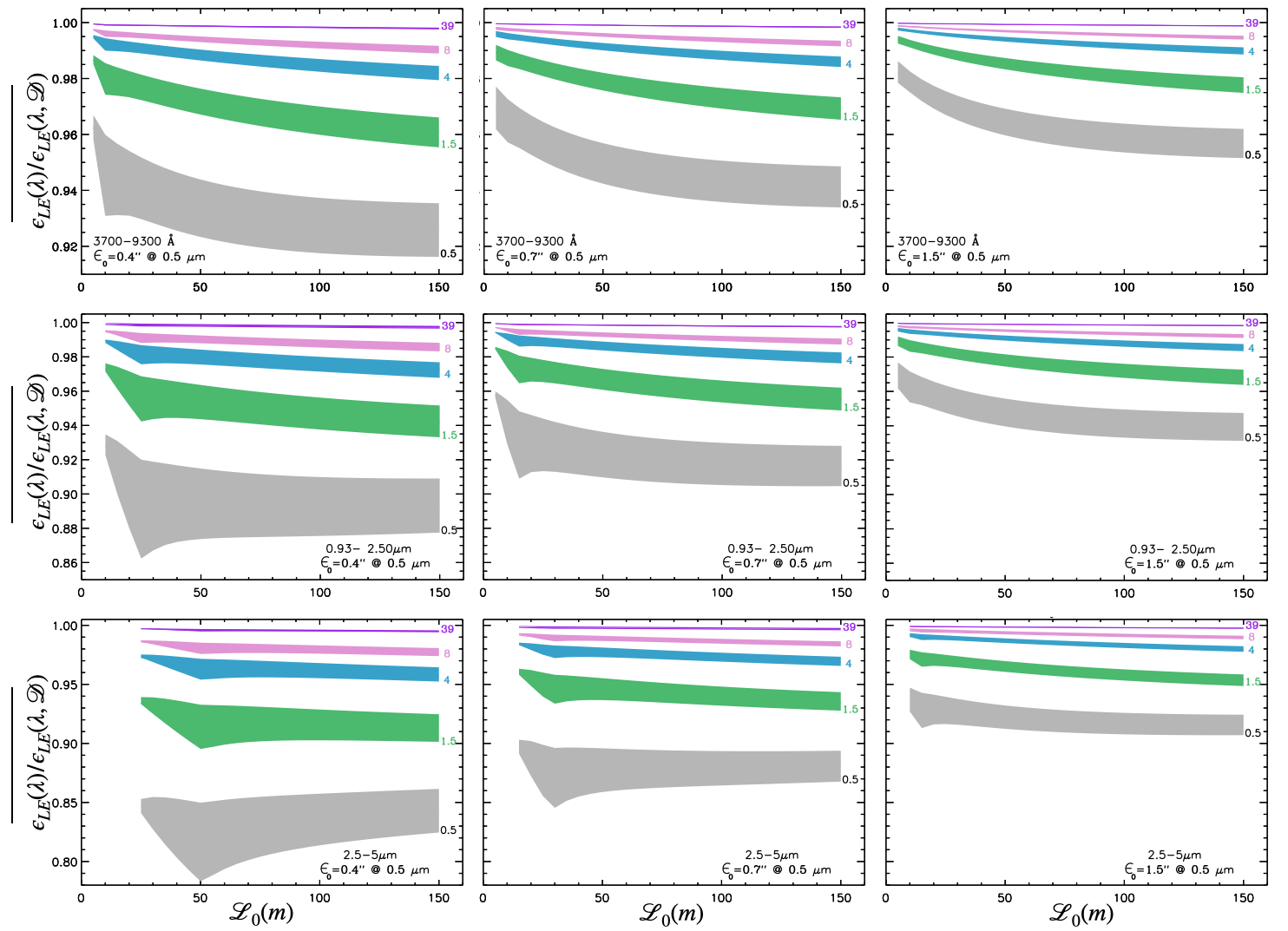}
 \caption{Wavelength average ratio of the two approaches for estimating the image quality at the focal plane of a telescope (i.e., $\epsilon_{LE}$($\lambda$)/$\epsilon_{LE}$($\lambda$, $\mathcal{D}$ from Eqs. \ref{eq:ELE_lambda} and \ref{eq:ELE_lambda_D}) as a function of the \LO of the turbulence for the spectral ranges (top-panels) 3700-9300 \AA; (middle-panels) 0.93-2.5 $\mu$m; and (bottom-panels) 2.5-5.0 $\mu$m at excellent (left-panels), typical (central-panels), and fair (right-panels) atmosheric turbulence conditions. We have considered only those conditions with $\mathcal{L}_{0}$/r${_0}$>20, ensuring errors are within $\pm$1\%. The shaded area marks the one-sigma of the mean. Colors indicate different telescope diameters as indicated at the end-right. 
}
\label{IQ_ratio_mean}
\end{center}
\end{figure*}

Figure \ref{IQ_ratio_mean} shows the wavelength average ratio of $\epsilon_{LE}$($\lambda$) and $\epsilon_{LE}$($\lambda$, $\mathcal{D}$) as a function of the \LO for different $\epsilon_{0}$ conditions and telescope diameters. The diameter becomes relevant for small telescopes (i.e., $\mathcal{D}$ < 1.5 m), underestimating the image quality by more than $\sim$2\%, $\sim$4\%, and $\sim$6\% in the optical (i.e., 3700-9300 \AA ), 0.93-2.5 $\mu$m, and 2.5-5 $\mu$m spectral ranges, respectively. For middle to large telescopes (4m to 10m-class telescopes), the image quality is underestimated by a factor always < 5\% in the entire near-infrared range (i.e., 0.93-5 $\mu$m), being <2\% at the optical for any turbulence condition. For extremely large telescopes (i.e., $\mathcal{D}$ > $\sim$20 m), the $\epsilon_{LE}$($\lambda$)/$\epsilon_{LE}$($\lambda$, $\mathcal{D}$) is always almost 1.

\twocolumn
\section{Filter-band images recovered from IFS data}
\label{appenB}

In this appendix, we use a simple model to analyze the putative impact of the filter-band width selected to recover filter-band images from IFS data cubes on the expected IQ of the image. We also include the parameters derived for the Moffat profiles fit to model the IQ of filter-band images recovered from MUSE observations.

\subsection{Impact of the filter-band width}
\label{filter-impact}

\begin{figure}
\begin{center}
\includegraphics[trim={0cm 2.cm 1cm 1.75cm},clip,width=9.25cm]{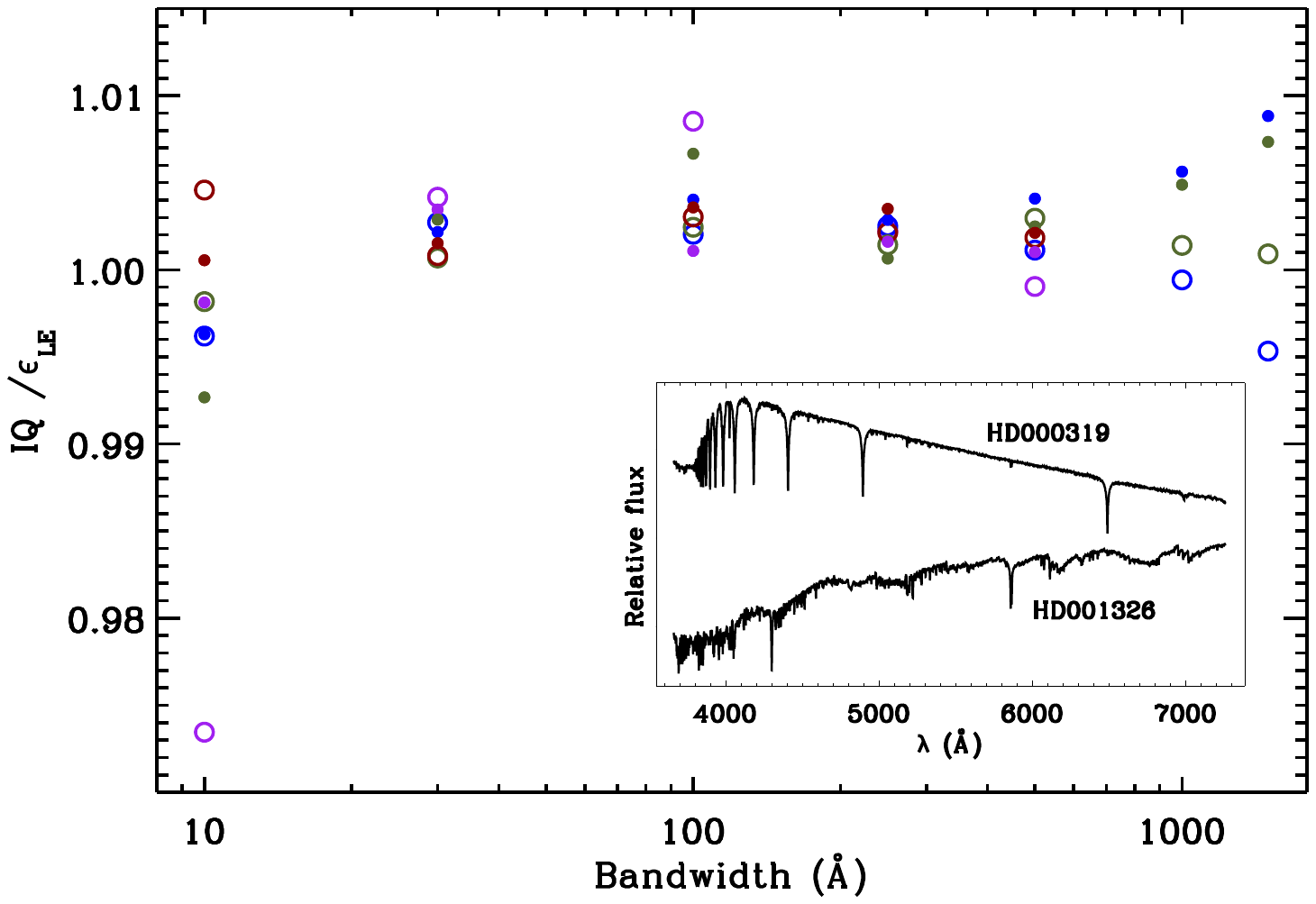}

 \caption{IQ of band-filter images recovered from mock IFS observations of the stars in the inset plot, relative to the input $\epsilon_{LE}$ for the simulation, plotted as a function of the bandwidths used for image recovery. Turbulence conditions were set to $\epsilon_{0}$=0.83 arcsecs and \LO=22 m to compute $\epsilon_{LE}$ from equation \ref{eq:ELE_lambda}. Filled and open circles correspond to mock IFS observations of the A1V (HD000136) and M6 V (HD001326) type stars, respectively. Colors correspond to the central wavelength of the band filters: $\lambda_{1}=4000 \AA \rightarrow$ purple, $\lambda_{2}=5000 \AA \rightarrow$ blue, $\lambda_{3}=6000 \AA \rightarrow$ green, and $\lambda_{4}=7000 \AA \rightarrow$ red. 
 }
\label{gaussian_model_filter}
\end{center}
\end{figure}

We performed a simple simulation of seeing-limited IFS observations in the optical range spanning from 3500 to 7500 \AA \ for two point-like sources to explore the effect of bandwidth on the measured IQ. We chose the spectra of HD000319 (type A1V) and HD001326B (type M6 V) stars (see Fig. \ref{gaussian_model_filter}) from the MILES stellar library \citep{2006Sanchez-Blazquez}. We selected two stars with different spectral emissions to also assess the influence of the source spectral behavior on the IQ measurements from filter-band images.

We generated a mock datacube for a $1\arcmin\times1\arcmin$ field-of-view, using spaxels with dimensions ($\Delta\alpha,\Delta\delta,\Delta\lambda$)=($0\farcs1,0\farcs1,1$\AA). We modeled the seeing-limited PSF using a circular Gaussian distribution, with the FWHM varying with wavelength in accordance with the predicted $\epsilon_{LE}$($\lambda)$ derived from equation \ref{eq:ELE_lambda} for a $\epsilon_{0}$ and \LO of 0.83 arcsecs and 22 m, respectively \citep{2010Martinez}, typical values at the Paranal observatory. Each wavelength slice was perturbed by normally-distributed pseudo-random noise with a mean of zero and a standard deviation of one, ensuring a minimum S/N of 5 per slice.

From the mock data cubes, we chose four central wavelengths ($\lambda_{1}=4000$, $\lambda_{2}=5000$, $\lambda_{3}=6000$, and $\lambda_{4}=7000$ \AA) to recover five band-filter images at each wavelength, using bandwidths of 10, 30, 100, 250, and 500 \AA. Additionally, we recovered broad-band filter images with widths of 1000 and 1500 \AA, centered on $\lambda_{2}$ and $\lambda_{3}$. 

Figure \ref{gaussian_model_filter} presents a comparison between the expected $\epsilon_{LE}$ at $\lambda_{i}$ and the measured FWHM${\lambda_{i}}$ of recovered filter images within the spectral bands [$\lambda_{i}-\frac{N}{2},\lambda_{i}+\frac{N}{2}$], with i=1,2,3,4, and N=30,100,250,500, and N=1000, 1500 for i=2,3. Note that we selected $\Delta\lambda$=1 \AA \ in the mock data cube to simplify the assessment. The resulting IQ varies less than $\pm1\%$ with respect to the input $\epsilon_{LE}$($\lambda_{i})$ of the slice at $\lambda_{i}$ corresponding to the central wavelength of the bands in almost all the recovered images.

For 10 \AA \ bandwidth filter images, the S/N strongly contributes to the discrepancies between IQ and $\epsilon_{LE}$. A minimum S/N of 25 seems to be an adequate threshold for the measurement of the image PSF widths. The spectral energy distribution of the observed object also impacts the outcomes, with more notable deviations occurring when the central wavelength of the band filter coincides with a spectral region of weak emission in the source. Therefore, in general, deviations tend to increase with wider bands, although the S/N of recovered band-filter images and the spectrum of the observed object could also contribute to these deviations. Henceforth, for band-filter images obtained from seeing-limited IFS observations with S/N$\geq$25, we assume uncertainties less than 1\% of the PSF FWHM, specifically associated with the chosen filter bandwidths for image recovery.

\subsection{Parameters of the Moffat models fitted to filter-band images of isolated stars in MUSE data cubes} 
\label{appenA}

\begin{figure*}[ht]
\begin{center}
\includegraphics[trim={3cm 10.cm 3cm 3cm},clip,width=8cm]{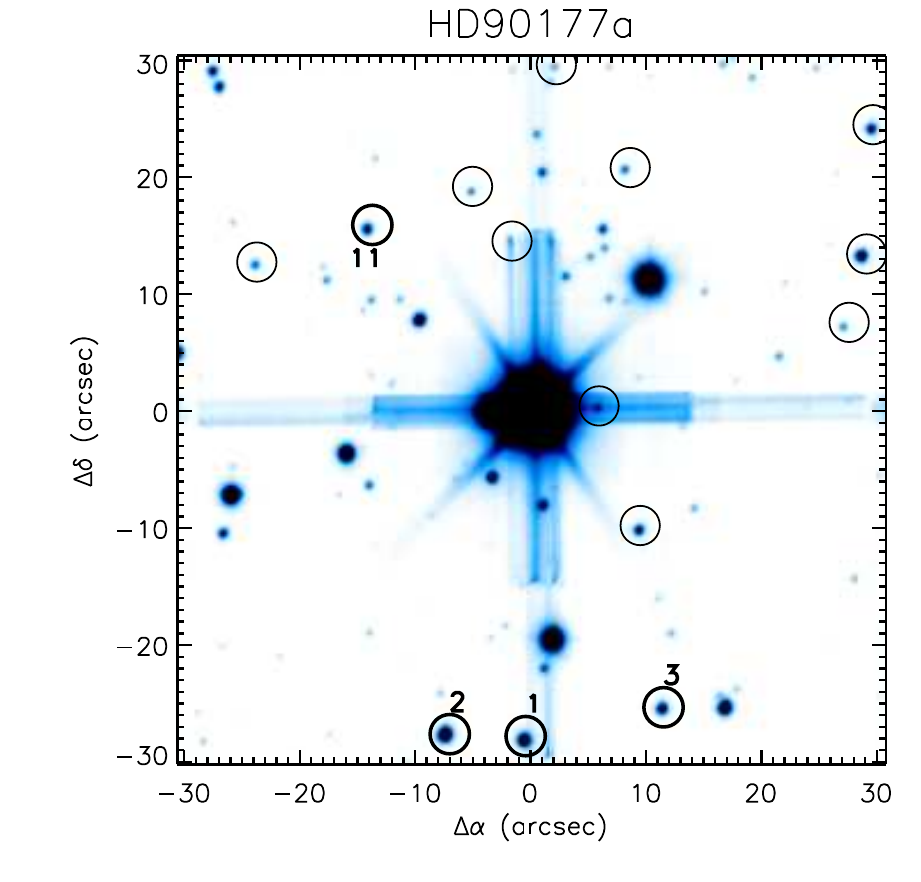}
\includegraphics[trim={3cm 10.cm 3cm 3cm},clip,width=8cm]{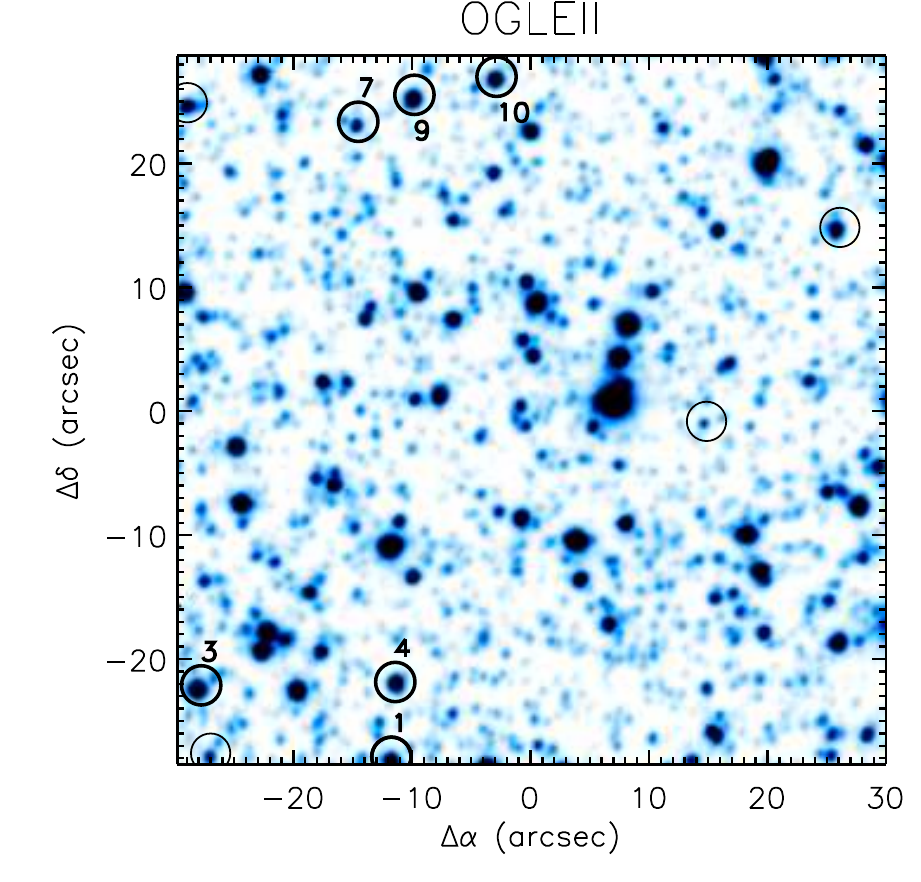}
\caption{Broad-band filter images recovered from MUSE observations of the fields: (left) HD90177, and (right) OGLEII}. 
\label{MUSE_fields}
\end{center}
\end{figure*}

We present here the images of the MUSE observed fields identifying the stars analyzed in this study to trace image quality across wavelengths (see Fig. \ref{MUSE_fields}). We also present the wavelength variation of the centroids and widths (i.e., x$_{0}$, y$_{0}$, $\sigma_1$, and $\sigma_2$ parameters in section \S\ref{IFS_section}) for the Moffat fitting profiles models to the 44 narrow-band filter images recovered for each selected star for each observed field within the MUSE retrieved data cubes.
Differences in FWHMs derived from the $\sigma_{i}$ (i=1,2) parameters along the two axes are almost always smaller than the MUSE spatial sampling (i.e., 0.2 arcsec). Therefore, we define the IQ($\lambda_i$) as the average of the FWHM$_1$ and FWHM$_2$ at each $\lambda_i$. Following the same approach as outlined in section \S\ref{negligible_dome}, we determine a combination of $\epsilon_{0}^{b}$ and \LO$^{b}$ values that best fit the IQ($\lambda$) for each of the stars analyzed in each MUSE field. Below, we provide a case-by-case description of the MUSE fields, Moffat fittings, and the optimal $\epsilon_{0}^{b}$ and \LO$^{b}$ combinations that match the measured IQ($\lambda$) for each star, and Table \ref{tabla_parameters_stars} summarizes the results. The dispersion of the derived atmospheric parameters from individual stars in the same field provides the typical error of the proposed methodology.

\begin{table*}
	\centering
	\caption{Atmospheric parameters derived from the measured image quality at different wavelengths from MUSE data cubes.}
	\label{tabla_parameters_stars}
	\begin{tabular}{ccc|ccc|ccc|ccc|ccc} 
       \multicolumn{3}{c|}{\bf HD90177a} &  \multicolumn{3}{c|}{\bf HD90177b} &  \multicolumn{3}{c|}{\bf HD90177c} &  \multicolumn{3}{c|}{\bf HD90177d} &  \multicolumn{3}{c}{\bf OGLEII}  \\ [3pt] 
       Star & $\epsilon_{0}^{b}$ & $\mathcal{L}_{0}^{b}$ & Star & $\epsilon_{0}^{b}$ & $\mathcal{L}_{0}^{b}$ & Star & $\epsilon_{0}^{b}$ & $\mathcal{L}_{0}^{b}$ & Star & $\epsilon_{0}^{b}$ & $\mathcal{L}_{0}^{b}$ & Star & $\epsilon_{0}^{b}$ & $\mathcal{L}_{0}^{b}$ \\ [3pt]
        (1) & (2) & (3) & (1) & (2) & (3) & (1) & (2) & (3) & (1) & (2) & (3) & (1) & (2) & (3) \\ 
        \hline
	     &     &     &     &     &     &     &     &     &     &     &  & & &    \\ [0pt]

        1 & 0.79 & 18.1$\pm$1.3  &  1 & 0.78 & 14.0$\pm$0.9 & 1 & 0.81 & 15.3$\pm$1.0 & 1 & 0.71 & 21.3$\pm$1.7 & 1 & 0.99 & 15.9$\pm$1.0 \\  [3pt]
         2 & 0.82 & 15.5$\pm$1.0  &  2 & 0.78 & 17.6$\pm$1.3 & 2 & 0.82 & 14.8$\pm$1.0 & 2 & 0.69 & 32.9$\pm$3.2 & 2 & 0.98 & 10.3$\pm$0.6 \\  [3pt]
         3 & 0.79 & 28.9$\pm$2.5  &  3 & 0.76 & 31.4$\pm$3.0 & 3 & 0.83 & 19.5$\pm$1.4 & 3 & 0.77 & 21.1$\pm$1.6 & 4 & 0.96 & 10.4$\pm$0.6 \\  [3pt]
         11 &0.77 & 22.0$\pm$1.8  & 11 & 0.72 & 29.8$\pm$2.6 &11 & 0.79 & 17.5$\pm$1.3 &11 & 0.63 & 47.9$\pm$6.2 & 7 & 0.93 & 29.1$\pm$2.5       \\  [3pt]
           & &   &  &  &  & &  & & &  &   &                                                                        9 & 0.87 & 21.4$\pm$1.6        \\  [3pt]
           & &   &  &  &  & &  & & &  &   &                                                                       10 & 0.91 & 11.2$\pm$0.6        \\  [3pt] \hline 
	     &     &     &     &     &     &     &     &     &     &     &  & & &    \\ [-3pt]
 $\overline{P}$  & 0.79 & 21.1  &  & 0.76 & 23.2 & & 0.81 & 16.8 & & 0.70 & 30.8  &  & 0.94 & 16.3        \\  [3pt]
 $\sigma_{P}$    & 0.02 &  5.8   &  & 0.03 &  8.7 & & 0.02 &  2.2 & & 0.06 & 12.7  &  & 0.05 & 7.6        \\  [3pt]
\hline

	\end{tabular}
	\tablefoot{Columns indicate the MUSE observed field, and each sub-column corresponds to: (1) the number label of the star in the field (see Fig. \ref{MUSE_fields}), and (2) and (3) are the best combination of $\epsilon_{0}$ and \LO values minimizing residuals when comparing the predicted $\epsilon_{LE}$ using Eq. \ref{eq:ELE_lambda} with these values to the measured IQ at any wavelength, assuming a negligible dome turbulence contribution. The last two lines present the average ($\overline{P}$) and standard deviation ($\sigma_{P}$) of the estimated parameters. These values are derived by averaging the individual estimations from each star in the field.}
\end{table*}

\subsubsection{HD90177 field}

HD90177 is a luminous blue variable star in the Carina Nebula, one of the largest diffuse nebulae in the southern sky, hosting numerous massive and young stars. HD90177 is known for intense mass-loss episodes, and it is surrounded by a unique stellar population distribution suggesting a potential association with a moving group of B-type stars in a spiral arm \citep{2021Mehner}.

\begin{figure*}
\begin{center}
\includegraphics[trim={0.5cm 12.cm 2.5cm 3.cm},clip,width=8cm]{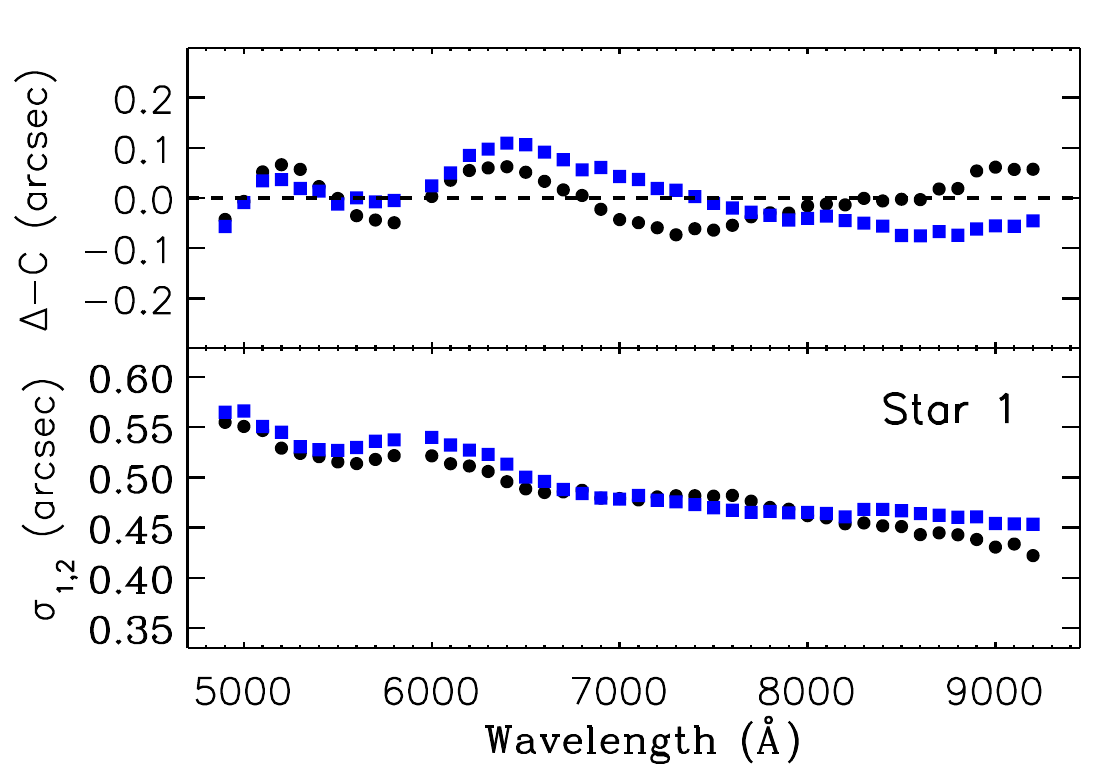}
\includegraphics[trim={0.5cm 12.cm 2.5cm 3.cm},clip,width=8cm]{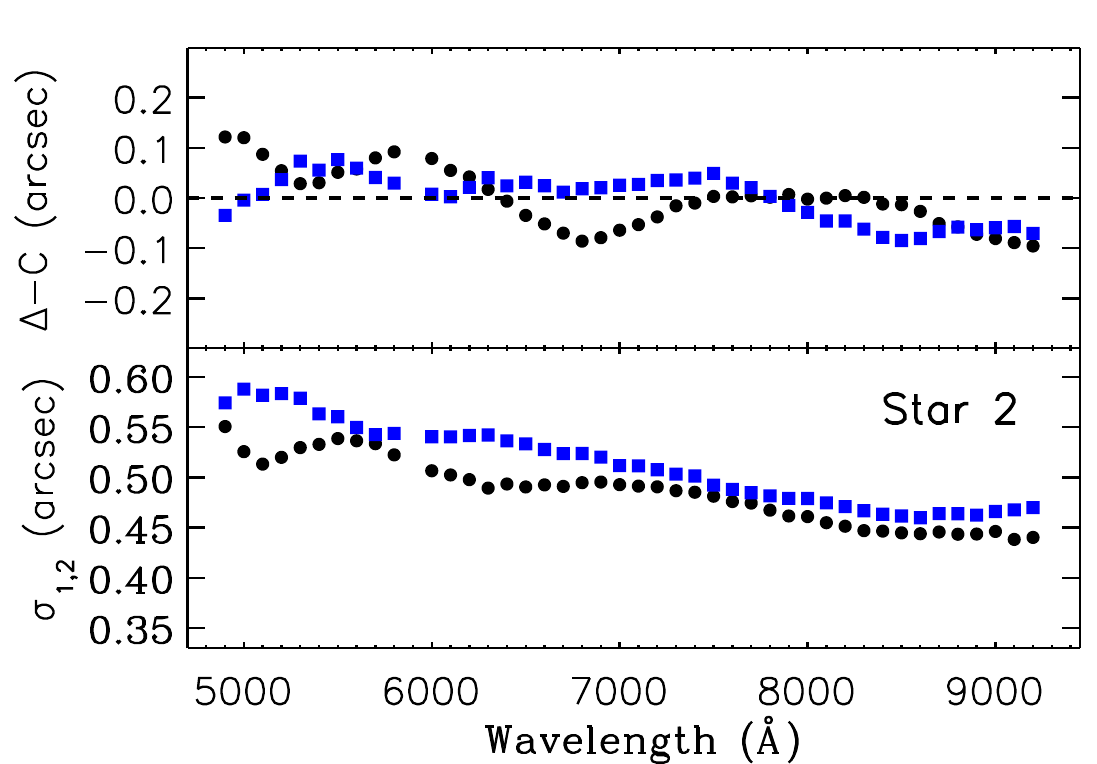}
\includegraphics[trim={0.5cm 12.cm 2.5cm 3.cm},clip,width=8cm]{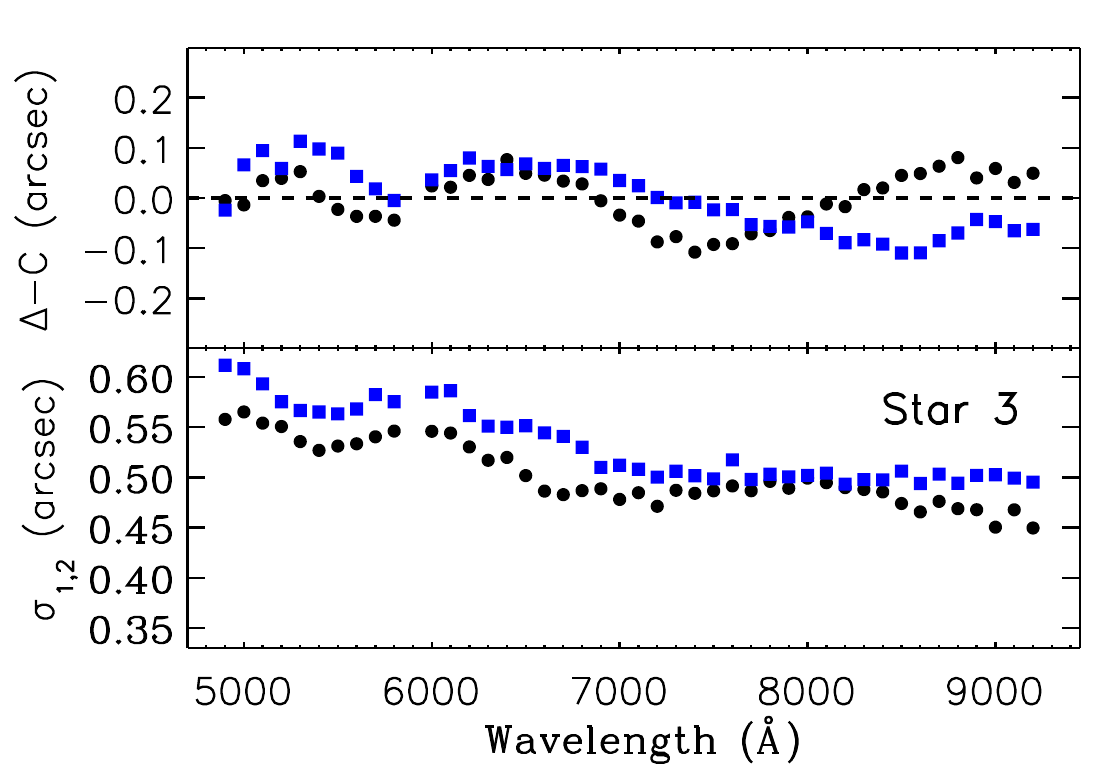}
\includegraphics[trim={0.5cm 12.cm 2.5cm 3.cm},clip,width=8cm]{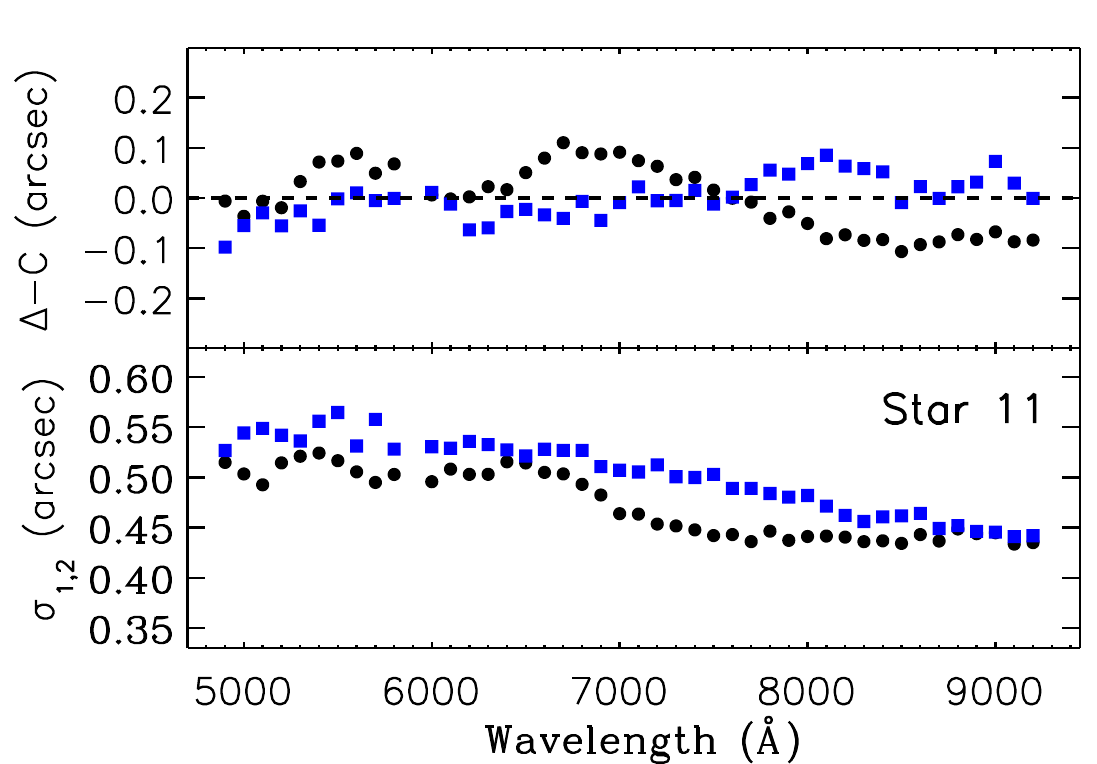}
\caption{Characteristic parameters of the Moffat profile model with $\beta=2.5$ fitting the 44 narrow-band filter images of selected point-like sources in the MUSE Field of HD90177 labeled HD90177a (see Fig. \ref{MUSE_fields}(to-left)). The top panels show the wavelength-dependent variations of the centroid (x${0}$,y${0}$) with respect to the average star peak location (indicated by the dashed horizontal line). The bottom panels present the variation of the width parameter $\sigma_{i}$ (i=1,2) with wavelength. Measurements along axis i=1 (horizontal) and i=2 (vertical) are represented by filled black circles and blue squares, respectively. The star numbers correspond to the star labels in Fig. \ref{MUSE_fields}(top-left).}
\label{HD90177a_parameters}
\end{center}
\end{figure*}

\begin{figure*}
\begin{center}
\includegraphics[trim={0.5cm 12.cm 2.5cm 3.cm},clip,width=8cm]{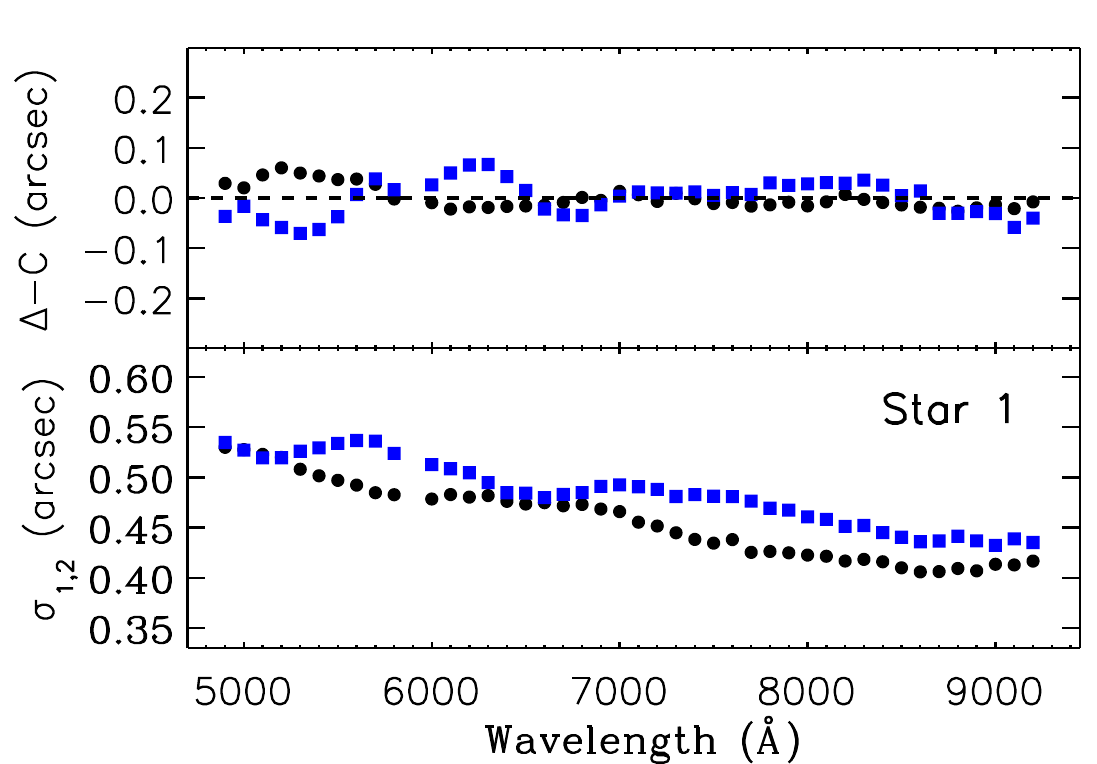}
\includegraphics[trim={0.5cm 12.cm 2.5cm 3.cm},clip,width=8cm]{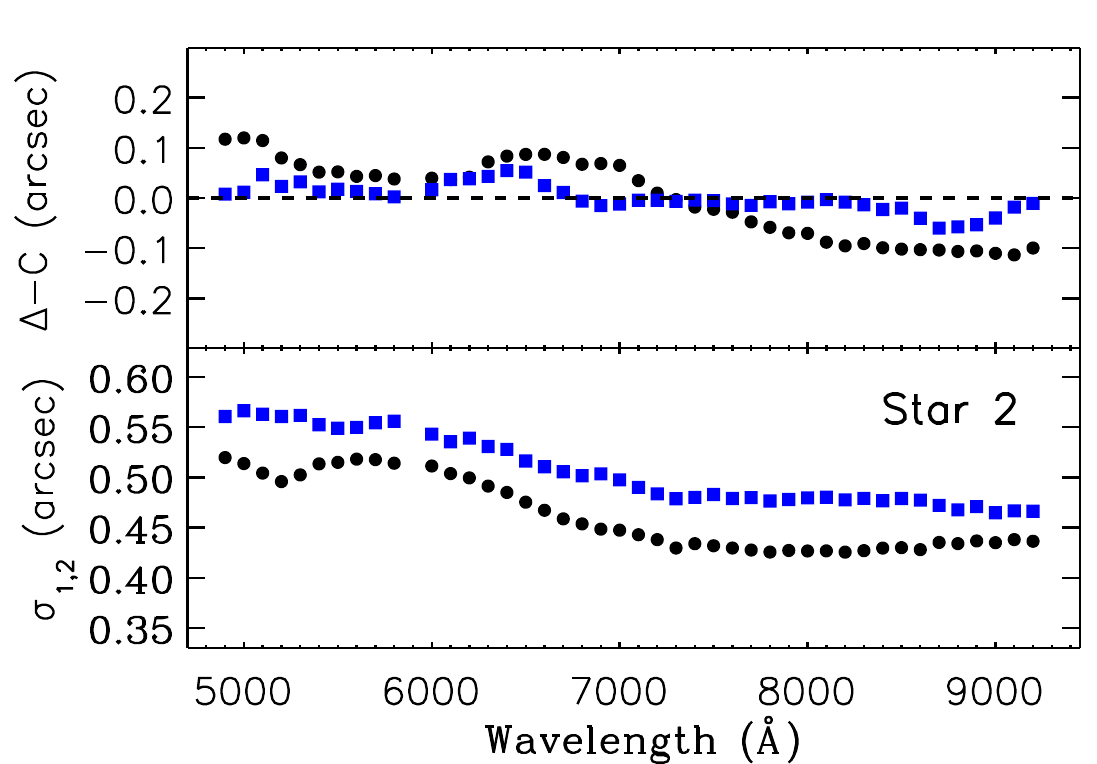}
\includegraphics[trim={0.5cm 12.cm 2.5cm 3.cm},clip,width=8cm]{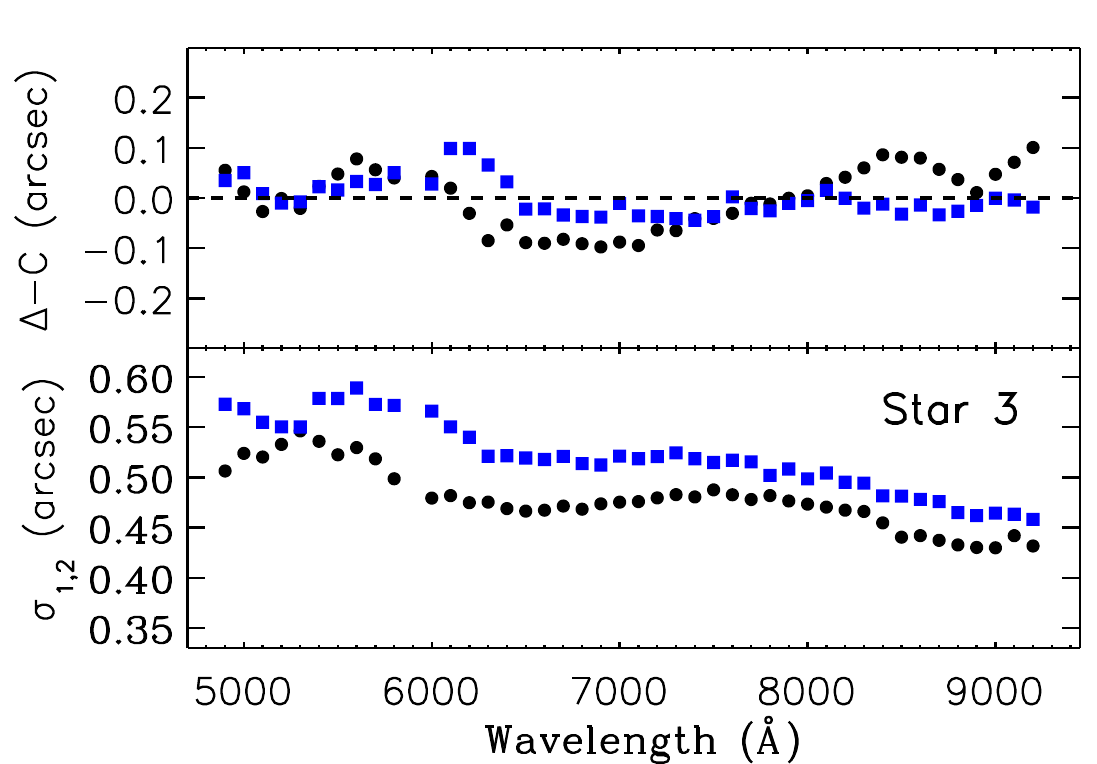}
\includegraphics[trim={0.5cm 12.cm 2.5cm 3.cm},clip,width=8cm]{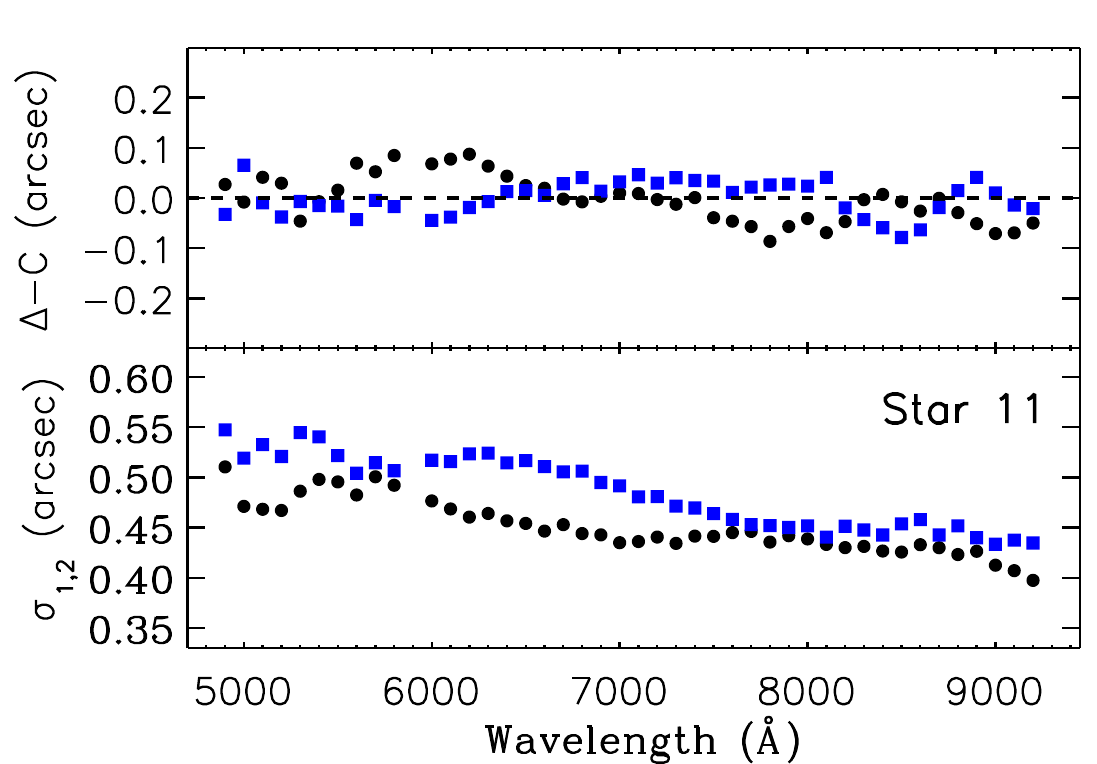}
\caption{Same as Fig. \ref{HD90177a_parameters} but for the MUSE data cube labeled HD90177b}.
\label{HD90177b_parameters}
\end{center}
\end{figure*}

\begin{figure*}
\begin{center}
\includegraphics[trim={0.5cm 12.cm 2.5cm 3.cm},clip,width=8cm]{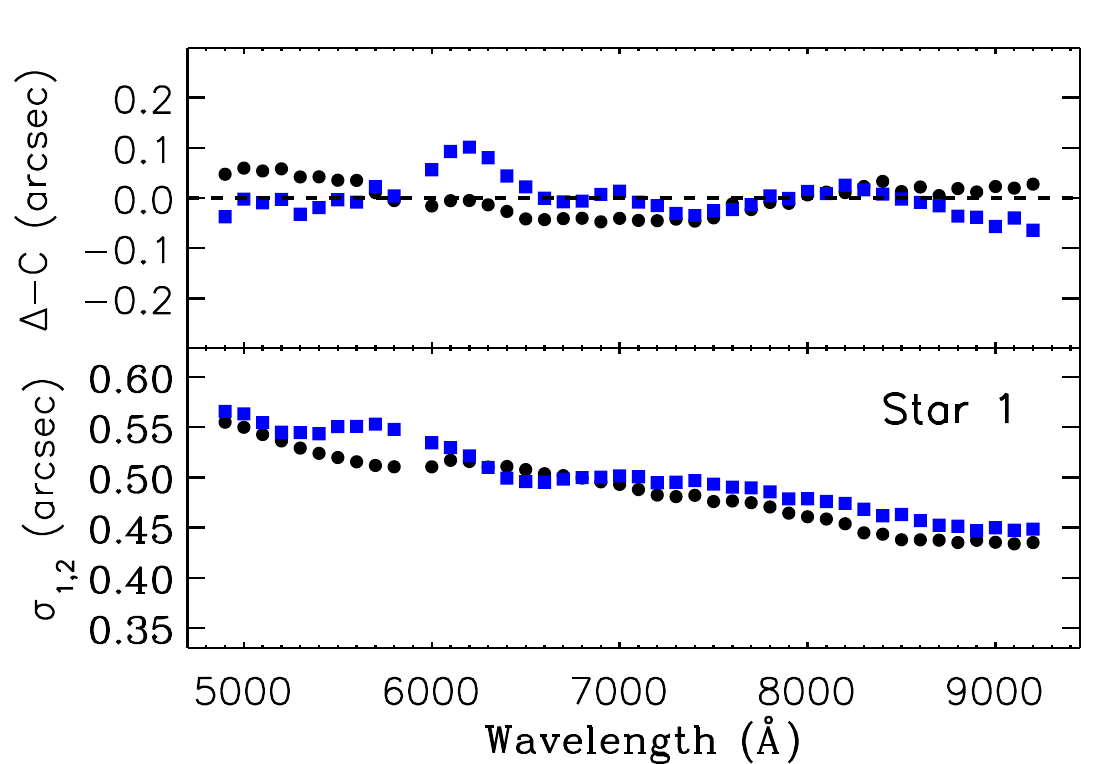}
\includegraphics[trim={0.5cm 12.cm 2.5cm 3.cm},clip,width=8cm]{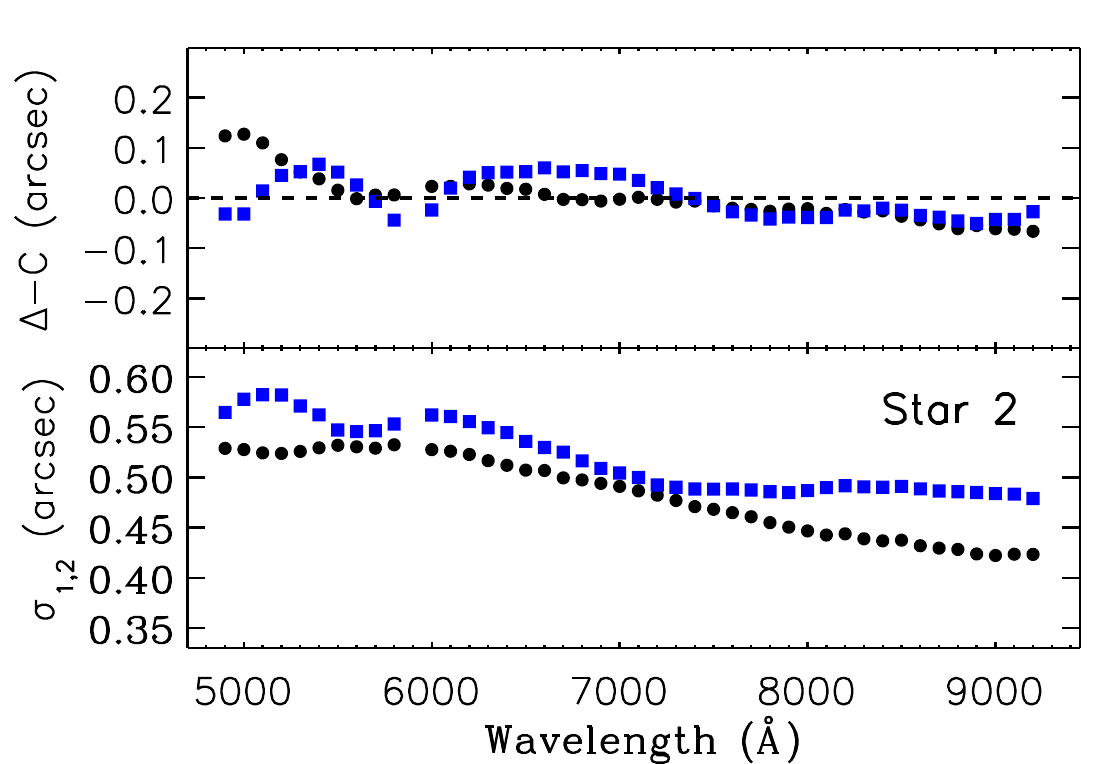}
\includegraphics[trim={0.5cm 12.cm 2.5cm 3.cm},clip,width=8cm]{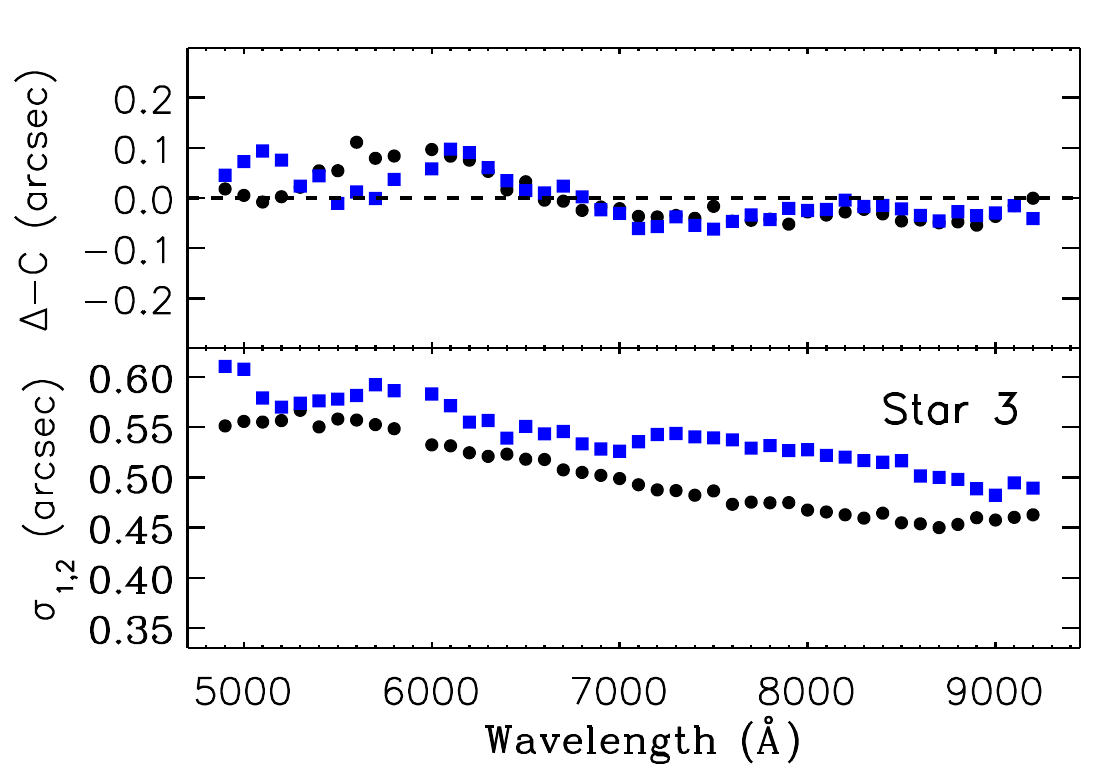}
\includegraphics[trim={0.5cm 12.cm 2.5cm 3.cm},clip,width=8cm]{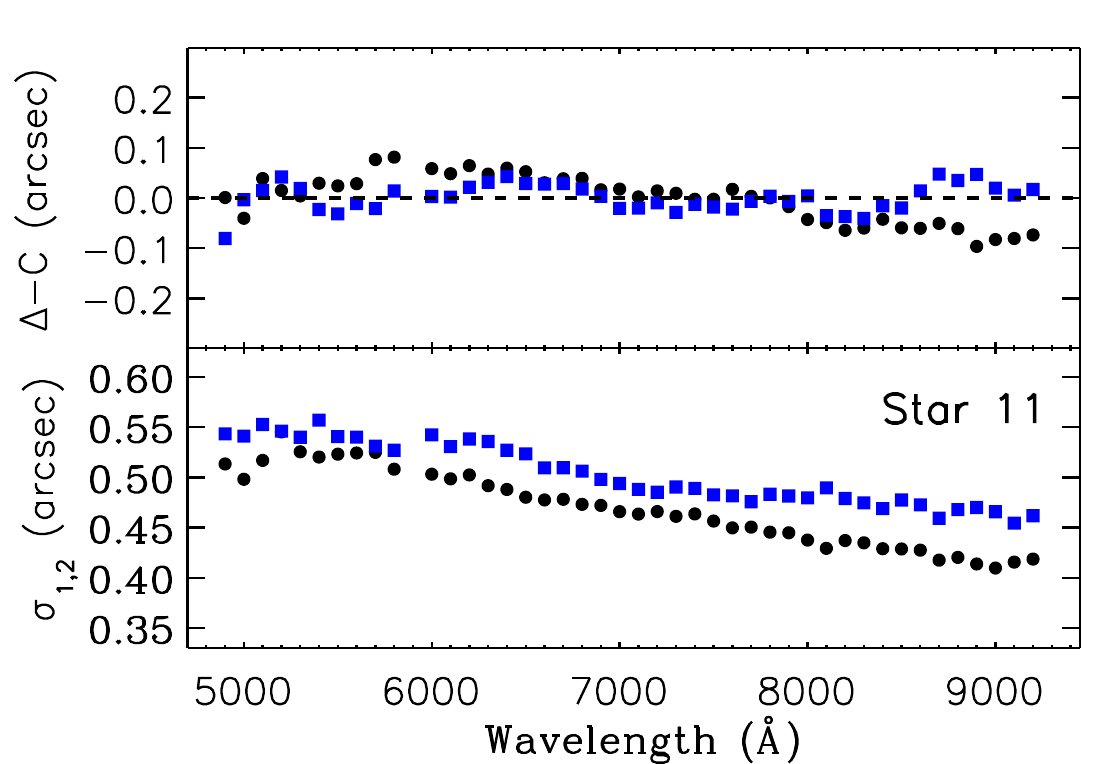}
\caption{Same as Fig. \ref{HD90177a_parameters} but for the MUSE data cube labeled HD90177c}.
\label{HD90177c_parameters}
\end{center}
\end{figure*}

\begin{figure*}
\begin{center}
\includegraphics[trim={0.5cm 12.cm 2.5cm 3.cm},clip,width=8cm]{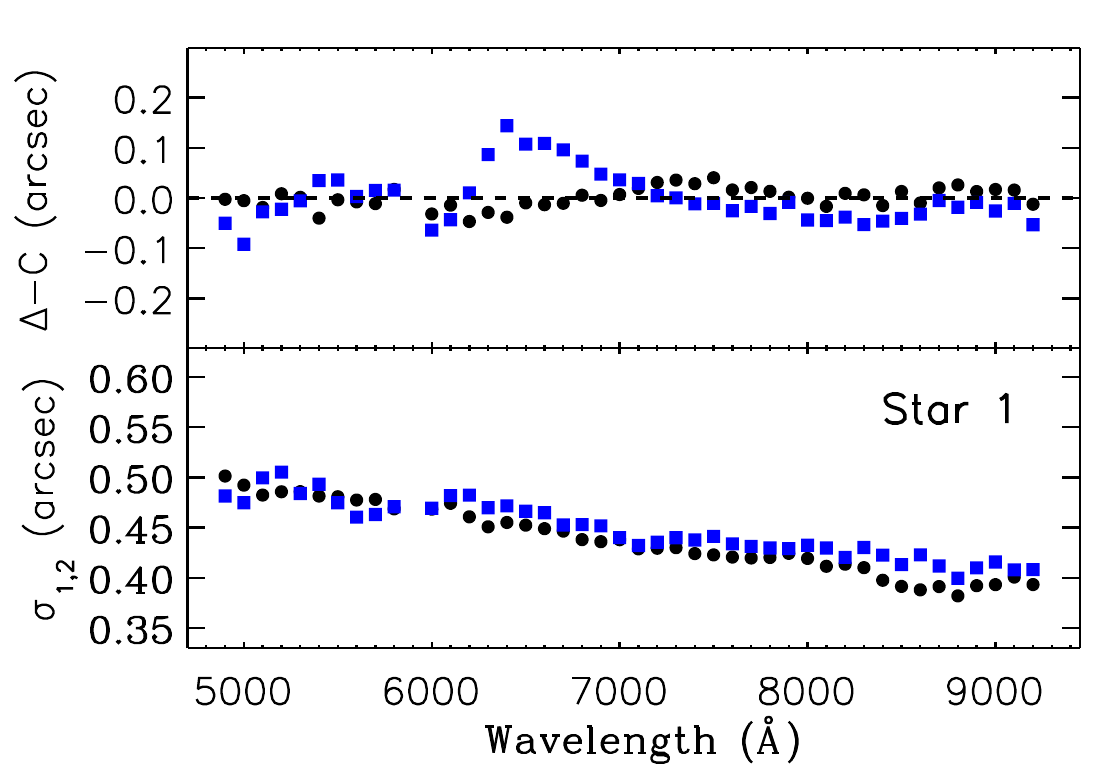}
\includegraphics[trim={0.5cm 12.cm 2.5cm 3.cm},clip,width=8cm]{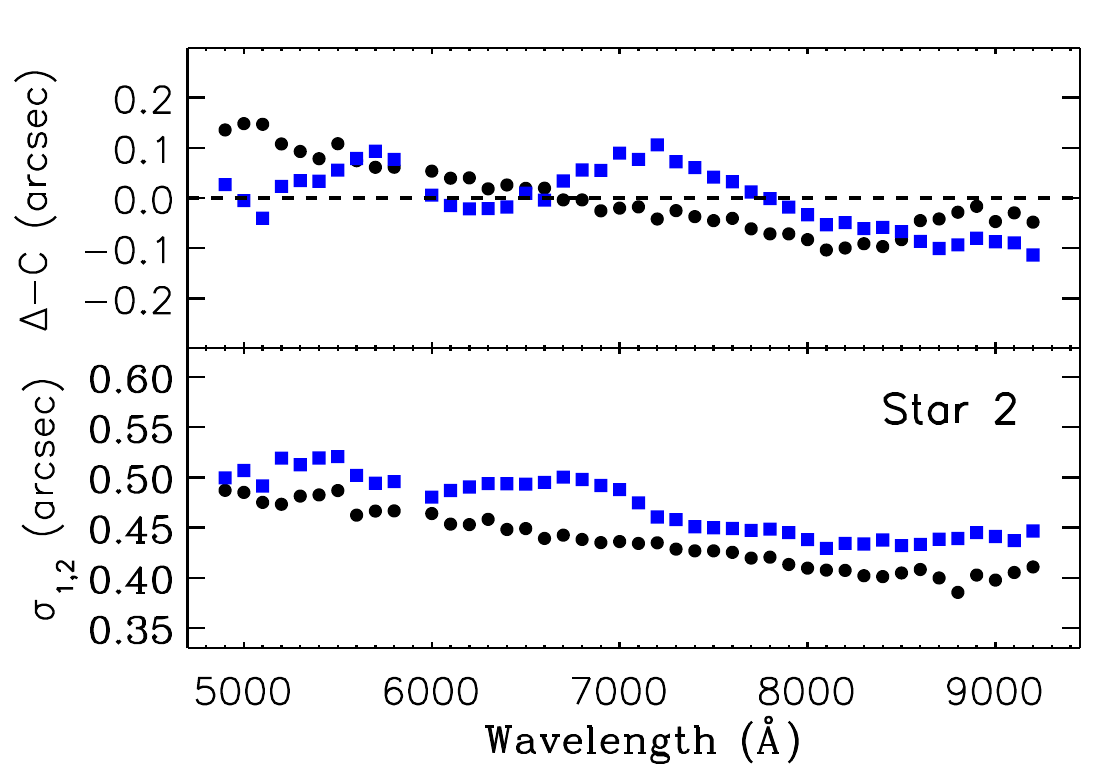}
\includegraphics[trim={0.5cm 12.cm 2.5cm 3.cm},clip,width=8cm]{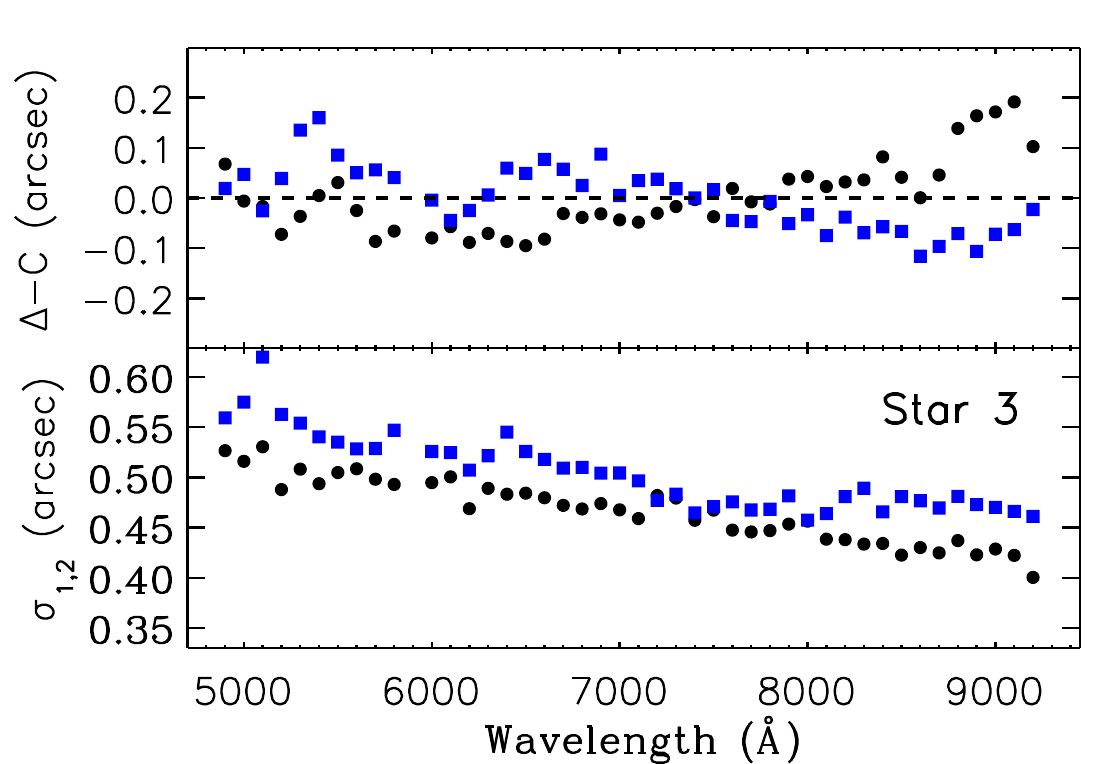}
\includegraphics[trim={0.5cm 12.cm 2.5cm 3.cm},clip,width=8cm]{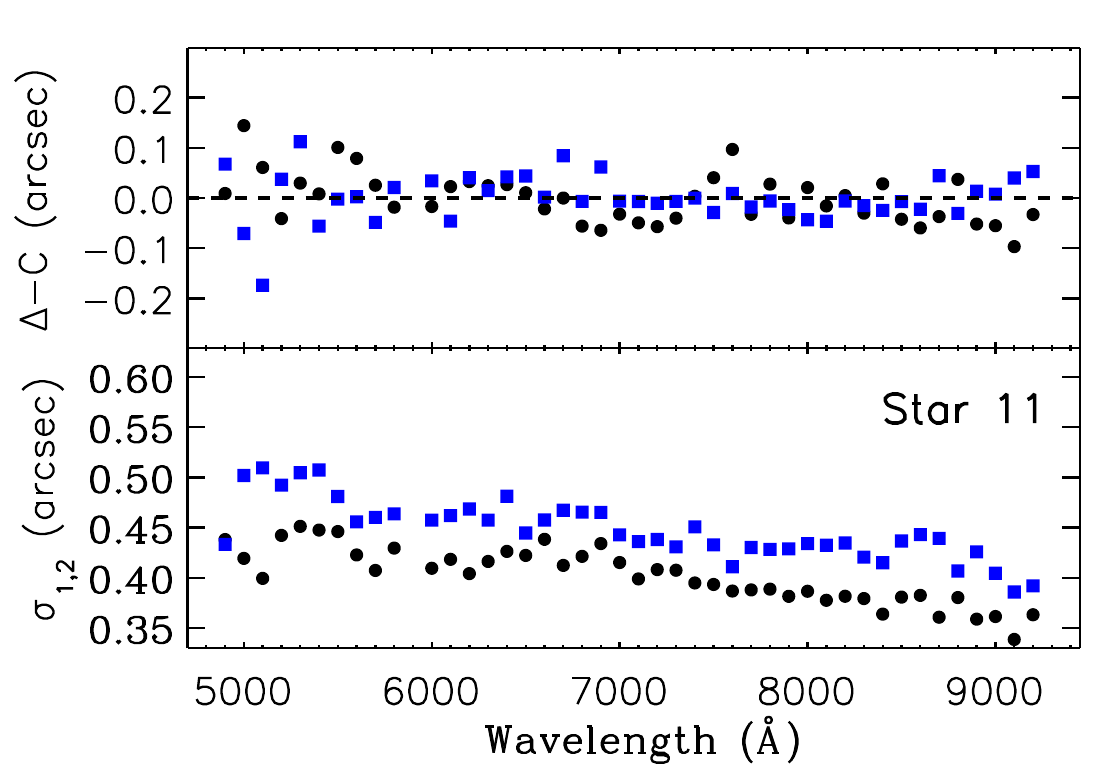}
\caption{Same as Fig. \ref{HD90177a_parameters} but for the MUSE data cube labeled HD90177d}.
\label{HD90177d_parameters}
\end{center}
\end{figure*}

We obtained four MUSE data cubes labeled as HD90177a, HD90177b, HD90177c, and HD90177d (see Table \ref{night_parameters}) for the HD90177 field. All were observed on the same night within a temporal window of just under three hours. The HD90177 star, located at the center of the MUSE data cubes, generates strong stray light artifacts due to its brightness, affecting some stars in the observed MUSE field, which includes many other fainter stars than HD90177 (Fig. \ref{MUSE_fields}-left). We identified 14 stars in the field without detected companions within a 5-arcsec-radius circular aperture. Among them, we discarded four stars due to contamination by stray light from the central bright star. Additionally, five stars had low S/N in at least 50\% of the defined filter bands in any of the four MUSE data cubes, and one star was too close to the field border, preventing a proper fitting of the Moffat model. Figures \ref{HD90177a_parameters}, \ref{HD90177b_parameters}, \ref{HD90177c_parameters}, and \ref{HD90177d_parameters} present the parameters for the Moffat model profiles fitted to the filter-band images of the remaining four stars in this field for the four MUSE data cubes, plotted against the central wavelengths of the respective bands.

The centroids of the Moffat models follow wavelength-dependent variations, which are different for each star but consistently within $\pm$0.2 arcsec, corresponding to one MUSE spatial pixel. The $\sigma_{i}$ (i=1,2) parameters for the Moffat models decrease for increasing wavelengths similarly for the four stars.  

The values for $\epsilon_{0}^{b}$ derived from the analysis of the IQ for each star in the four MUSE data cubes for the HD90177 field are compatible, with an average and standard deviation of $0.79\pm0.02$, $0.76\pm0.03$, $0.81\pm0.02$, and $0.70\pm0.06$ arcsec for HD90177a, HD90177b, HD90177c, and HD90177d data cubes, respectively (see Table \ref{tabla_parameters_stars}). The average and standard deviation of the $\mathcal{L}_{0}^{b}$ values derived from each star are $21.12\pm5.83$, $23.20\pm8.69$, $16.77\pm2.16$, and $30.80\pm12.66$ meters, respectively, for HD90177a, HD90177b, HD90177c, and HD90177d data cubes.

\begin{figure*}
\begin{center}
\includegraphics[trim={3cm 13.cm 2.5cm 1.75cm},clip,width=8cm]{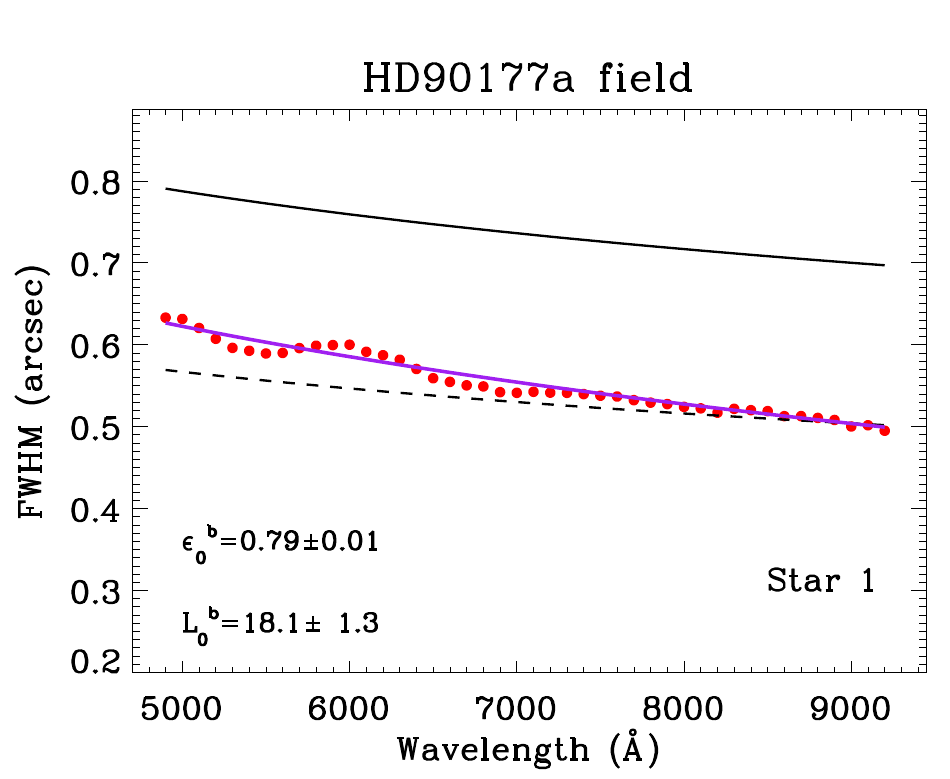}
\includegraphics[trim={3cm 13.cm 2.5cm 1.75cm},clip,width=8cm]{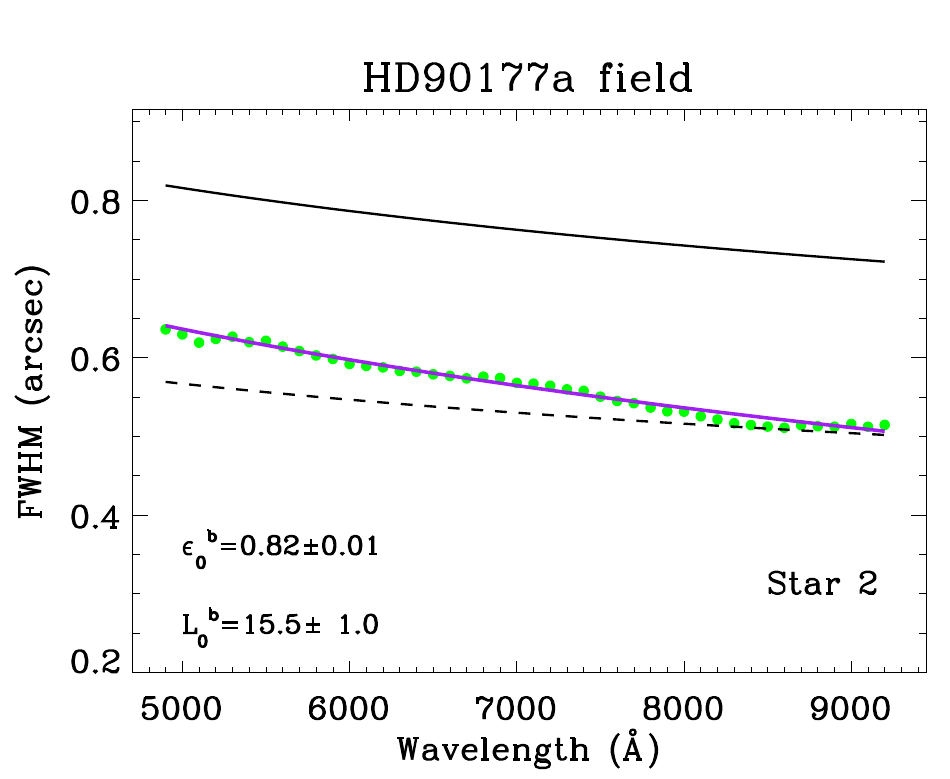}
\includegraphics[trim={3cm 13.cm 2.5cm 1.75cm},clip,width=8cm]{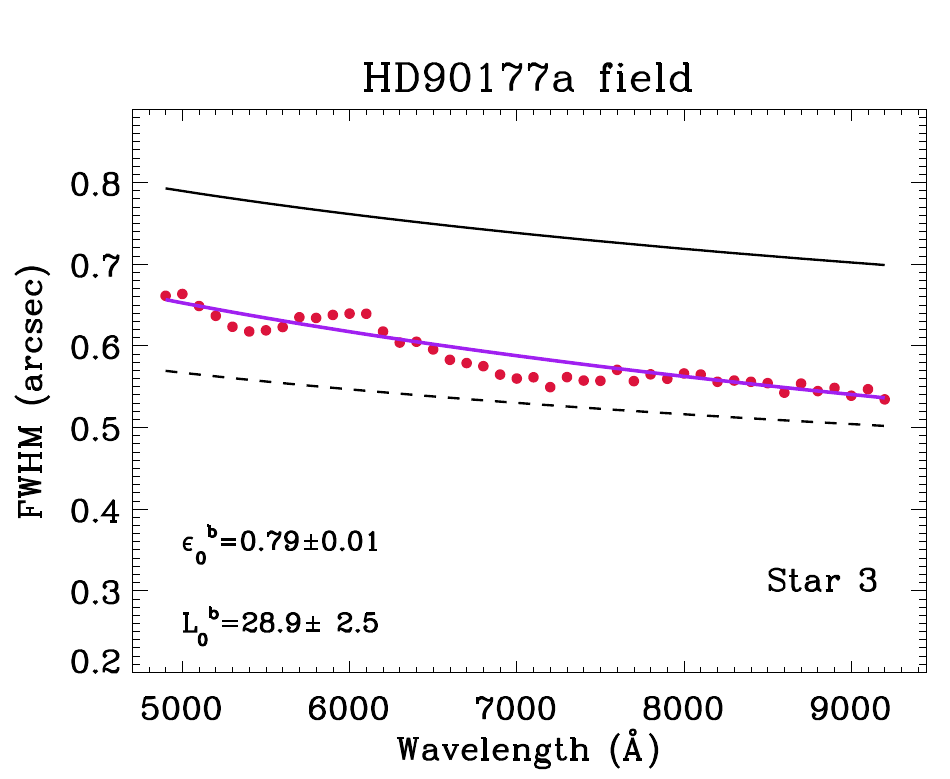}
\includegraphics[trim={3cm 13.cm 2.5cm 1.75cm},clip,width=8cm]{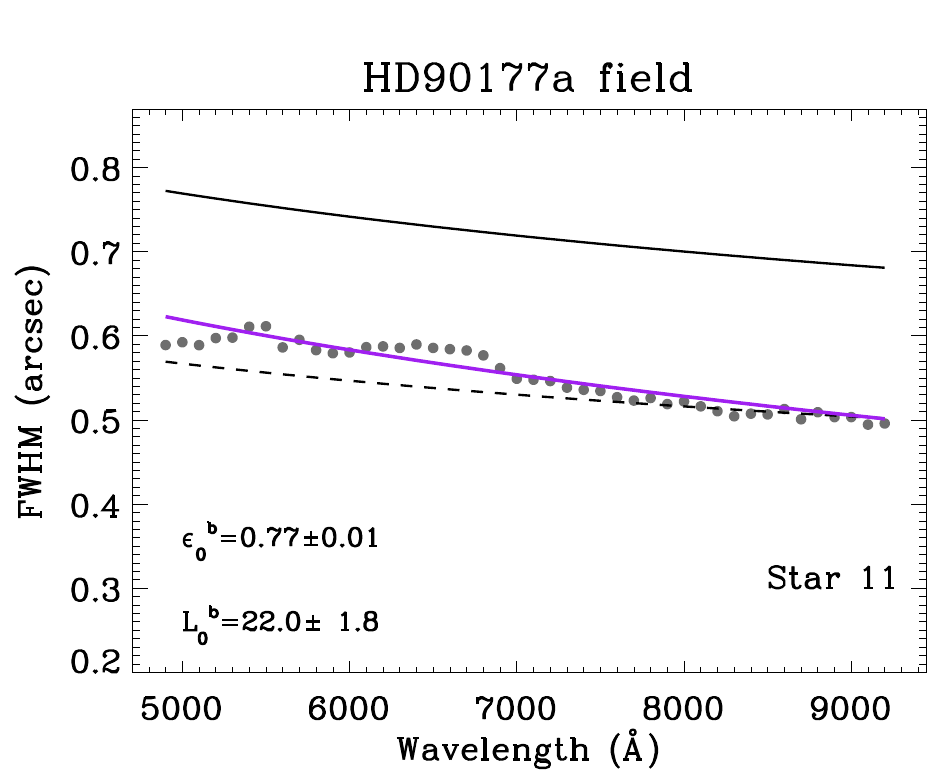}
 \caption{ Wavelength variation of the average FWHM for the Moffat fitting profile to the narrow-band filter images recovered for each of the selected stars in the MUSE HD90177 field labeled HD90177a (see Fig. \ref{MUSE_fields}(left). The dashed-black curve corresponds to the wavelength behavior of the $\epsilon_{0}$ derived from the average DIMM values during the MUSE observations (see Table \ref{selection_criteria}). The combination of $\epsilon_{0}$ and \LO best fitting the observations is indicated in the left corner of each panel, values obtained by minimizing the residuals while varying $\epsilon_{0}$ and \LO values. The purple line corresponds to the wavelength $\epsilon_{LE}$ variation using $\epsilon_{0}^{b}$ and \LO$^{b}$.}
\label{HD90177a_stars_fit}
\end{center}
\end{figure*}

\begin{figure*}
\begin{center}
\includegraphics[trim={3cm 13.cm 2.5cm 1.75cm},clip,width=8cm]{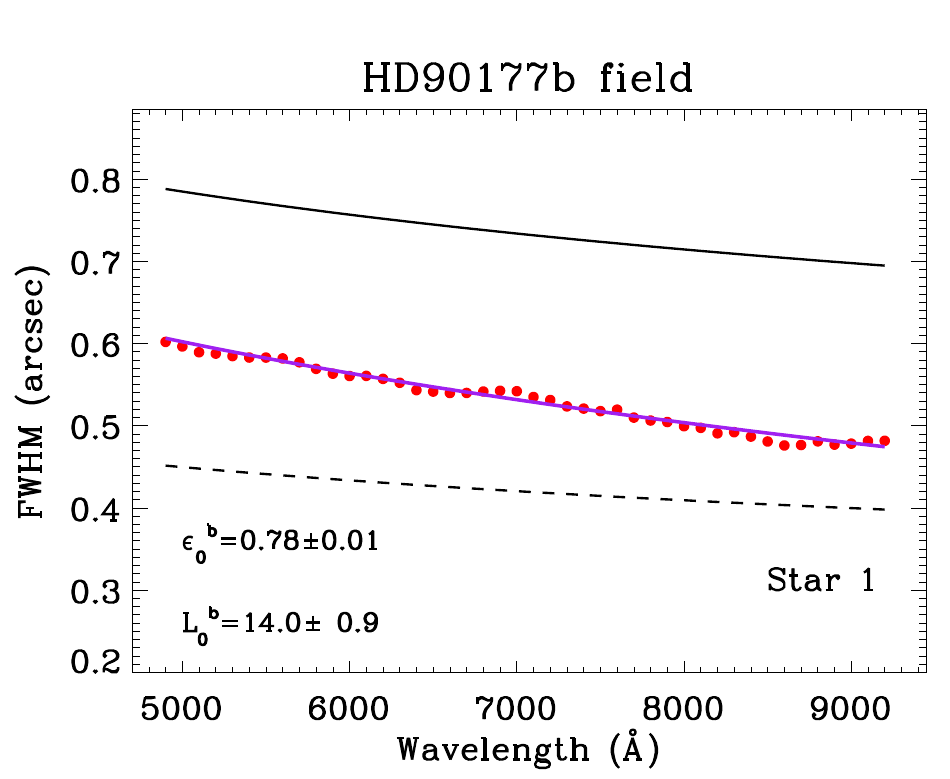}
\includegraphics[trim={3cm 13.cm 2.5cm 1.75cm},clip,width=8cm]{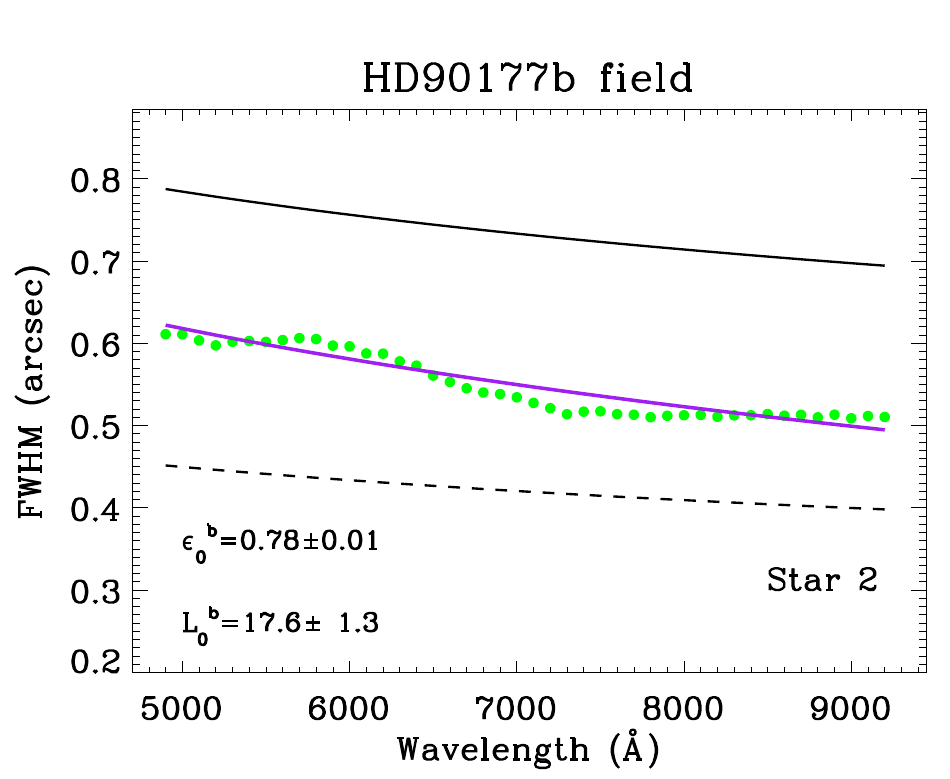}
\includegraphics[trim={3cm 13.cm 2.5cm 1.75cm},clip,width=8cm]{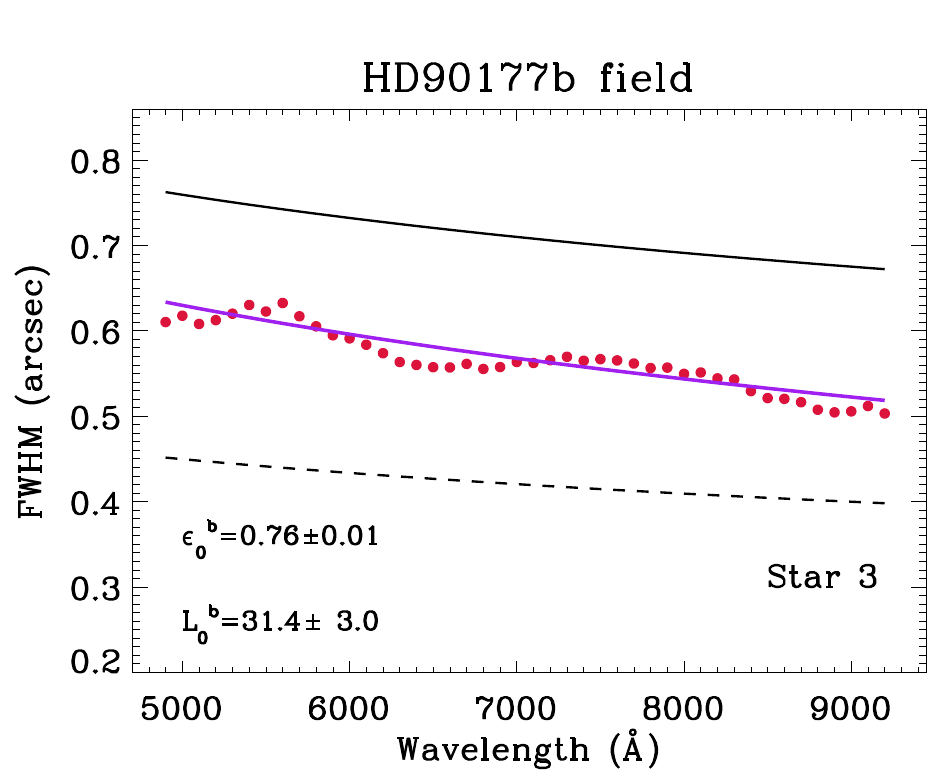}
\includegraphics[trim={3cm 13.cm 2.5cm 1.75cm},clip,width=8cm]{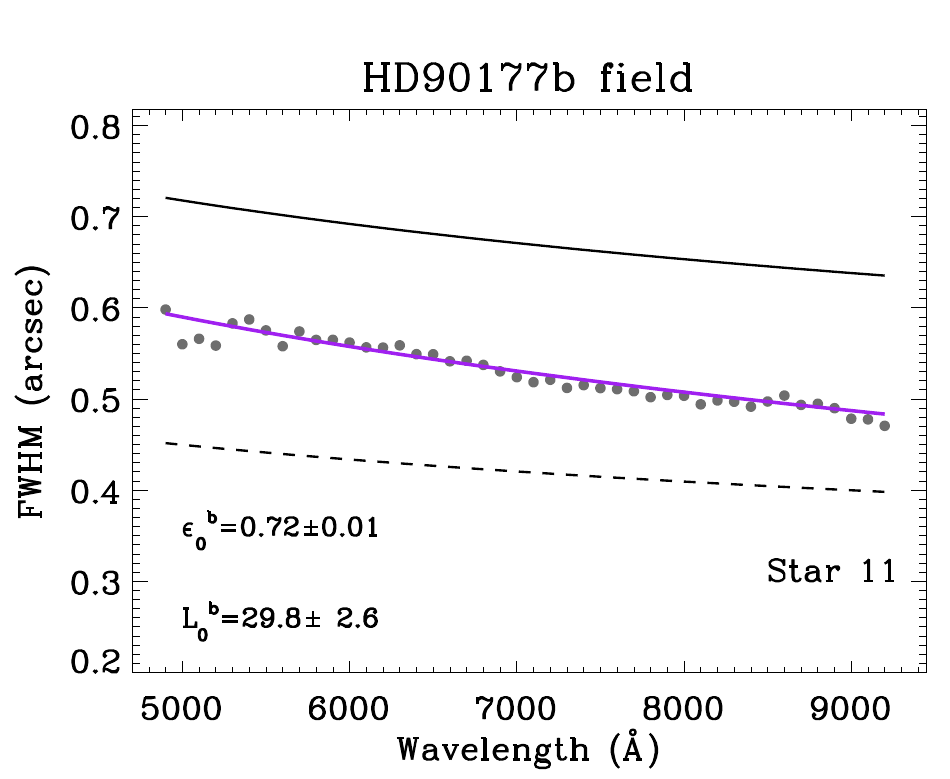}
\caption{Same as Fig. \ref{HD90177a_stars_fit} but for the MUSE data cube labeled HD90177b.  }
\label{HD90177b_stars_fit}
\end{center}
\end{figure*}

\begin{figure*}
\begin{center}
\includegraphics[trim={3cm 13.cm 2.5cm 1.75cm},clip,width=8cm]{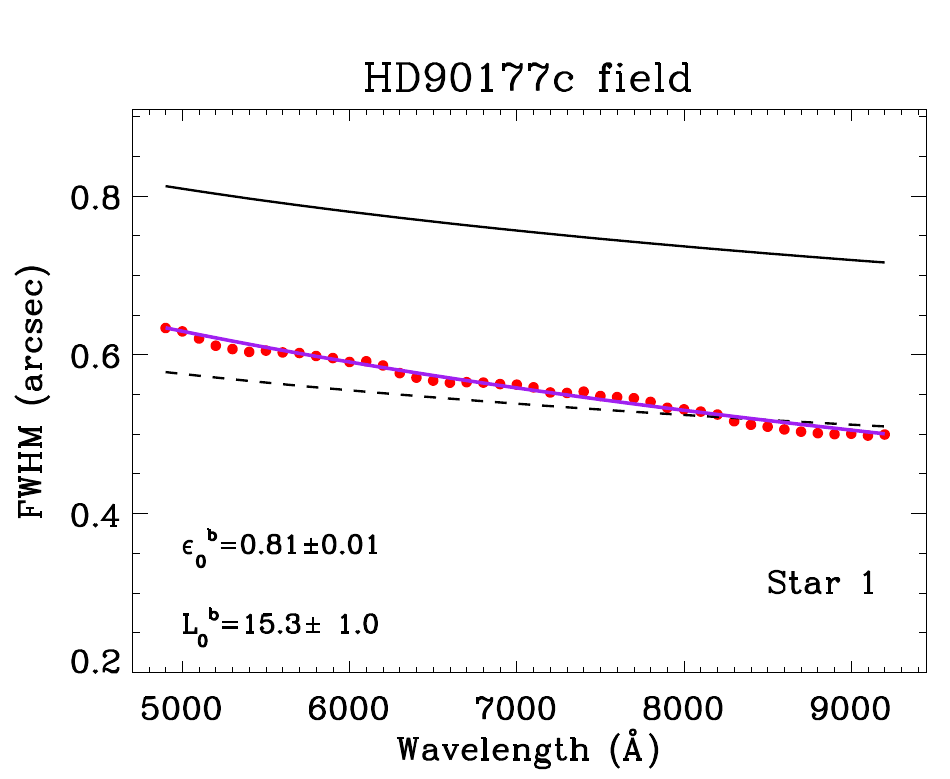}
\includegraphics[trim={3cm 13.cm 2.5cm 1.75cm},clip,width=8cm]{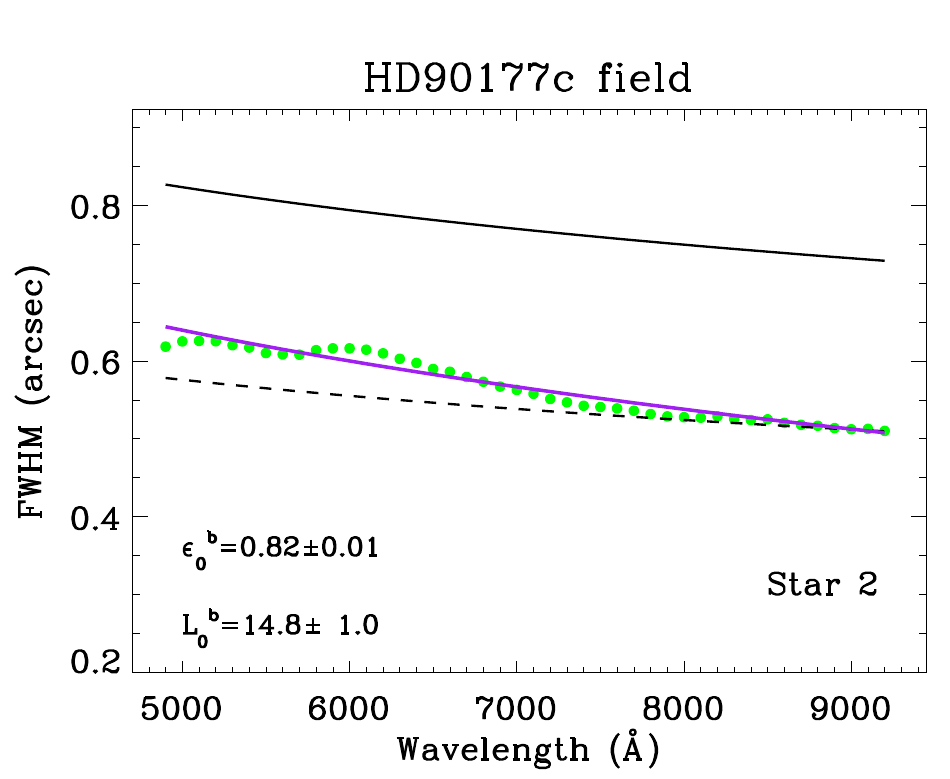}
\includegraphics[trim={3cm 13.cm 2.5cm 1.75cm},clip,width=8cm]{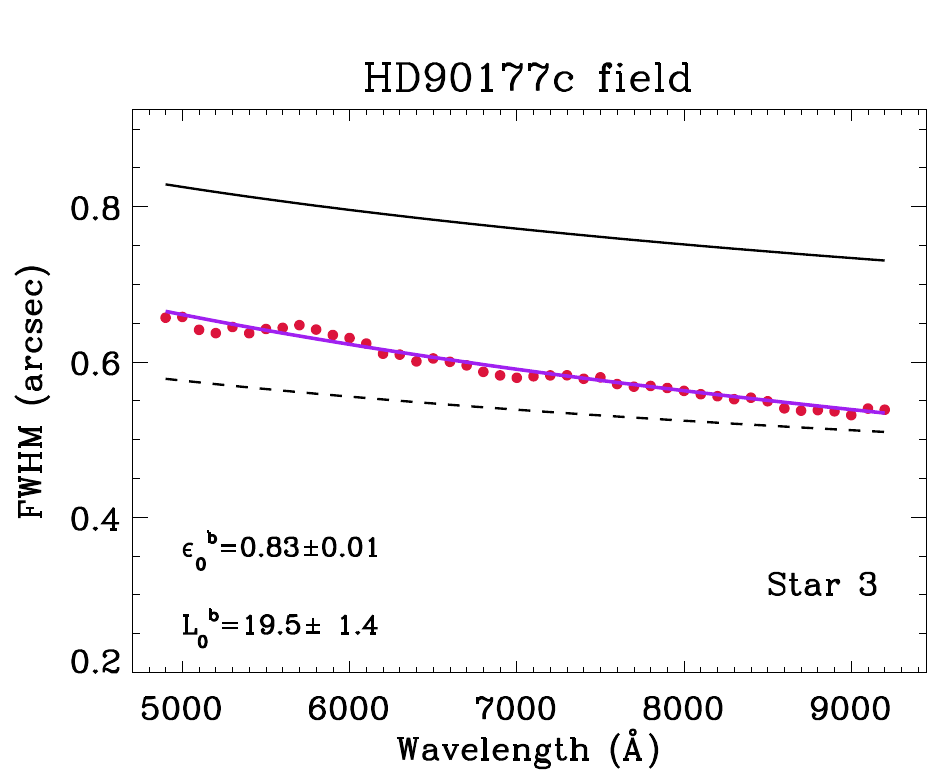}
\includegraphics[trim={3cm 13.cm 2.5cm 1.75cm},clip,width=8cm]{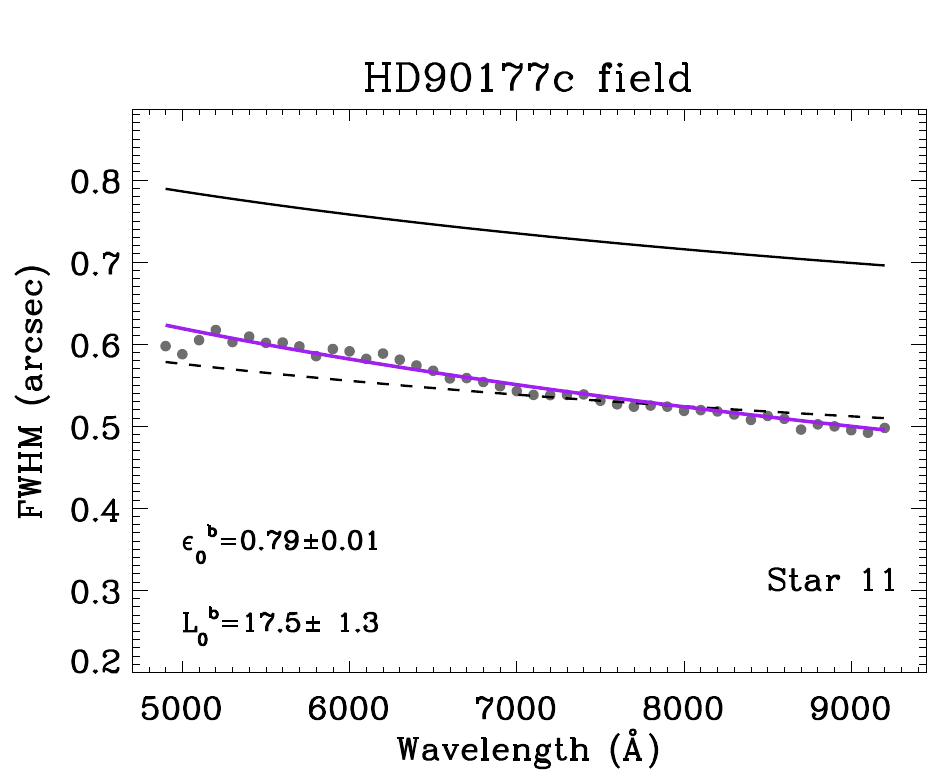}
\caption{Same as Fig. \ref{HD90177c_stars_fit} but for the MUSE data cube labeled HD90177c.  }
\label{HD90177c_stars_fit}
\end{center}
\end{figure*}

\begin{figure*}
\begin{center}
\includegraphics[trim={3cm 13.cm 2.5cm 1.75cm},clip,width=8cm]{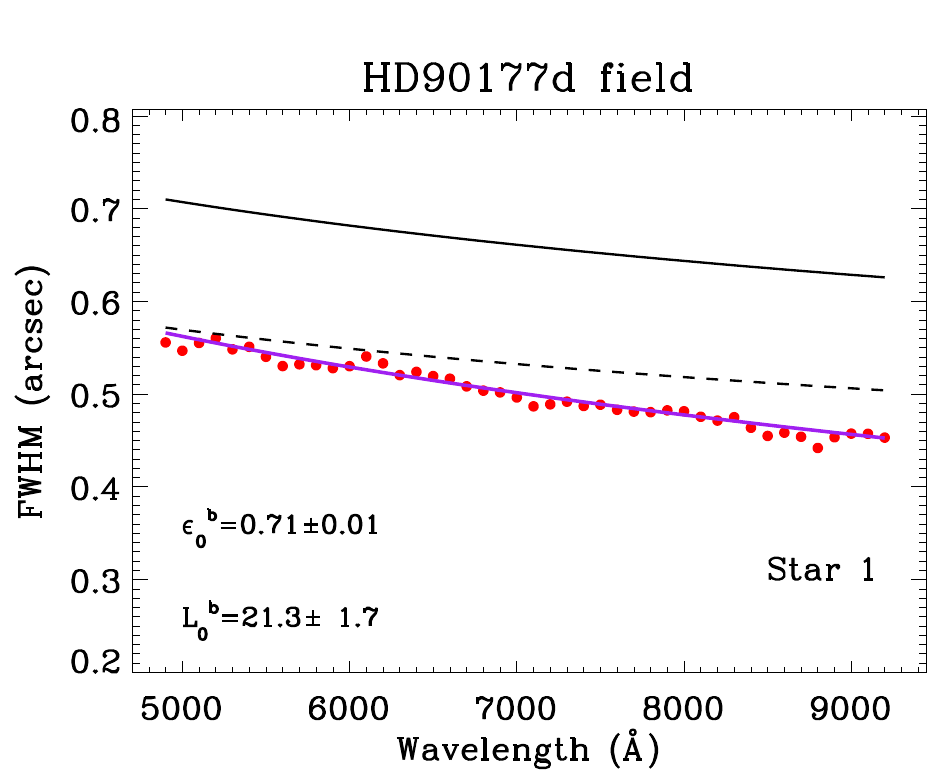}
\includegraphics[trim={3cm 13.cm 2.5cm 1.75cm},clip,width=8cm]{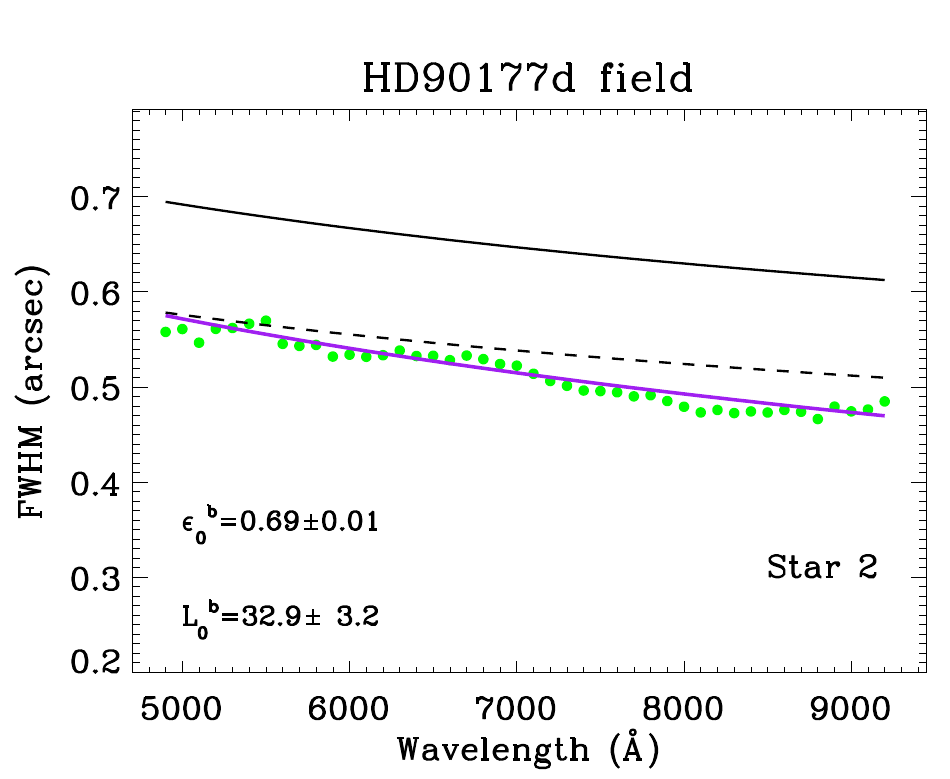}
\includegraphics[trim={3cm 13.cm 2.5cm 1.75cm},clip,width=8cm]{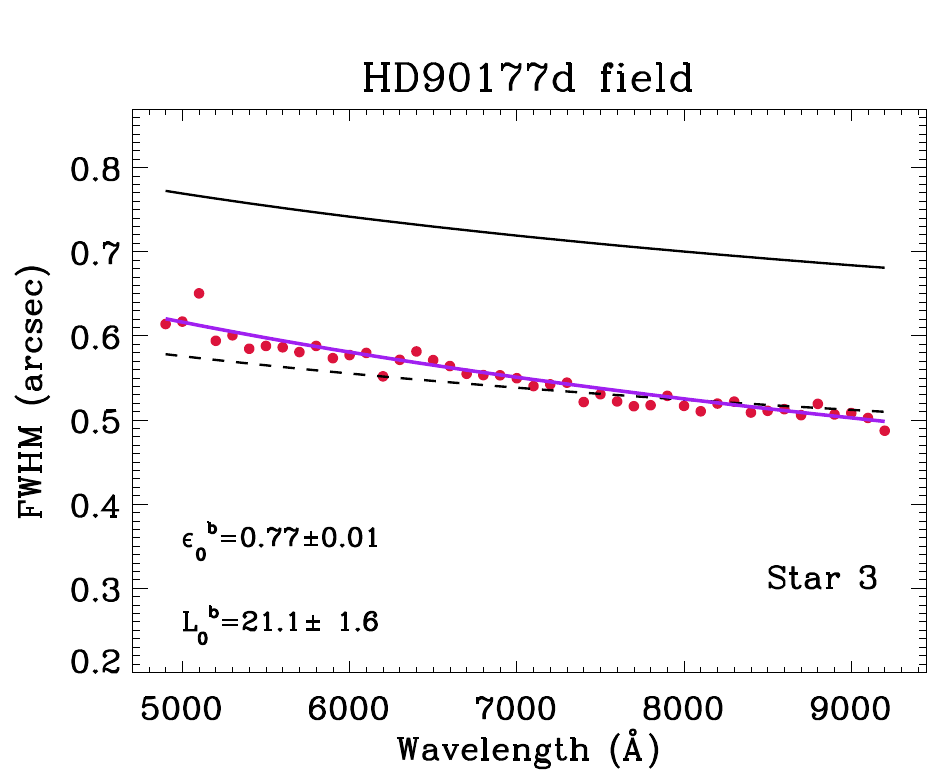}
\includegraphics[trim={3cm 13.cm 2.5cm 1.75cm},clip,width=8cm]{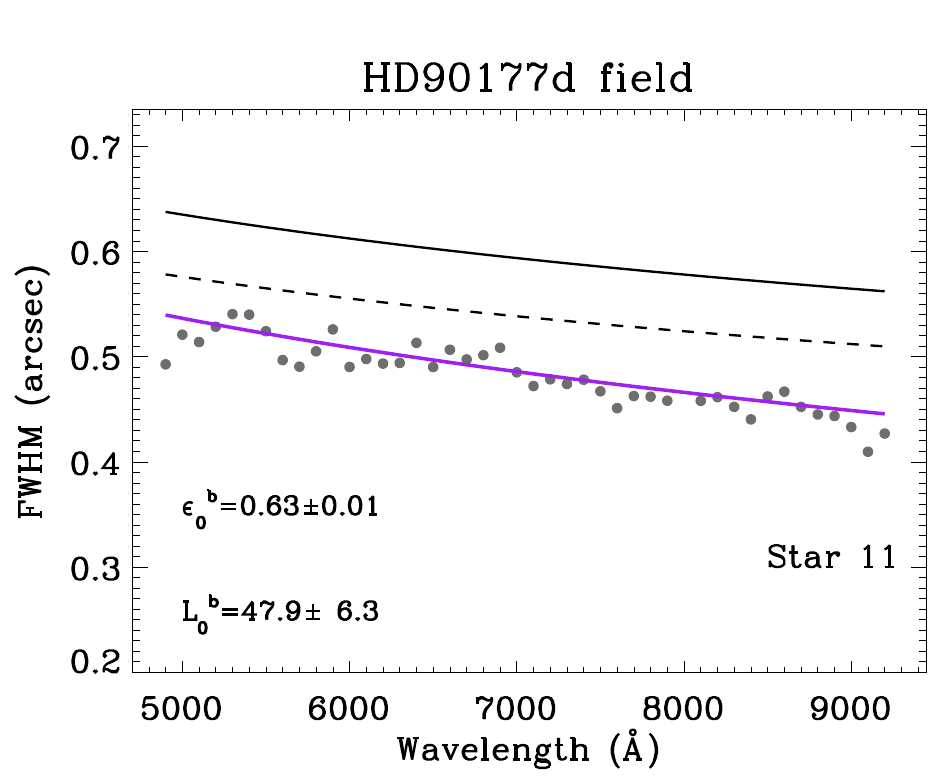}
\caption{Same as Fig. \ref{HD90177d_stars_fit} but for the MUSE data cube labeled HD90177d.  }
\label{HD90177d_stars_fit}
\end{center}
\end{figure*}

\subsubsection{OGLEII}

\begin{figure*}
\begin{center}
\includegraphics[trim={0.5cm 12.cm 2.5cm 3.cm},clip,width=8cm]{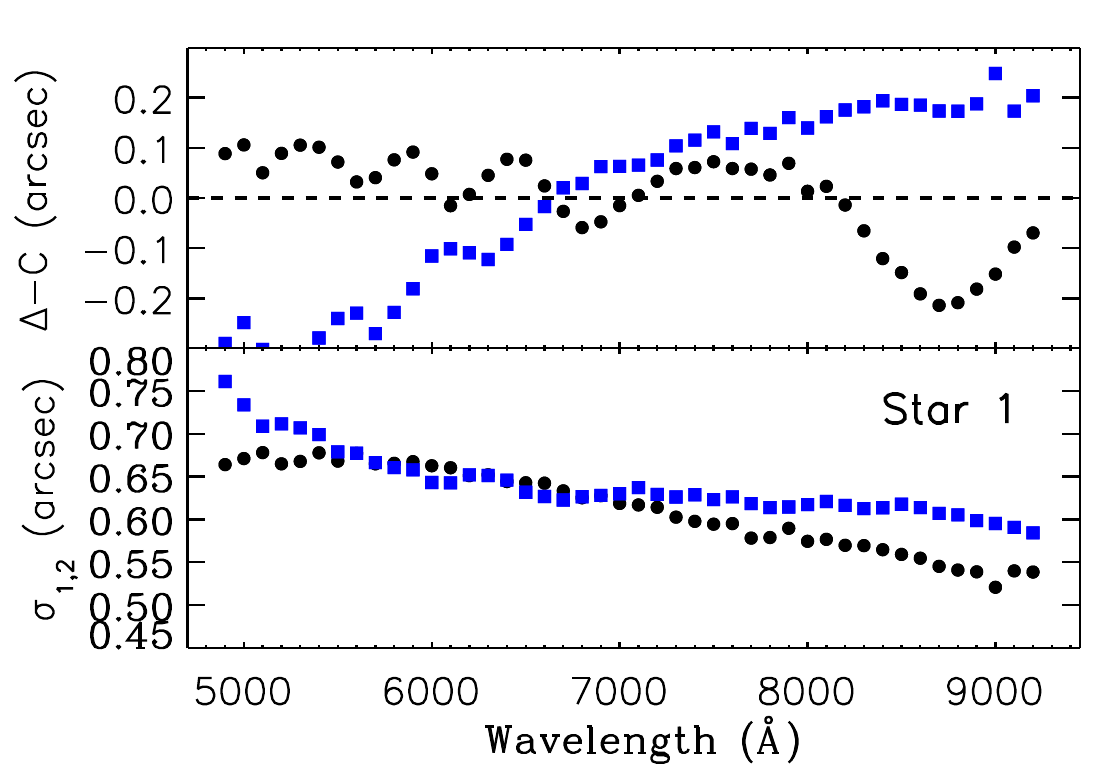}
\includegraphics[trim={0.5cm 12.cm 2.5cm 3.cm},clip,width=8cm]{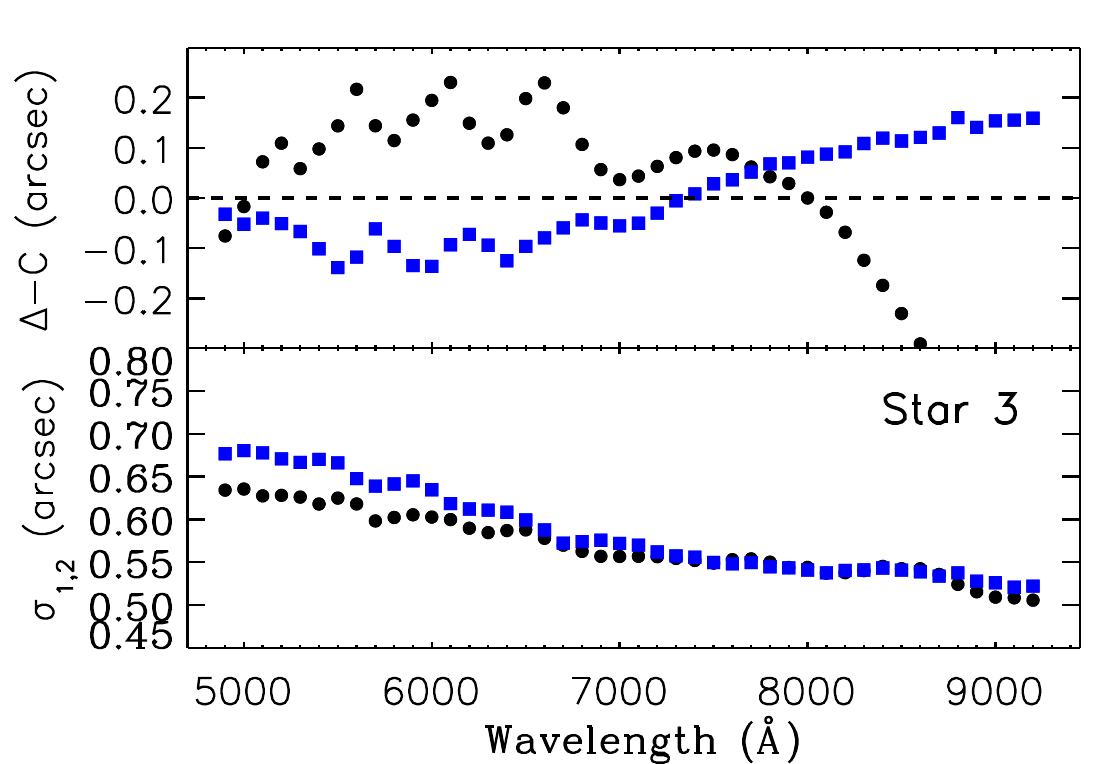}
\includegraphics[trim={0.5cm 12.cm 2.5cm 3.cm},clip,width=8cm]{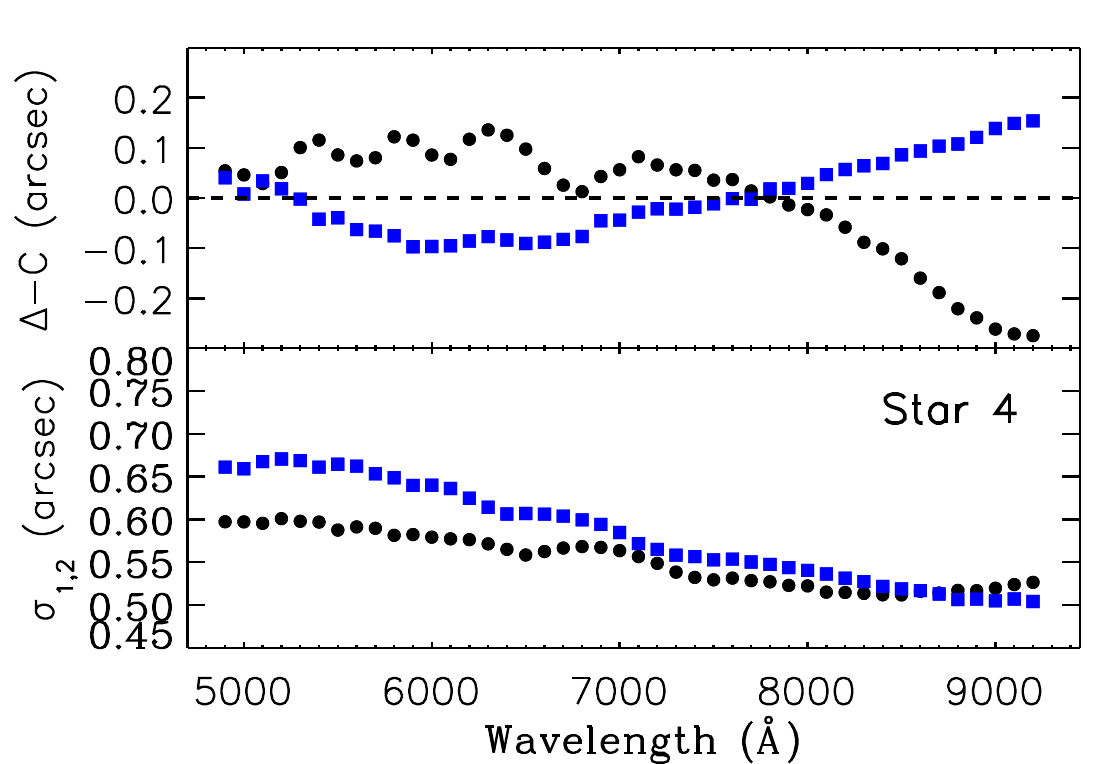}
\includegraphics[trim={0.5cm 12.cm 2.5cm 3.cm},clip,width=8cm]{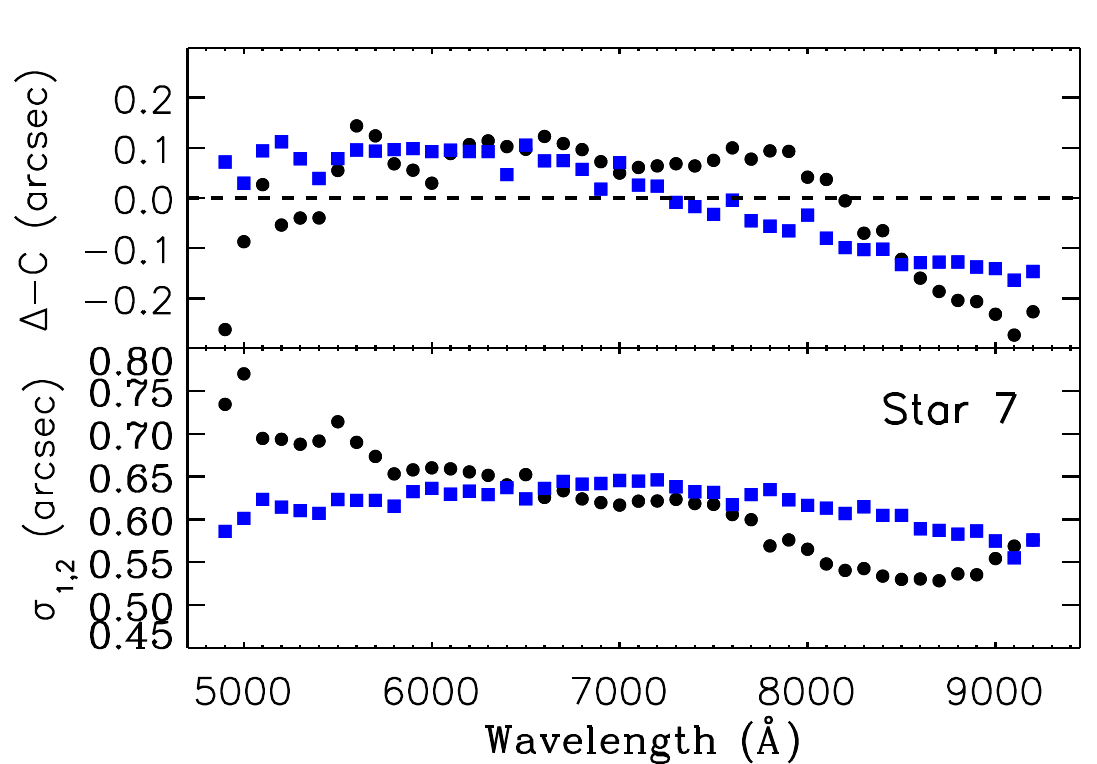}
\includegraphics[trim={0.5cm 12.cm 2.5cm 3.cm},clip,width=8cm]{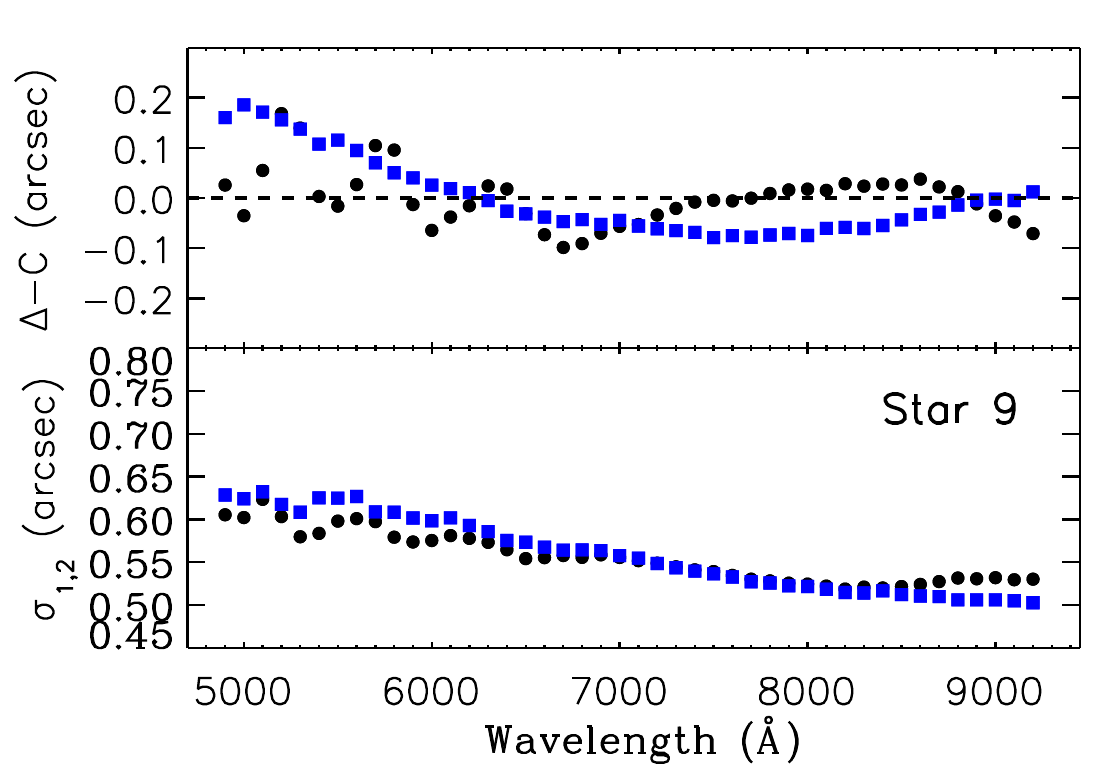}
\includegraphics[trim={0.5cm 12.cm 2.5cm 3.cm},clip,width=8cm]{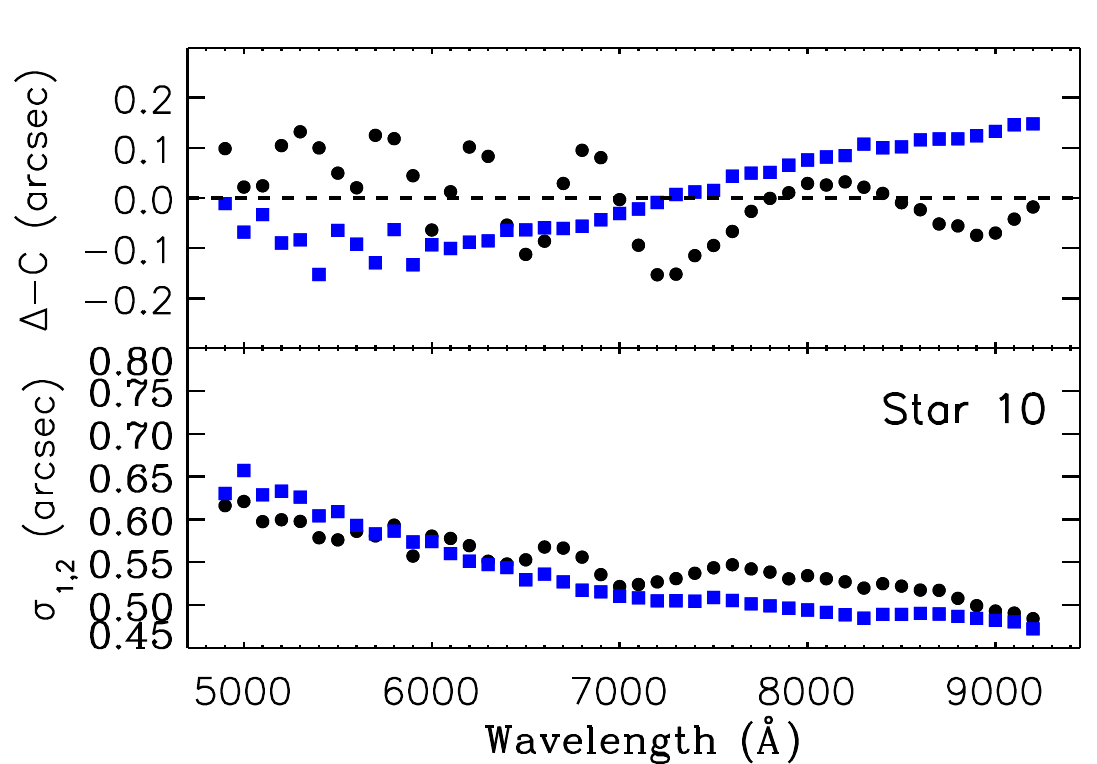}

 \caption{Same as Fig. \ref{HD90177a_parameters} but for the MUSE data cube of the OGLEII field.}
\label{OGLEII_parameters}
\end{center}
\end{figure*}

The MUSE field labeled as 'OGLEII' seems to correspond to an observation of the 'OGLEII dia bul-sc01 1821' star according to the MUSE data cube header. This star is a Mira variable that is included in the X-Shooter spectral library \citep{2020Gonneau}. Located in the galactic bulge region, the MUSE data cube shows a high density of stars of different brightness (Fig. \ref{MUSE_fields}-right). We found 10 stars in the MUSE field without companions brighter than a 5-sigma background within a 5 arcsec-radius region. Among them, we excluded three stars because the surrounding background revealed the presence of faint stars, impacting the fitting of the Moffat model. Additionally, we discarded one star due to a poor S/N in at least 50\% of the defined filter bands. In Fig. \ref{OGLEII_parameters}, we show the parameters against the central wavelengths of the respective filter for the Moffat model profiles fitted to the filter-band images of the remaining six stars. We observe wavelength-dependent variations in the centroids within the $\pm$0.2 arcsecs of the MUSE spatial pixel for most of the measurements, with the largest deviations along the horizontal axis. As expected, the Moffat model's widths decrease as wavelengths increase. For the analyzed stars in this field, Star-7 displays the most significant deviation from a circular Moffat model on average.

The values for $\epsilon_{0}^{b}$ derived from the analysis of the IQ for each star in the OGLEII field are compatible, with an average and standard deviation of $0.94\pm0.05$ arcsec (see Table \ref{tabla_parameters_stars}). The average and standard deviation of the $\mathcal{L}_{0}^{b}$ values derived from each star are $16.3\pm7.6$ m.

\begin{figure*}
\begin{center}
\includegraphics[trim={3cm 13.cm 2.5cm 1.75cm},clip,width=8cm]{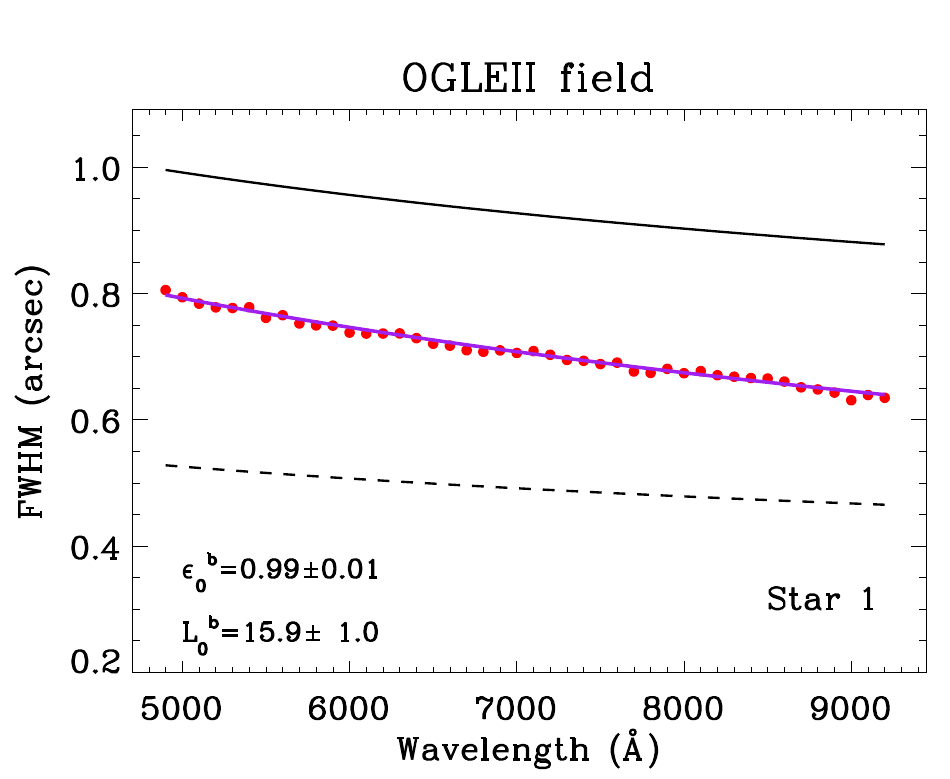}
\includegraphics[trim={3cm 13.cm 2.5cm 1.75cm},clip,width=8cm]{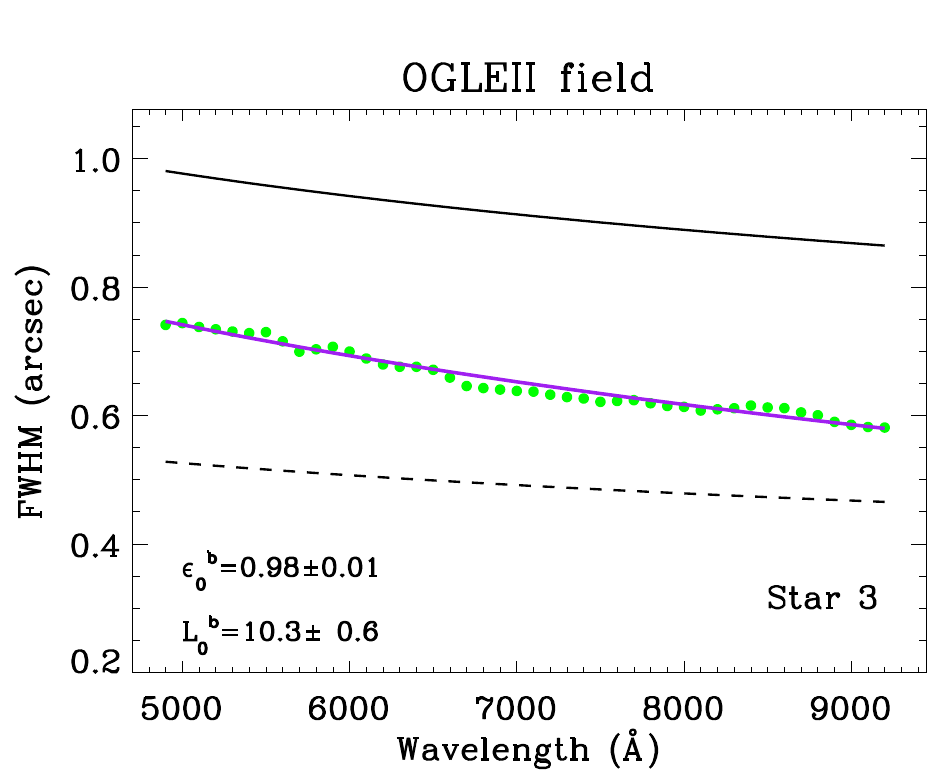}
\includegraphics[trim={3cm 13.cm 2.5cm 1.75cm},clip,width=8cm]{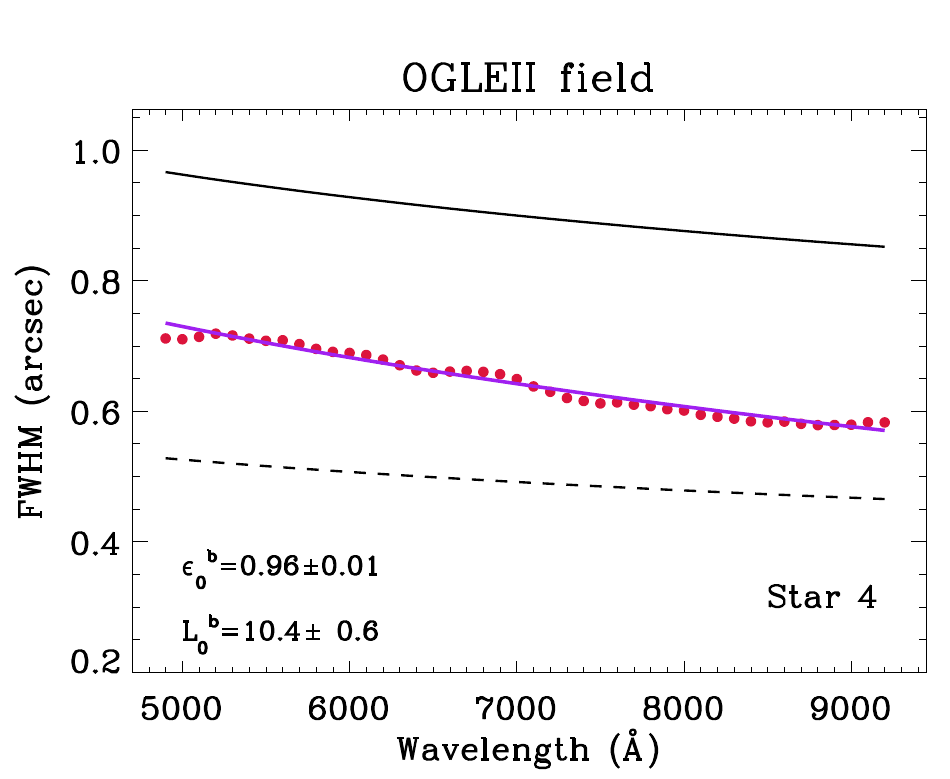}
\includegraphics[trim={3cm 13.cm 2.5cm 1.75cm},clip,width=8cm]{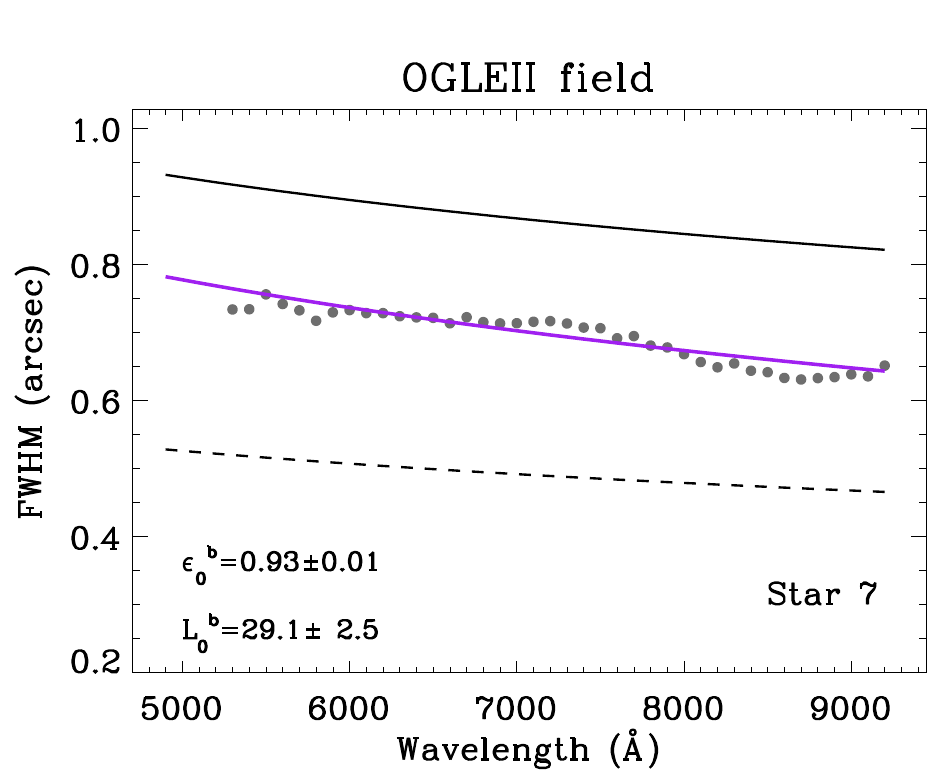}
\includegraphics[trim={3cm 13.cm 2.5cm 1.75cm},clip,width=8cm]{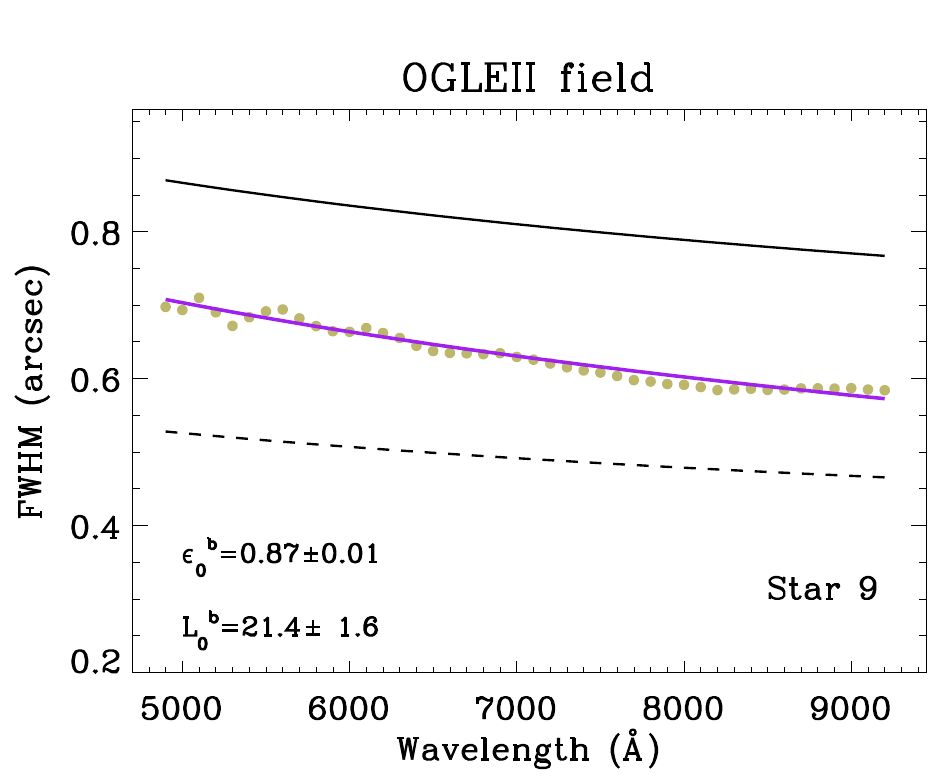}
\includegraphics[trim={3cm 13.cm 2.5cm 1.75cm},clip,width=8cm]{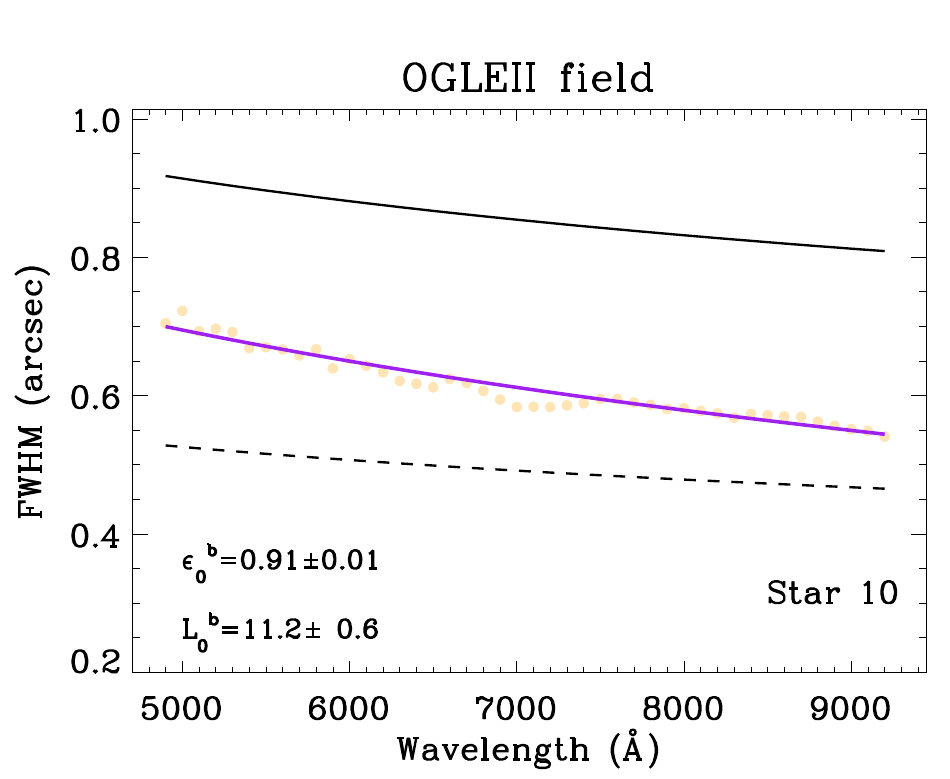}

 \caption{Same as Fig. \ref{HD90177a_stars_fit}  but for the MUSE data cube of the OGLEII field.}
\label{OGLEII_stars_fit}
\end{center}
\end{figure*}

\end{appendix}

\end{document}